\documentclass[usenatbib]{mn2e}
\usepackage{graphicx}
\usepackage{epstopdf}
\usepackage{natbib}
\usepackage{verbatim}
\usepackage{amssymb,amsmath}
\usepackage{ulem}
\usepackage[draft]{hyperref}
\usepackage{color}
\usepackage{rotating}
\usepackage[mathscr]{eucal}
\usepackage{float}
\usepackage{subcaption}  
\extrafloats{100}

\pdfoutput=1

\renewcommand{\thefootnote}{\fnsymbol{footnote}}

\newcommand\aj{AJ}
\newcommand\araa{ARA\&A}
\newcommand\apj{ApJ}
\newcommand\apjl{ApJ}
\newcommand\apjs{ApJS}
\newcommand\aap{A\&A}
\newcommand\mnras{MNRAS}
\newcommand\pasj{PASJ}

\title[The ISM in Interacting Galaxies]{Interacting galaxies on FIRE-2: The connection between enhanced star formation and interstellar gas content}

\author[J. Moreno et al.]
       {\parbox{18cm}{Jorge Moreno$^{1,2,3}$\footnotemark[1], Paul~Torrey$^{4,5,\dagger}$, Sara~L. Ellison$^{6}$, David~R. Patton$^{7}$, Philip F. Hopkins$^{3}$, Michael Bueno$^{2,8,9,10}$, Christopher C. Hayward$^{11}$, Desika Narayanan$^5$,  Du\v{s}an Kere\v{s}$^{12}$, Asa F.~L.~Bluck$^{13,14,15}$ and Lars Hernquist$^2$
       }\vspace{0.3cm}\\ 
         $^1$Department of Physics and Astronomy, Pomona College, Claremont, CA 91711, USA\\
         $^2$Harvard-Smithsonian Center for Astrophysics, 60 Garden Street, Cambridge, MA, 02138, USA\\ 
         $^3$TAPIR, Mailcode 350-17, California Institute of Technology, Pasadena, CA 91125, USA \\
         $^4$MIT Kavli Institute for Astrophysics \& Space Research, Cambridge, MA, 02139, USA\\
         $^5$Department of Astronomy, University of Florida, 211 Bryant Space Sciences Center, Gainesville, FL, USA\\
         $^6$Department of Physics \& Astronomy, University of Victoria, Finnerty Road, Victoria, British Columbia, V8P 1A1, Canada\\
         $^7$Department of Physics \& Astronomy, Trent University, 1600 West Bank Drive, Peterborough, Ontario, K9J 7B8, Canada\\
         $^8$Department of Physics and Astronomy, Haverford College, 370 Lancaster Ave, Haverford, PA 19041\\
         $^9$Department of Physics \& Astronomy, Northwestern University, Evanston, IL 60202, USA\\
         $^{10}$Center for Interdisciplinary Exploration \& Research in Astrophysics (CIERA), Evanston, IL 60202, USA\\
         $^{11}$Center for Computational Astrophysics, Flatiron Institute, 162 Fifth Avenue, New York, NY 10010, USA\\   
         $^{12}$Department of Physics, Center for Astrophysics and Space Sciences, University of California at San Diego, 9500 Gilman Drive, La Jolla,\\ \,\,\,\, \,CA 92093, USA\\
         $^{13}$Kavli Institute for Cosmology \& Cavendish Astrophysics, University of Cambridge, Madingley Road, Cambridge, CB3 0HA, UK\\
         $^{14}$Hughes Hall College, University of Cambridge, Wollaston Road, CB1 2EW, UK \\
         $^{15}$Swiss Federal Institute of Technology: ETH Zurich, Department of
Physics, Wolfgang-Pauli Strasse 27, CH-8097 Zurich, Switzerland \\
         $\dagger$Hubble Fellow
         }         

\begin{document}

\maketitle

\begin{abstract}
We present a comprehensive suite of high-resolution (parsec-scale), idealised (non-cosmological) galaxy merger simulations (24 runs, stellar mass ratio $\sim$2.5:1) to investigate the connection between interaction-induced star formation and the evolution of the interstellar medium (ISM) in various temperature-density regimes. We use the {\small GIZMO} code and the second version of the ``Feedback in Realistic Environments" model ({\small FIRE-2}), which captures the multi-phase structure of the ISM. Our simulations are designed to represent galaxy mergers in the local Universe. In this work, we focus on the `galaxy-pair period' between first and second pericentric passage. We split the ISM into four regimes: {\it hot}, {\it warm}, {\it cool} and {\it cold-dense}, motivated by the hot, ionised, atomic and molecular gas phases observed in real galaxies. We find that, on average, interactions enhance the star formation rate of the pair ($\sim$30$\%$, merger-suite sample average) and elevate their cold-dense gas content ($\sim$18$\%$). This is accompanied by a decrease in warm gas ($\sim$11$\%$), a negligible change in cool gas ($\sim$4$\%$ increase), and a substantial increase in hot gas ($\sim$400$\%$). The amount of cold-dense gas with densities above 1000 cm$^{-3}$ (the {\it cold ultra-dense} regime) is elevated significantly ($\sim$240$\%$), but only accounts for $\sim$0.15$\%$ (on average) of the cold-dense gas budget. 
\end{abstract}


\renewcommand{\thefootnote}{\fnsymbol{footnote}}
\footnotetext[1]{E-mail: jorge.moreno@pomona.edu}

\section{Introduction}
\label{sec:intro} 
Galaxy mergers and interactions are an integral component of any galaxy formation model. In the currently accepted $\Lambda$CDM cosmology, structures grow hierarchically, with small objects merging with each other and with larger objects to form the massive galaxies, groups, and clusters that exist in the Universe today \citep{White1991}. Specifically, observations suggest that galaxy-galaxy interactions enhance star formation rates \citep{SloanClosePairs,Ellison2013,Patton2013,Knapen2015}, decrease nuclear metallicities \citep{KGB06, Rupke2010, DilutionPeak2010, Torrey2012, Scudder2012}, and drive AGN activity~\citep{Ellison2011,Khabiboulline2014,Satyapal2014,Weston2017,Goulding2018}. On the numerical side, idealised (non-cosmological) binary merger simulations continue to play an important role in shaping our understanding of the merger process and its relative importance to galaxy formation \citep[e.g.,][]{BH91, BH96, MH96, Iono04,DiMatteo2007,DiMatteo2008,Teyssier2010,Torrey2012,Hayward2014,Moreno2015,Blumenthal2018}. 

\begin{figure*}
\centerline{\vbox{
\hbox{
\includegraphics[width=1.7in]{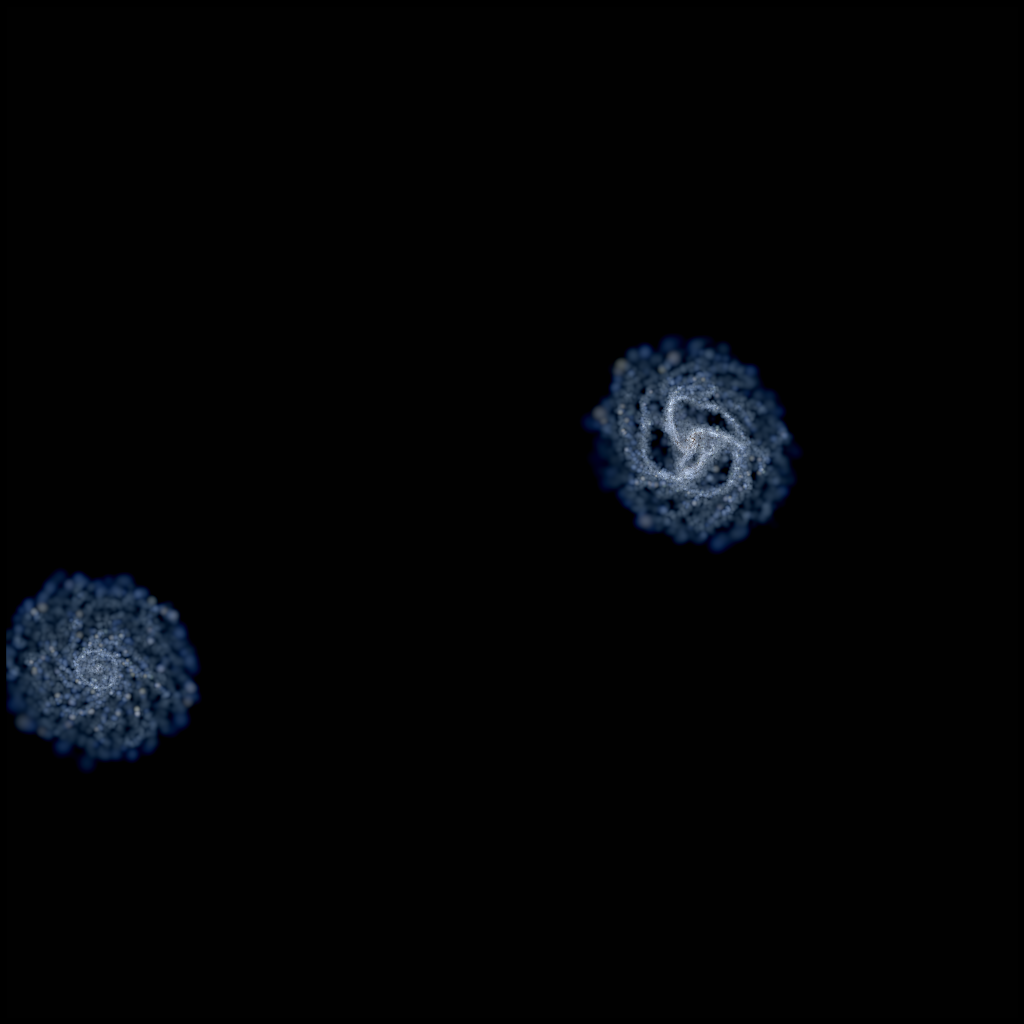}
\includegraphics[width=1.7in]{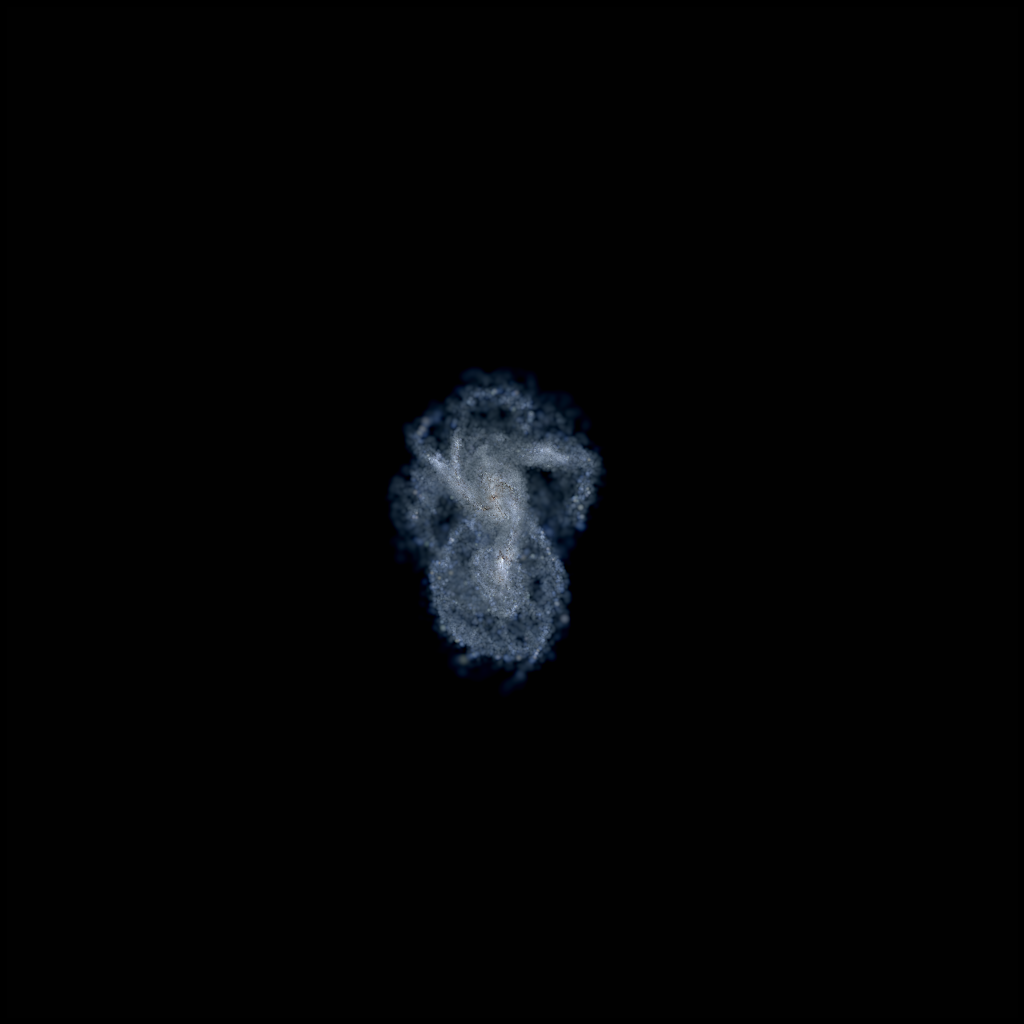}
\includegraphics[width=1.7in]{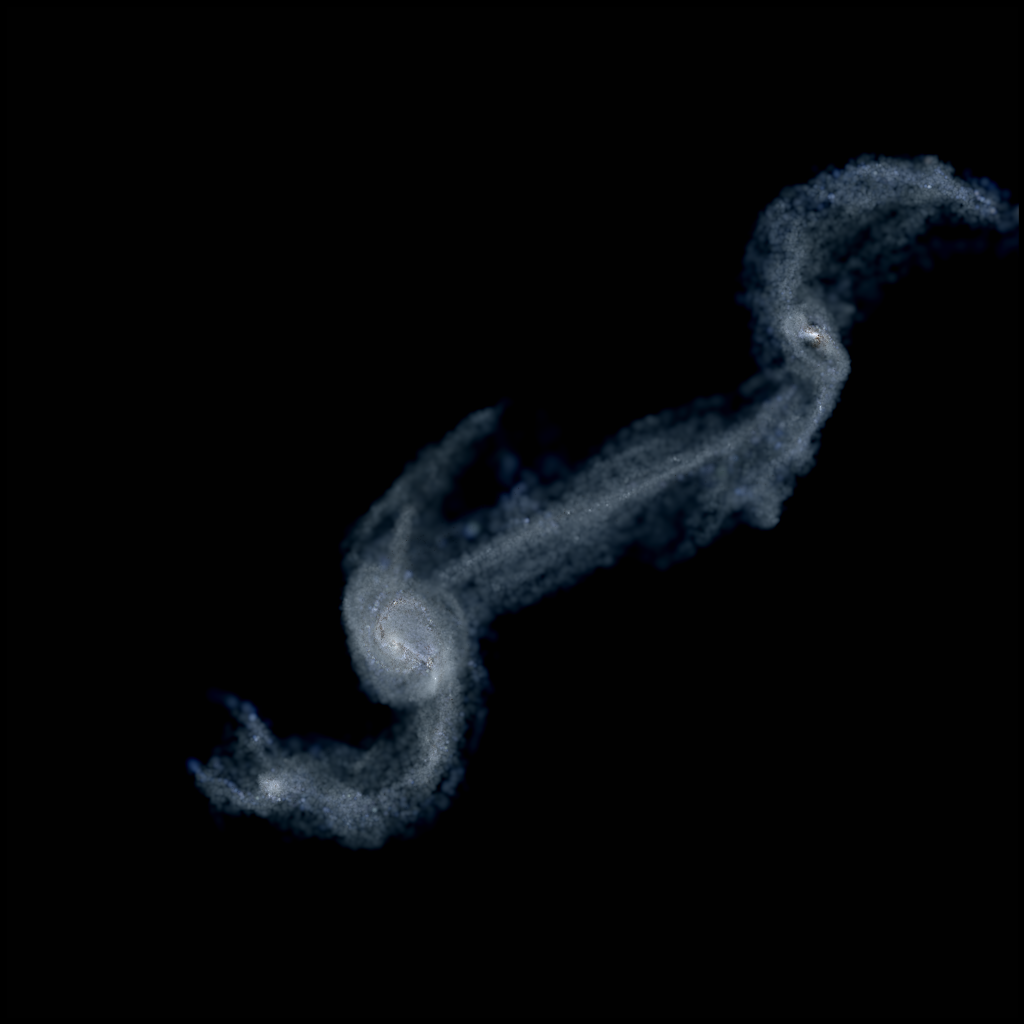}
\includegraphics[width=1.7in]{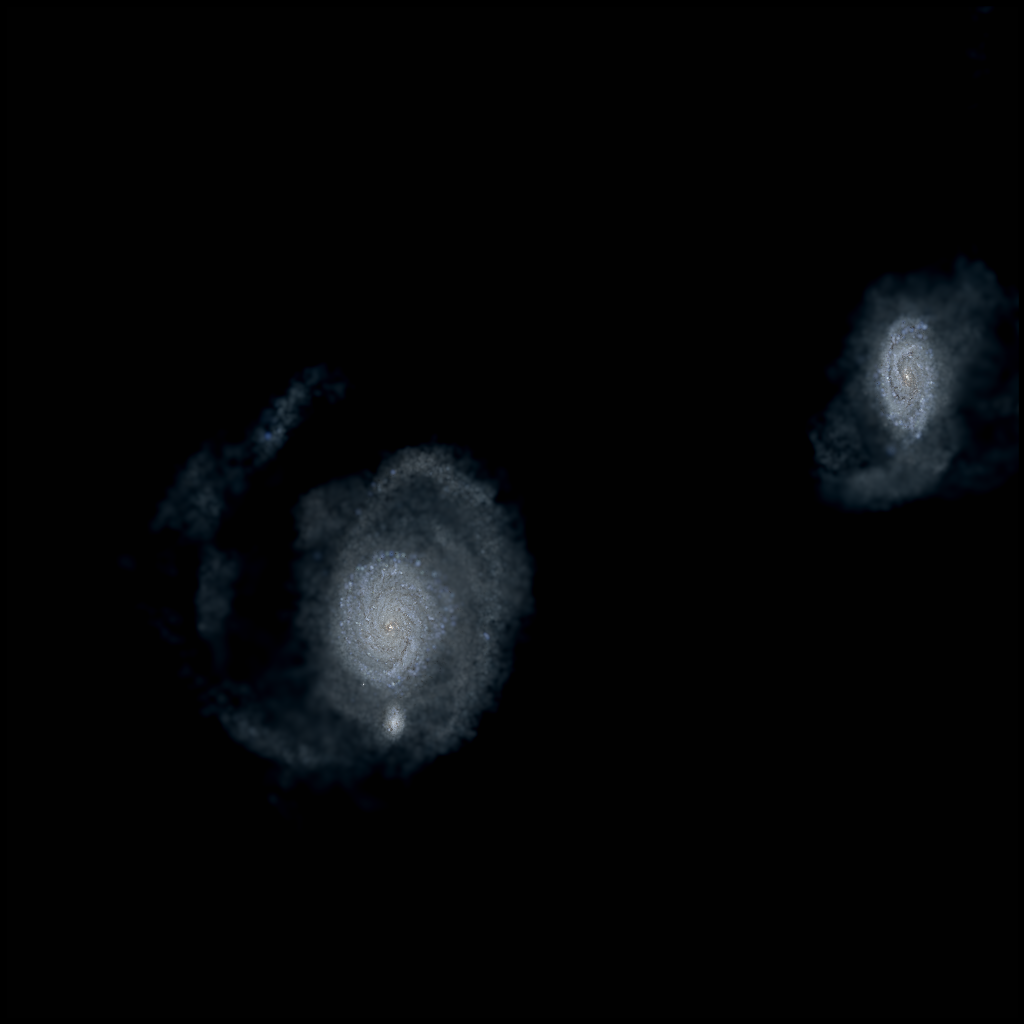}
}
\vspace{.02in}
\hbox{
\includegraphics[width=1.7in]{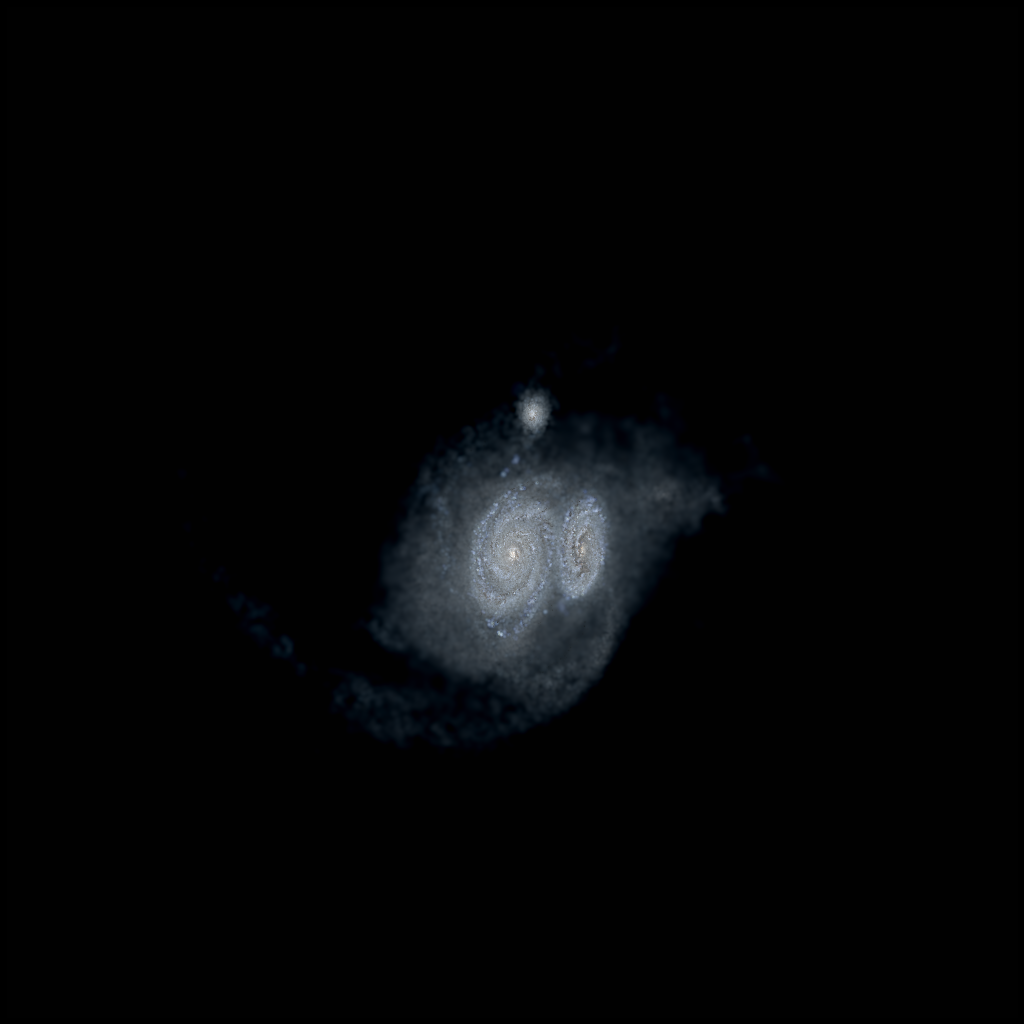}
\includegraphics[width=1.7in]{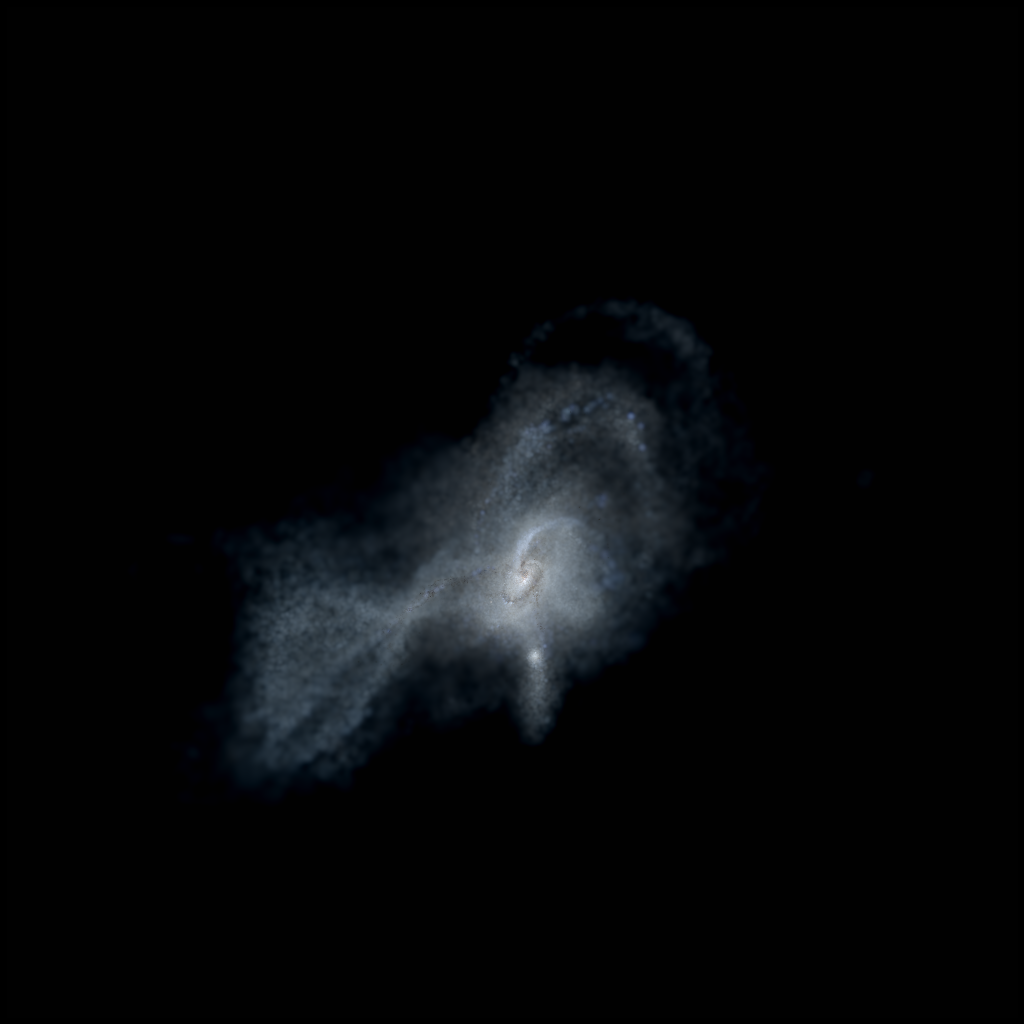}
\includegraphics[width=1.7in]{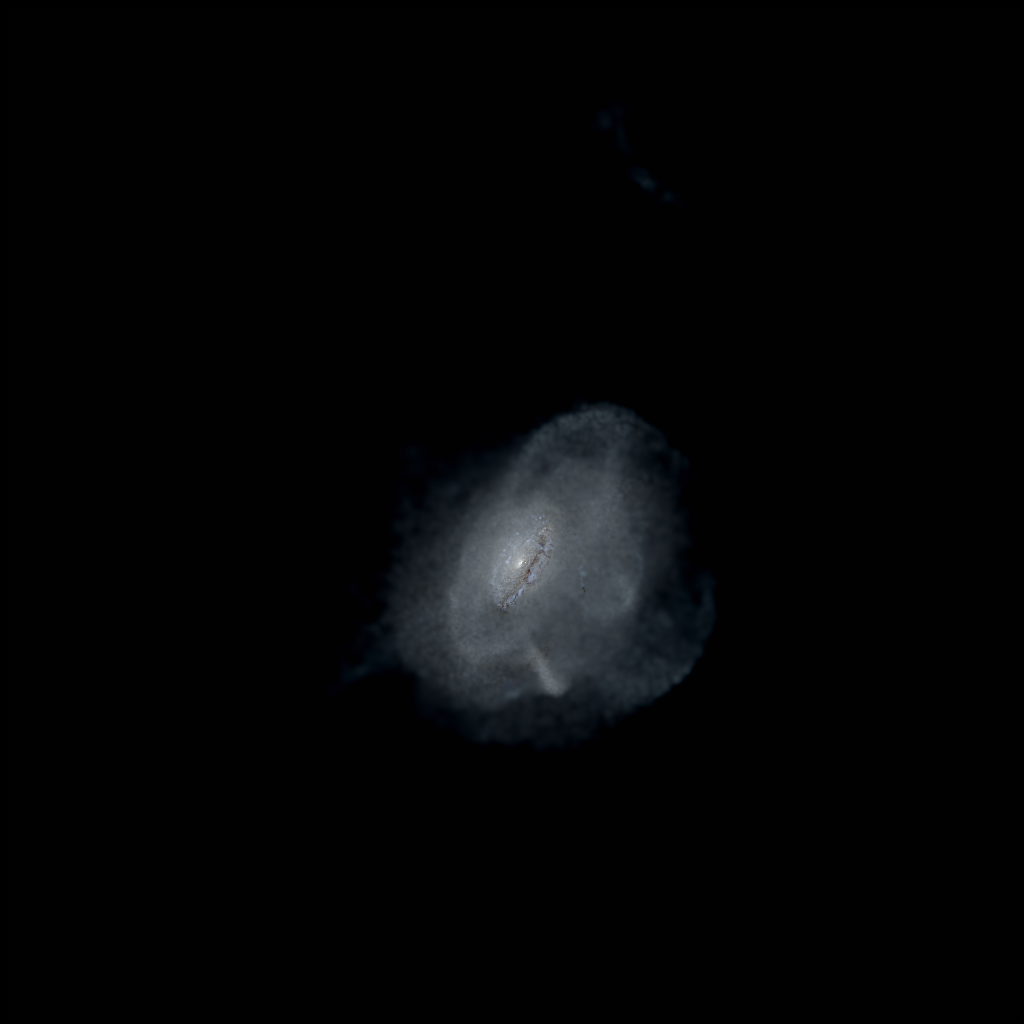}
\includegraphics[width=1.7in]{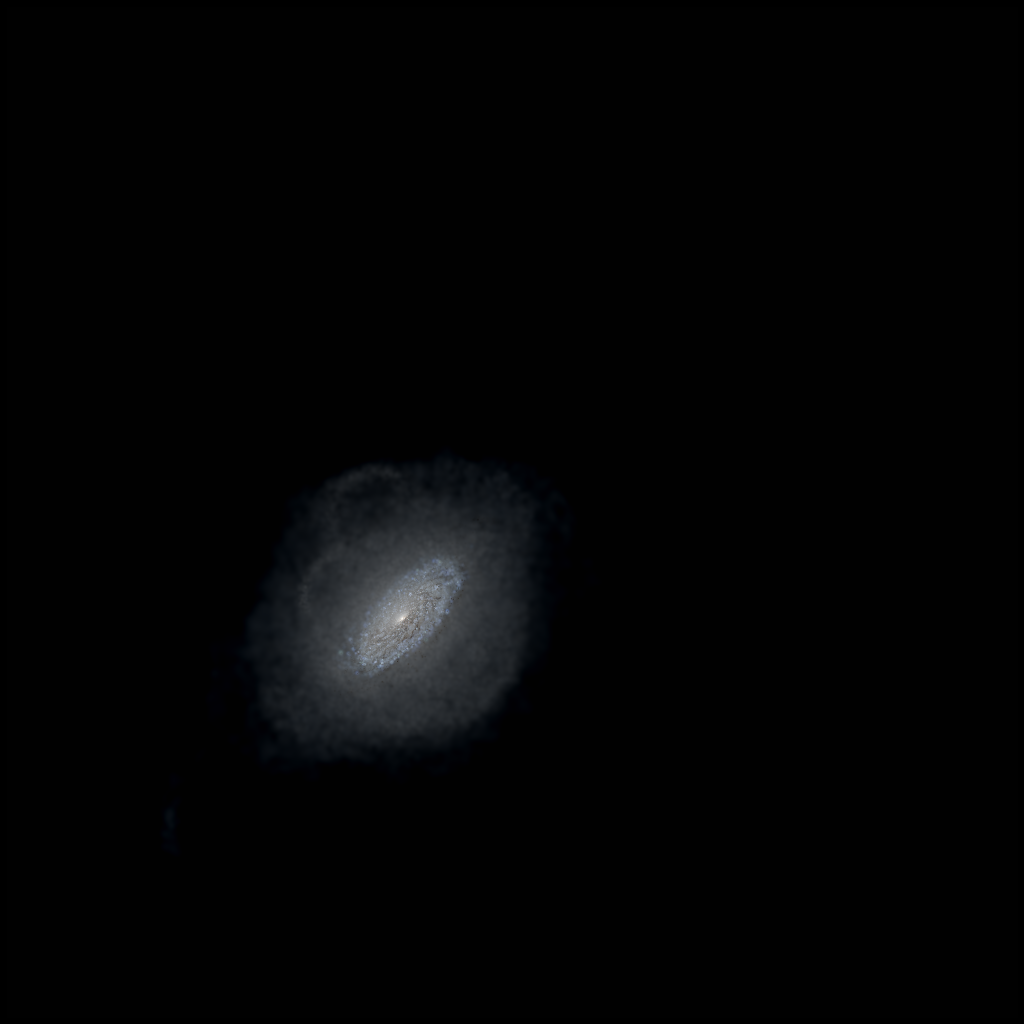}
}
\vspace{.1in}
\hbox{
\includegraphics[width=1.7in]{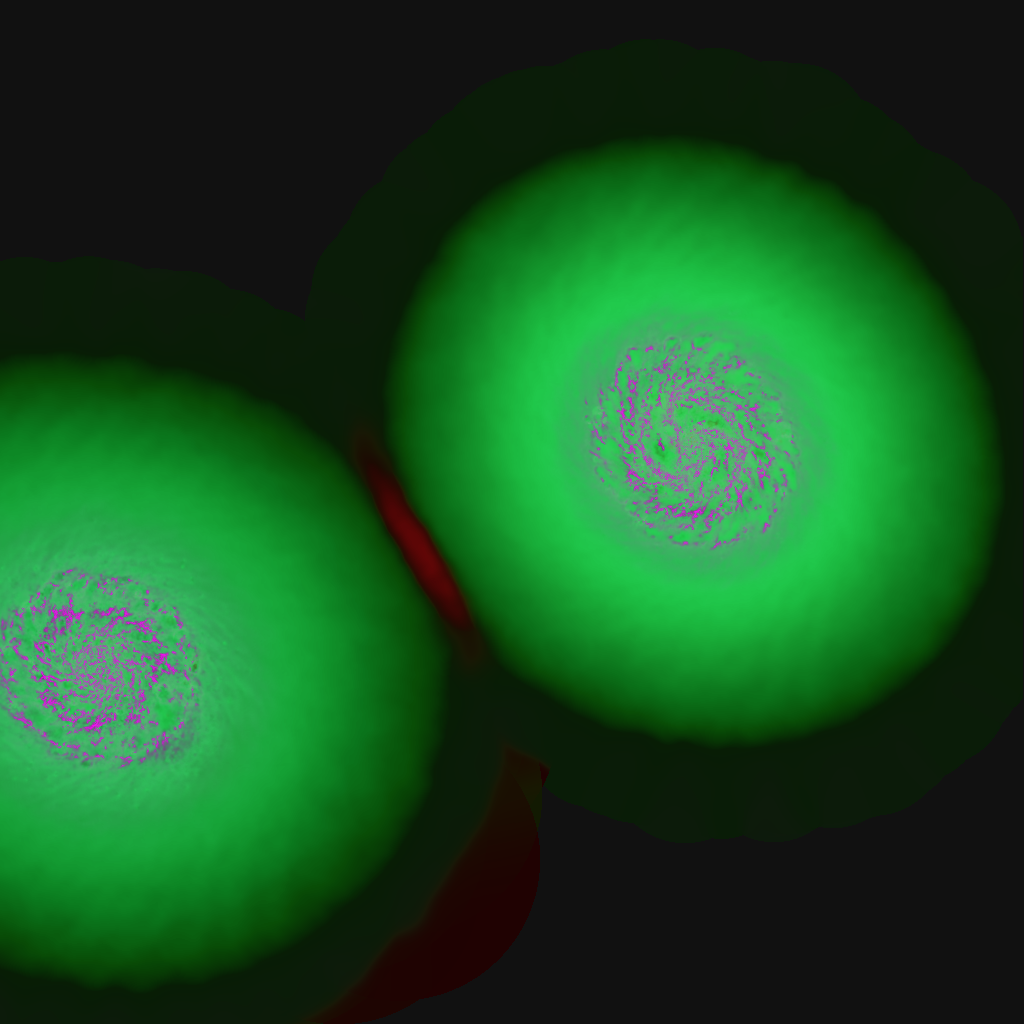}
\includegraphics[width=1.7in]{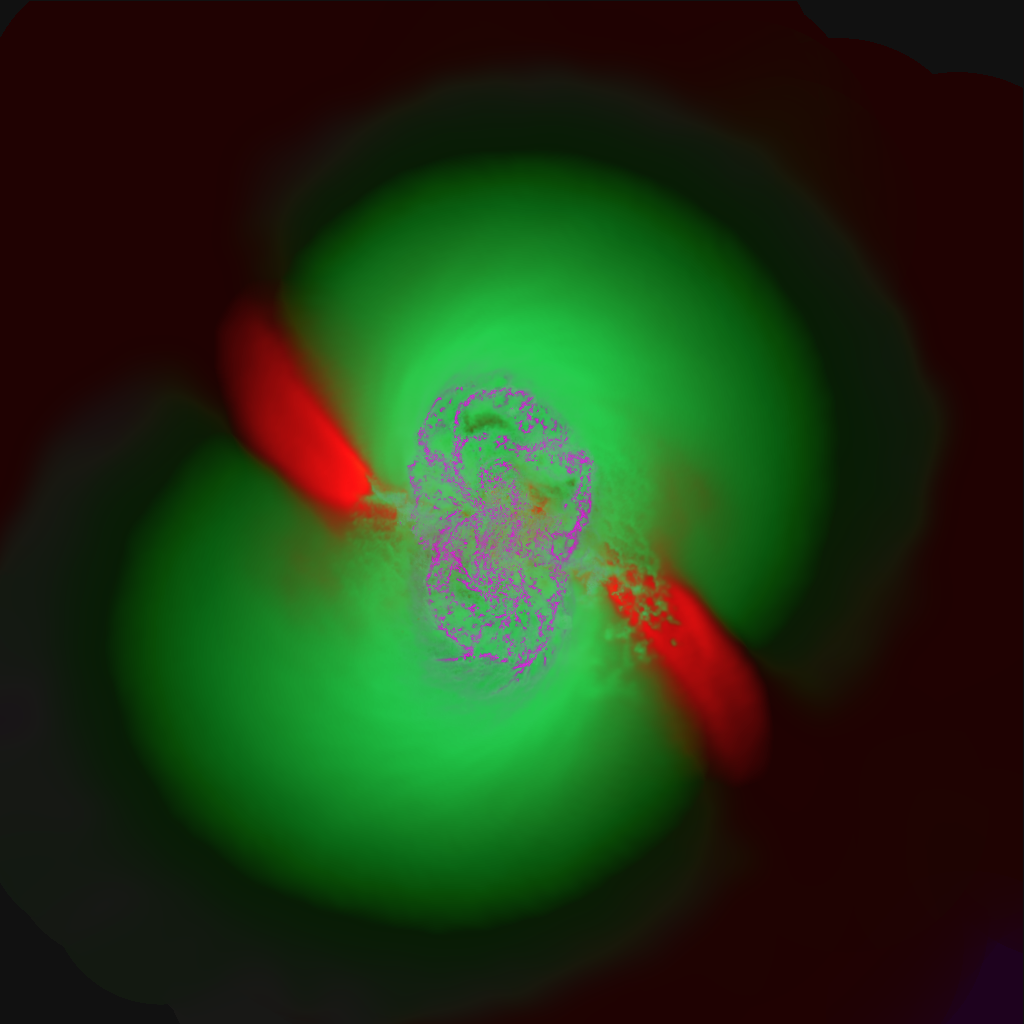}
\includegraphics[width=1.7in]{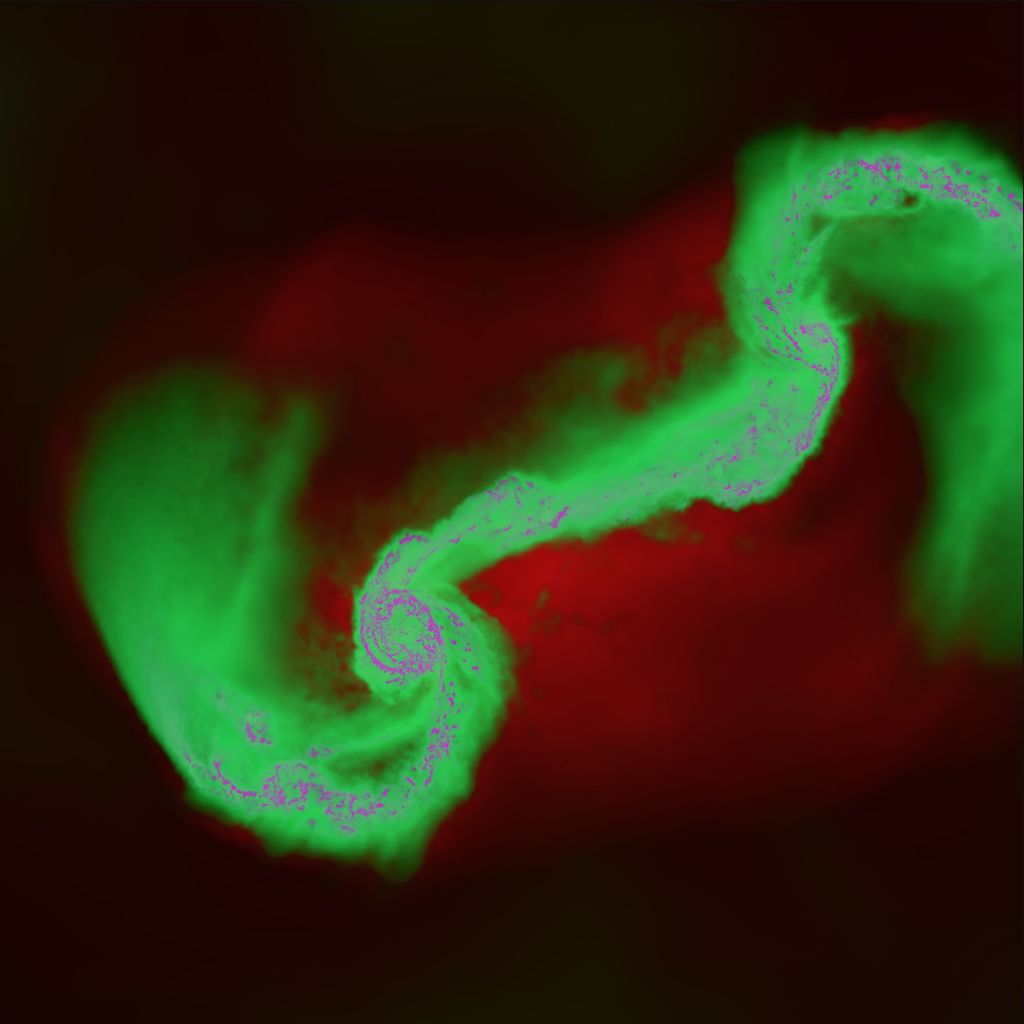}
\includegraphics[width=1.7in]{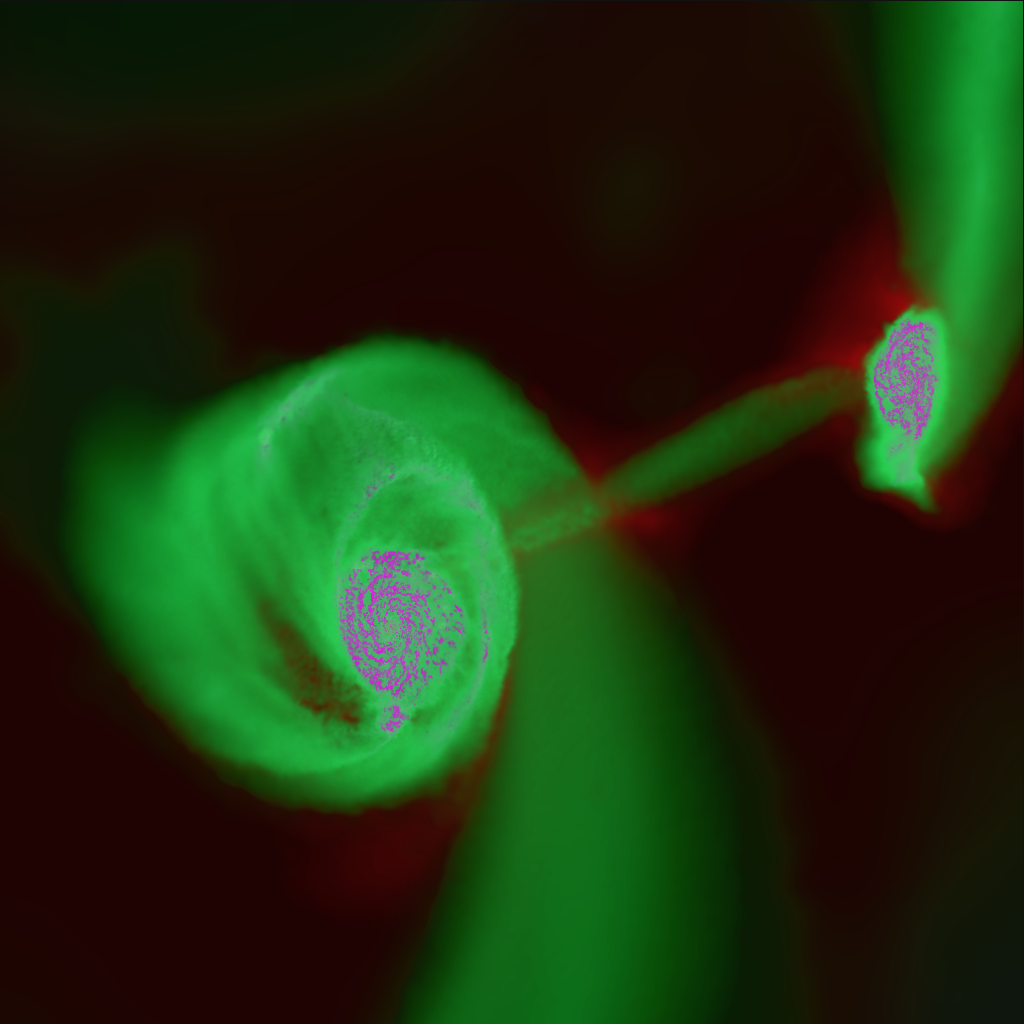}
}
\vspace{.02in}
\hbox{
\includegraphics[width=1.7in]{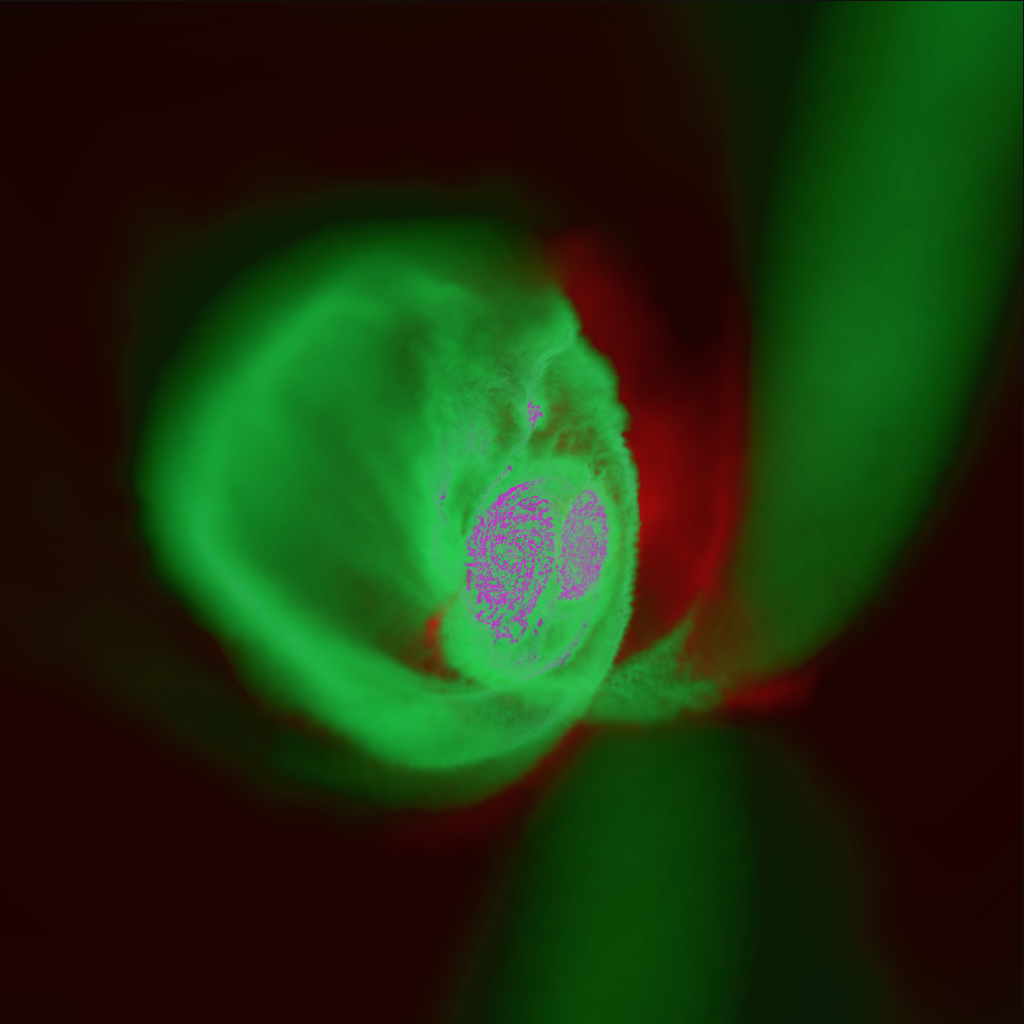}
\includegraphics[width=1.7in]{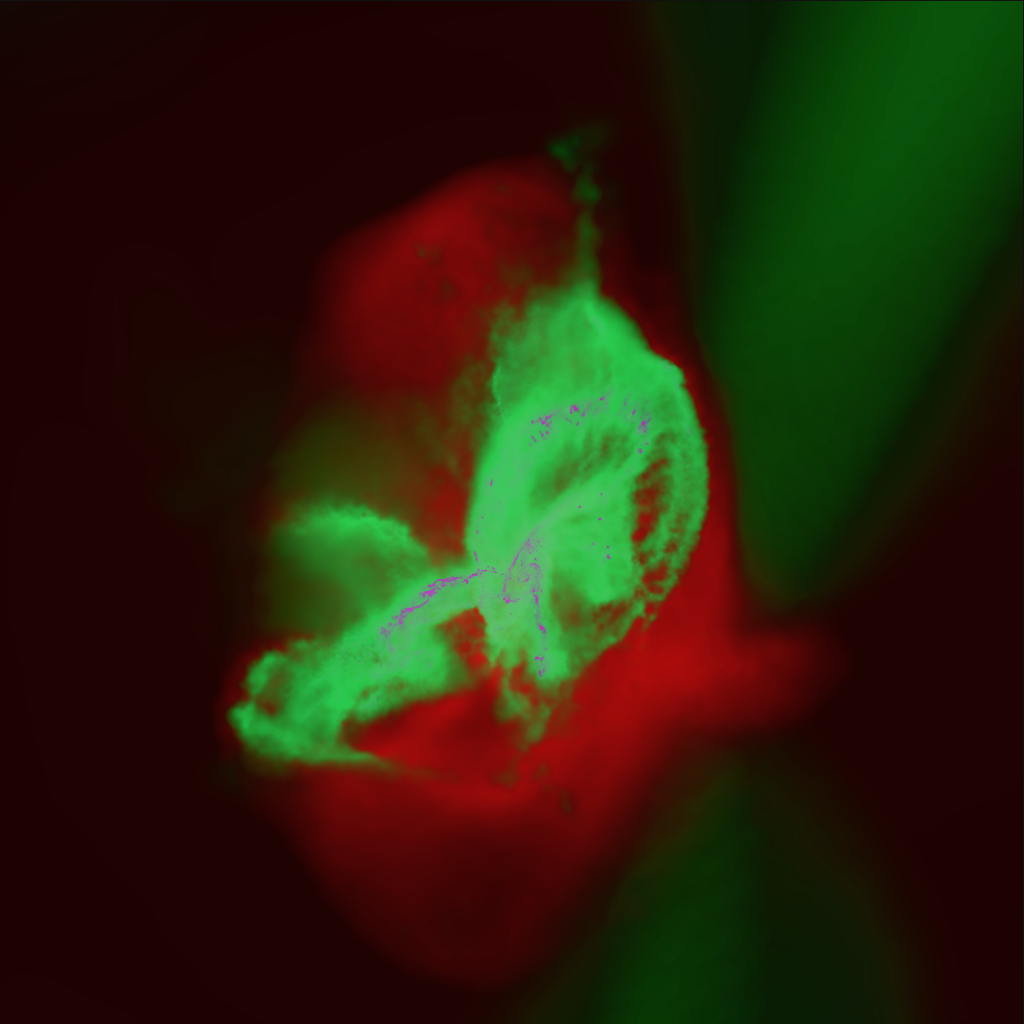}
\includegraphics[width=1.7in]{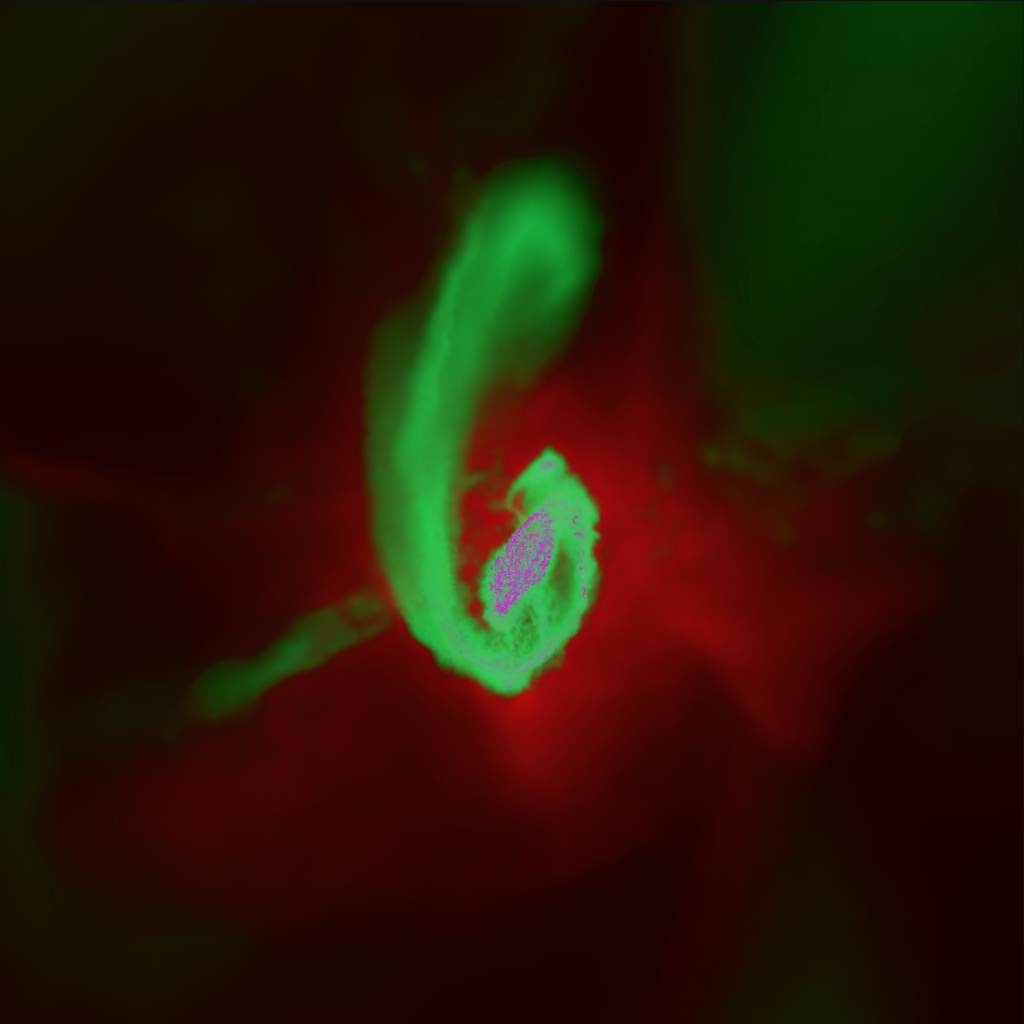}
\includegraphics[width=1.7in]{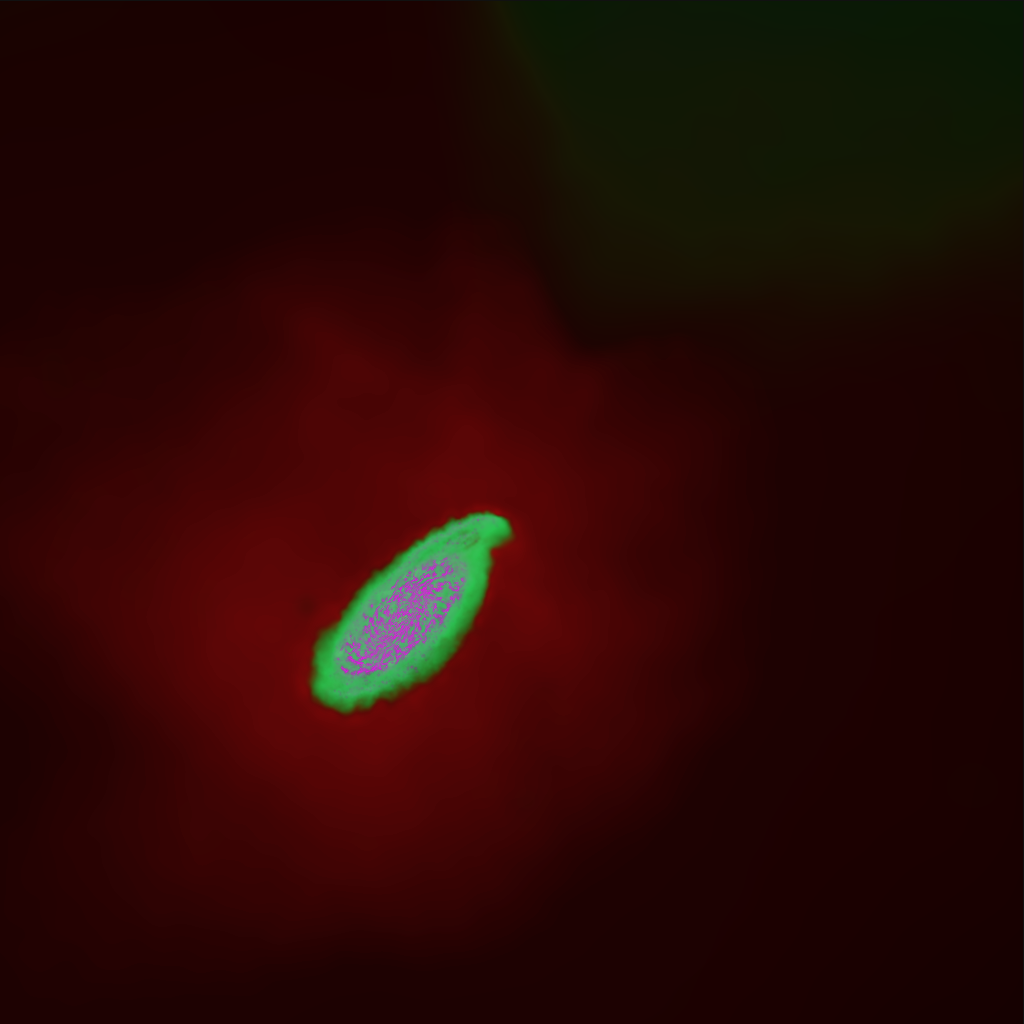}
}
\vspace{.0in}
}}
\caption{The galaxy interaction sequence. Field of view $=$ 100 kpc $\times$ 100 kpc. {\it Top two rows:} Stellar mock {\it ugr} (SDSS-band) composite images. {\it Bottom two rows:} Gas surface density, colour coded by temperature: magenta indicates gas temperatures $T < 1000\,K$, green refers to $T\sim 10^3-10^4\, K$, and red corresponds to $T>10^6 \,K$. After their first close encounter, the galaxies exhibit a bridge and tidal tails. After second pericentric passage and during coalescence, the morphology is disturbed until the remnant settles into a disk galaxy with faint stellar features and a hot gas halo. Videos are available at \url{http://research.pomona.edu/galaxymergers/}.
}
\label{fig:interaction_sequence}
\end{figure*}

A number of the previously published merger simulation suites employ fairly simple models for the interstellar medium (ISM). One traditional approach is to treat the multi-phase ISM as a single, pressurised fluid~\citep[e.g.,][]{SH03, Schaye2008, Murante2015}. This has the advantages of being well behaved numerically (i.e. convergent), computationally inexpensive, and of being based on known empirical relations \citep[e.g., the Schmidt-Kennicutt relation,][]{Schmidt, Kennicutt}, which can be imposed with care. The principal disadvantage of this approach is that the predictive power of the numerical models is restricted by requiring strict assumptions to be made. For example, in~\citet{Torrey2012}, the assumed stiffness of the ISM effective equation of state, as well as the mass loading factor for galactic winds, both impact the obtained results. 

Models employing a barotropic equation of state have recently pushed resolution to levels capable of resolving Giant Molecular Clouds (GMCs) and the structure of the ISM \citep{Bournaud2008,Bournaud2010,Renaud2008,Renaud2009, Teyssier2010,Renaud2013,Renaud2015}. These works highlight the importance of stellar feedback in regulating the ISM structure. Likewise, increasingly detailed models that capture the multi-phase structure of the ISM and adopt feedback-regulated star formation have been developed~\citep[e.g.,][]{Hopkins2011a, Agertz2012, FIRE, FIRE2}. A key characteristic of these models is a desire to resolve the scales where feedback from young stellar populations first have a global impact on the pressurization of the interstellar medium, whilst simultaneously adopting realistic physical prescriptions at those scales. Since turbulent pressure is believed to play a key role in preventing run-away ISM gas collapse~\citep{Ostriker2011,Hayward2017}, resolving the injection of momentum and energy sources that drives the supersonic turbulent ISM pressure is key to obtaining a self-regulating ISM. Moreover, it has been found that including local sources of stellar feedback (photoionisation, radiation pressure, supernovae energy/momentum injection, and stellar winds) naturally leads to the launching of large scale galactic outflows~\citep{HopkinsCox2013, HopkinsKeres2013}. The current disadvantage of these more detailed models is that they are computationally expensive, and so a large and comprehensive suite of galaxy merger simulations has not yet been published.

The goal of this paper is to employ modern, state-of-the-art, simulations to investigate the evolution of the ISM in different temperature-density regimes during the merger. We use the {\small GIZMO} simulation code \citep{GIZMO,GIZMO2} - in conjunction with the second version of the `Feedback In Realistic Environments' ({\small FIRE-2}) model \citep{FIRE2} - to build a comprehensive suite of high resolution galaxy merger simulations. This work expands on a smaller set of simulations by \citet{HopkinsCox2013,HopkinsKeres2013}, which use an earlier version of our model for only a handful of mergers. The suite we present here effectively replaces an earlier suite of GADGET-3 \citep{SH03} galaxy merger simulations created by our group \citep{Patton2013,Moreno2015}. 

Figure~\ref{fig:interaction_sequence} displays the various stages of the interaction sequence for one of our simulations. The top two rows show mock {\it ugr} SDSS-band composite images. The bottom two rows show the multi-phase structure of the ISM. This colour projection displays gas in terms of temperature in approximately three bins: $T < 1000\,K$ (magenta), $T\sim 10^3-10^4\, K$ (green), and $T>10^6 \,K$ (red). For this figure, we employ the same visualisation techniques as in \cite{FIRE2} and in other works by the FIRE collaboration. The interaction sequence unfolds as follows. The first pericentric passage produces tidal tails and a bridge (first and third rows), which are more evident in the gaseous component. The galaxies approach again and merge, displaying disturbed morphology first, and ultimately settling down into a disk galaxy with a prominent bulge. The post-merger remnant exhibits faint stellar features, such as shells and streams, as well as a hot gas halo. This work focuses on the early stages of interaction. In the pre-coalescence period, mergers can be relatively `cleanly' identified in observations by searching for galaxy pairs, whereas identifying and timing coalescencing galaxies is more challenging. Thus, by focusing on the pre-coalescence period (hereafter, the `galaxy-pair period'), we can more straightforwardly compare with observational constraints on the ISM structure of galaxy pairs. 

Section~\ref{sec:methods} reviews the {\small FIRE-2} model and describes our suite of galaxy merger simulations. Section~\ref{sec:results} describes the evolution of star formation, the various temperature-density gas regimes, and their interplay. We present our discussion and conclusions in Sections~\ref{sec:discussion} and~\ref{sec:conclusions}.\\

\section{Methods}
\label{sec:methods} 

\subsection{FIRE-2: The `Feedback In Realistic Environments' Physics Model (Version 2)}
\label{subsec:fire-2}
 
The {\small FIRE-2} model \citep{FIRE2} is very similar to its predecessor \citep[{\small FIRE},][]{FIRE}, but with updated hydrodynamic solvers and supernova coupling algorithms. Differences between these two versions of the {\small FIRE} model only have a minor impact on phase mixing in the circumgalactic medium, which is largely irrelevant for this paper. We provide a brief description of the model here, and refer the reader to the above two papers for details. Unlike most recent papers employing the {\small FIRE-2} model, this work is not the result of zoom-in simulations selected from a cosmological box. Instead, we use {\small FIRE-2} physics to run idealised (non-cosmological) galaxy merger simulations\footnote{For information on the \small{FIRE} Project, visit \url{https://fire.northwestern.edu}. Videos of our galaxy merger simulations are available at \url{research.pomona.edu/galaxymergers}.}.

The treatment of radiative heating and cooling includes free-free, photo-ionisation/recombination, Compton, photoelectric, dust-collisional, cosmic ray, molecular, metal-line and fine-structure processes -- and it tracks 11 species independently. The code accounts for UV background \citep{CAFG2009} and locally-driven photo-heating and self-shielding. Star formation is constrained to self-gravitating (at the resolution scale), self-shielded gas denser than 1000 cm$^{-3}$ \citep{Hopkins2013}. Stellar feedback mechanisms include momentum flux from radiation pressure; energy, momentum, mass and metal injection from SNe (Types Ia and II) and stellar mass loss (OB and AGB stars). {The location where supernovae go off can potentially have an impact on the ISM \citep{Gatto2015}. In our model, this is governed solely by the location of the star particle when the explosion occurs \citep{Hopkins2018sn}.} Other stellar feedback channels include photoionisation and photo-electric heating. Single stellar population properties are attributed to each star particle. We tabulate stellar masses, ages, metallicities, feedback event rates, luminosities, energies and mass-loss rates with \textsc{starburst99}, a stellar synthesis model \citep{Leitherer1999}. We assume a \cite{kroupa2001} initial mass function (IMF). Our simulations do not include feedback caused by supermassive black hole (SMBH) gas accretion because the coupling between this type of feedback and the circumnuclear ISM at the resolutions explored here is not well understood yet \citep[but see][]{Torrey2017}. For works on SMBH feedback with FIRE, see \citet{Hopkins2016} and \citet{Angles2017}. Due to the lack of observational constraints, we do not include hot gas atmospheres at the start of the simulation. In this paper we employ the Meshless Finite-Mass (MFM) version of {\small GIZMO}, which is useful to keep track of inter-phase gas transitions.

\subsection{Suite of Galaxy Merger Simulations}
\label{subsec:suites}

The suite in this paper is similar to a previous merger suite created by our group \citep{Patton2013,Moreno2015}, but with fewer runs, substantially higher resolution, and with a new physical model. Namely, {\small GIZMO/FIRE-2} \citep{FIRE2} replaces an effective equation-of-state model with {\small GADGET-3} \citep{SH03,GADGET}. For a thorough comparison of galaxy merger simulations using GADGET-3 and {\small FIRE}, we point the reader to Figure 9 of \cite{HopkinsCox2013}, which shows the bursty nature of SFR in this new model, in line with our work (Section~\ref{subsec:results_fiducial_sfr}).  Differences between the two versions of {\small FIRE} are largely insignificant in the regimes we explore in this paper. Table~\ref{table:specs} displays our adopted specifications. Our MFM particle masses are $1.9\times10^{5}$ $M_{\odot}$, $1.4\times10^{4}$ $M_{\odot}$ and $8.4\times10^{3}$ $M_{\odot}$ (dark matter, gas and stellar components). The highest gas density and spatial resolution achieved in our simulations are $5.8 \times10^{5}$ ${\rm cm}^{-3}$ and  $1.1$ pc, respectively. Our gravitational softening are 0.01 kpc for the dark matter and stellar components, and 0.001 kpc for the gaseous component.

\begin{table}
  \begin{center}
    \begin{tabular}{lllll} 
        \hline 
        Property & Value \\
        \hline   \hline 
   Mass resolution (dark matter)  & $1.9\times10^{5}$ $M_{\odot}$\\
   Mass resolution (gas)  & $1.4\times10^{4}$ $M_{\odot}$\\
   Mass resolution (stars)  & $8.4\times10^{3}$ $M_{\odot}$\\
   Highest gas density &  $5.8 \times10^{5}$ ${\rm cm}^{-3}$\\
   Highest spatial gas resolution &  $1.1$ pc\\
   Gravitational softening (collisionless) &  0.01 kpc\\
   Gravitational softening (gas) &  0.001 kpc\\
       \hline
    \end{tabular}
  \end{center}
\caption{Simulation specifications: mass resolution (dark matter, gas and stars), highest density and spatial resolution (gas), and gravitational softening (collisionless components and gas).
}
\label{table:specs}
\end{table}
\begin{table}
  \begin{center}
    \begin{tabular}{llllll|} 
        \hline 
        Property  & Primary & Secondary \\
        \hline   \hline 
   $M_{\rm halo}$ & $7.5\times10^{11} M_{\odot}$ & $3.5\times10^{11} M_{\odot}$\\
   $M_{\rm stellar}$ & $3.0\times10^{10} M_{\odot}$ & $1.2\times10^{10} M_{\odot}$\\
   $M_{\rm bulge}$ & $2.5\times10^{9} M_{\odot}$ & $7.0\times10^{8} M_{\odot}$ \\
   $M_{\rm gas}$ & $8.0\times10^{9} M_{\odot}$ & $7.0\times10^{9} M_{\odot}$ \\
   $\lambda_{\rm halo}$ & 0.05 & 0.05 \\
   $R_{\rm disk}$ & 2.85 ${\rm kpc}$ & 1.91 ${\rm kpc}$\\
   $h_{\rm disk \, (bulge)}/R_{\rm disk}$ & 0.14  (0.13) & 0.2 (0.136) \\
       \hline
    \end{tabular}
  \end{center}
\caption{Properties of progenitor galaxies: dark matter halo mass, stellar mass, bulge mass, total gas mass, halo spin parameter, stellar disk radial scale length, thickness of stellar disk in units of radial scale length and bulge scale length in units of disk scale length.
}
\label{table:iso}
\end{table}

\subsubsection{Isolated Galaxies}
\label{subsubsec:isosuite}

Isolated galaxies in our simulations are set up following analytic work by \citet{Mo1998}, employing the procedure outlined by \citet{Springel2005a}. Simulated bulges and dark matter haloes follow a \citet{Hernquist1990} profile. Table~\ref{table:iso} describes our two progenitor galaxies, hereafter labelled `primary' and `secondary' (stellar mass ratio $\sim$2.5:1). As we increase the stellar mass, we adjust the corresponding dark matter halo mass, following abundance matching results \citep{Moster2013}. We also adjust the bulge mass so that the bulge-to-total mass ratio decreases, in line with trends in the Sloan Digital Sky Survey (SDSS) \citep{Mendel2014}. We follow \cite{Saintonge2016} to set up our gas masses. For both galaxies, we set up our gaseous components to have disc lengths of $\sim$10 kpc, which yield diameters of $\sim$25 kpc (defined as the radial extents at which the gas surface density falls beneath 1 $M_{\odot}$ pc$^{-1}$), in line with observations by \citet{Broeils1997}. We also list values for the halo spin parameter, the stellar disk radial scale length, the thickness of stellar disk in units of radial scale length and bulge scale length in units of disk scale length. For a more thorough description on our adopted radial profiles (dark matter halo, stellar disc and bulge, gaseous disc) and the vertical structure of the gas disc, please read Section 2 of \citet{Springel2005a}. For reproducibility purposes, we point the interested reader to the following website for a complete list of parameters: \url{research.pomona.edu/galaxymergers/isolated-disks-initial-conditions/}.

\begin{figure}
\centerline{\vbox{
\vspace{-.12in}
\hbox{
\includegraphics[width=3.5in]{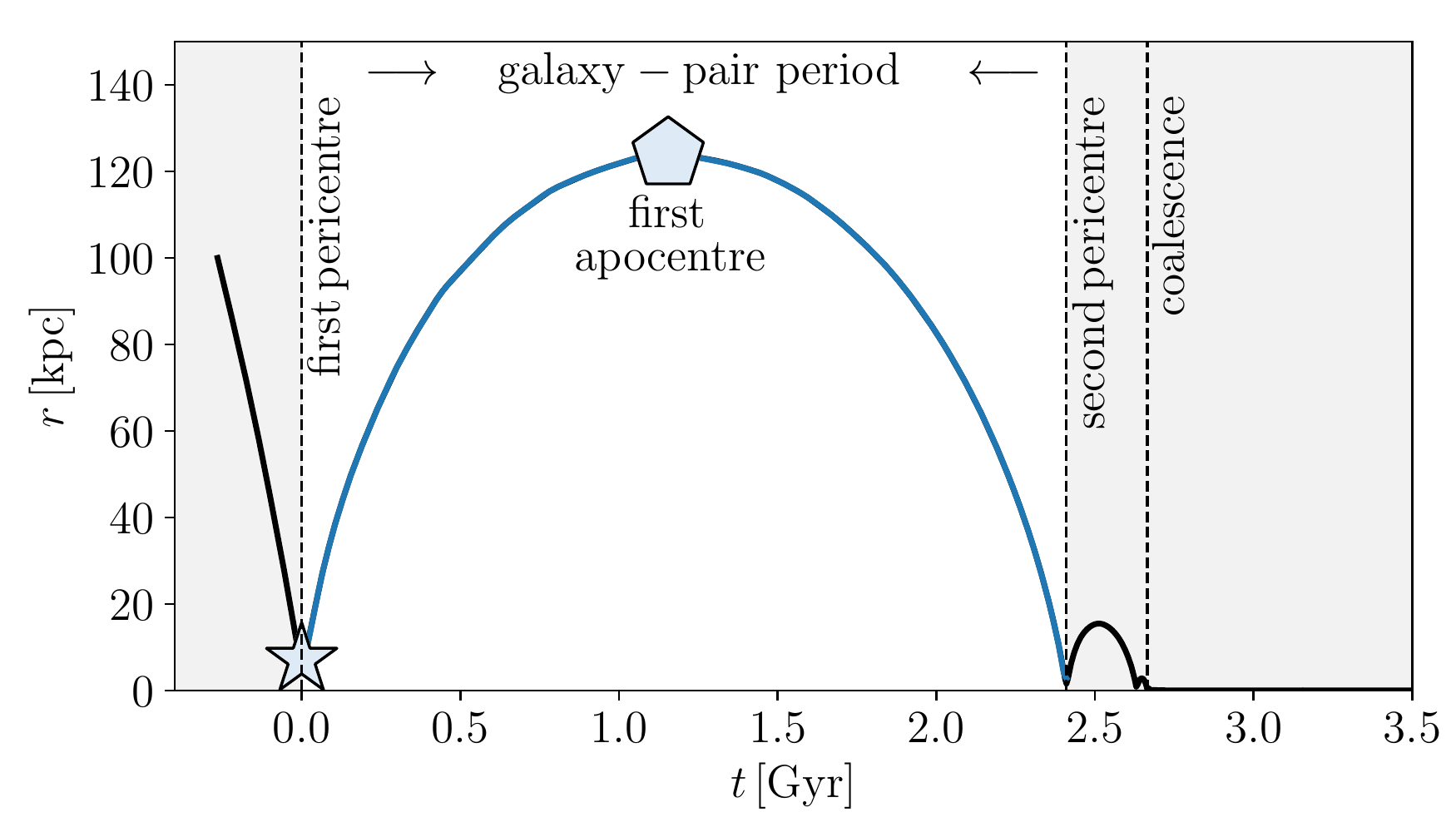}
}
\vspace{-.06in}
\hbox{
\includegraphics[width=3.5in]{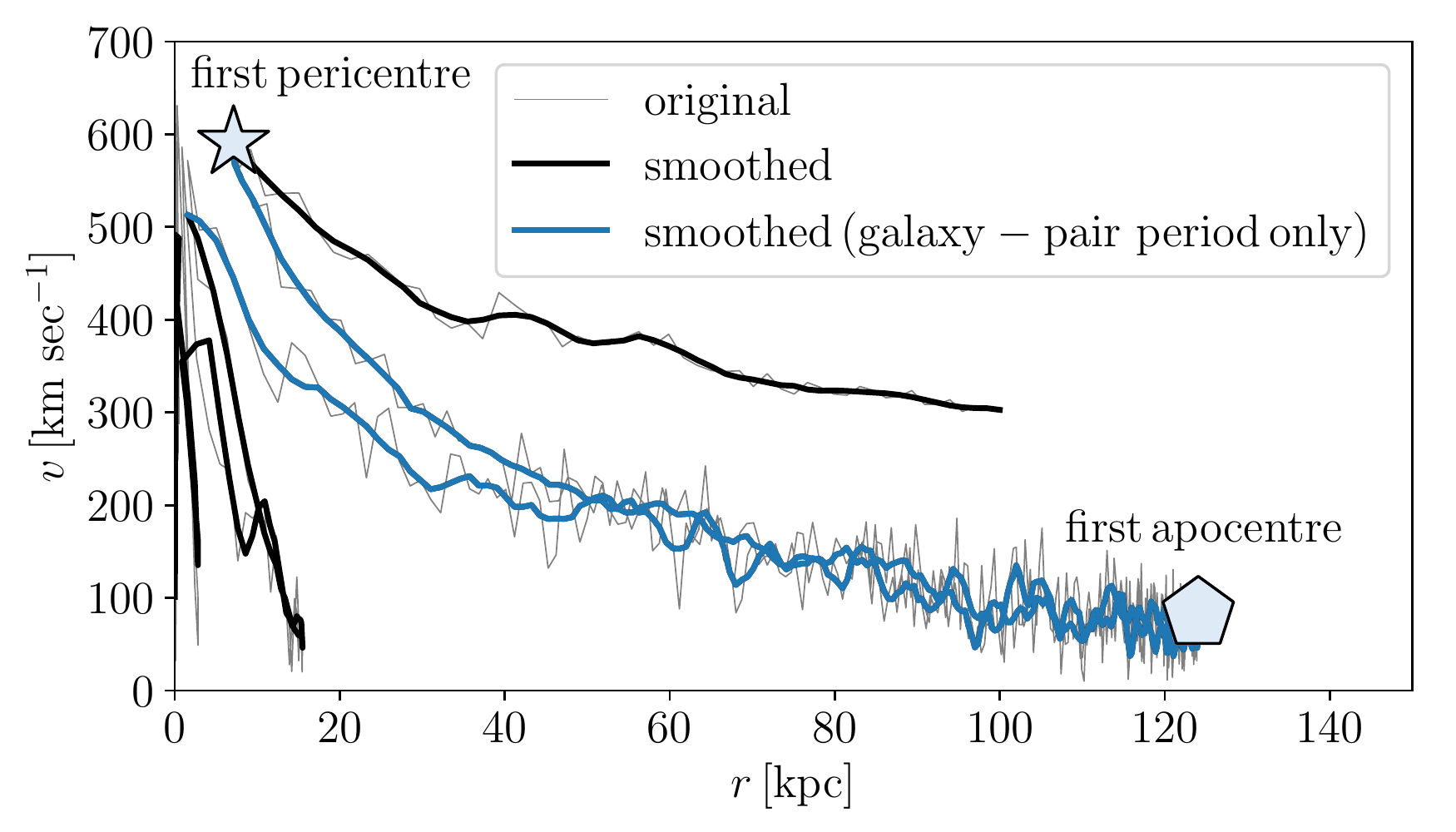}
}
\vspace{-.12in}
}}
\caption{Orbital coverage of our fiducial run, a nearly prograde merger with small separation and high relative velocity at first pericentric passage. {\it Top panel:} Galaxy-galaxy separation versus time. Vertical dashed lines correspond to the times of first and second pericentric passages, plus coalescence. Blue portions show the `galaxy-pair period', constrained to times between first and second pericentric passages. Periods outside the interacting period are masked in light grey. {\it Bottom panel:} Relative velocity versus separation. Galaxies have high relative velocities at small separations and spend a significant fraction of time at large separations. Lines are smoothed for display purposes. Results before smoothing are shown as thin grey lines. Large blue star and pentagon denote first pericentre and apocentre. See Figure~\ref{fig:pericoverage} for values at first pericentre and apocentre for other mergers in our suite.}
\label{fig:fiducialcoverage}
\end{figure}
 
Our isolated galaxies are stable and well-behaved on Gigayear timescales (Section~\ref{subsec:results_fiducial_phases}), in line with earlier tests using the same initial conditions \citep{Hopkins2012isolated,Hopkins2013,Torrey2017}. We note that, in our set up, we do not start with a multi-phase ISM. Instead, we initially set our gas to $10^5$ K with solar metallicity, from which a turbulent ISM emerges as a result of feedback produced by star formation, and in which subsequent metal enrichment is tracked self-consistently. We also note that our initial conditions do not include hot gas atmospheres, and the hot gas described in this paper arises solely from feedback heating the ISM. We elected not to include this hot component because of the lack of detailed hot gas profile measurements around spiral galaxies. \cite{Moster2011} attempt to include hot atmospheres in their merger simulations by adopting $\beta$-profiles from galaxy cluster observations \citep{Cavaliere1976,Jones1984,Eke1998}. However, it is not clear that such an extrapolation is valid at galactic scales. We refer the reader to Section~\ref{subsec:caveats}, where we expand on the suitability of this approximation. \\

 \subsubsection{The Fiducial Run}
\label{subsubsec:fiducial}

We study a fiducial run in detail to introduce terminology and to provide intuition. For this run, we adopt a nearly prograde orbit with small impact parameter ($\sim$7 kpc) to maximise the effects of the encounter. This limits the duration of the galaxy-pair period; i.e., the time between first and second pericentric passage. To compensate, we adopt a highly eccentric orbit. This choice is slightly different from the fiducial run in \cite{Moreno2015}, which only aims to maximise duration. Figure~\ref{fig:fiducialcoverage} shows the orbital evolution of this run. The top (bottom) panel shows the galaxy-galaxy separation (relative velocity) as a function of time {(separation)}. The vertical dashed lines in the top panel indicate the times at first and second pericentric passages, plus coalescence (defined here as the time at which the two central supermassive black holes are at a distance of 500 pc for the last time). Hereafter, we shift time so that $t = 0$ Gyr corresponds to first pericentric passage, unless stated otherwise. Black lines refer to the entire orbit and blue lines correspond to the `galaxy-pair period', corresponding to the times between first and second pericentric passage, the primary focus of this paper. We mask times outside the galaxy-pair period in light grey. The large blue star and pentagon denote orbital properties at first pericentre and apocentre respectively. These markers will be useful when comparing this fiducial run to other simulations in our merger suite. The jagged nature of the lines in the bottom panel is driven by the small-scale dynamical interaction between each the black hole and its neighbouring circumnuclear medium (we smooth our curves using a Savitzky-Golay filter for display purposes only). In this work we use the locations of the supermassive black holes as proxies for galactic centres. We verified the suitability of this approximation visually.

\subsubsection{Galaxy Merger Simulations}
\label{subsubsec:suite}

To explore the effects caused by variations in orbital merging configuration, we constructed a suite of 24 galaxy merger simulations involving our primary and secondary galaxies (Table~\ref{table:iso}). Following \citet{Moreno2015}, we adopt a choice of stellar masses (and mass ratio) representative of close galaxy pair catalogues drawn from the SDSS \citep{Patton2013} and samples extracted from cosmological simulations with abundance matching \citep{Moreno2012,Moreno2013}. Our suite is split into three families, corresponding to three distinct spin-orbit orientations: nearly `prograde', `polar' and `retrograde' (Table~\ref{table:angles}), corresponding to the ``e", ``f" and ``k" orientations from \citet{Robertson2006}, also used by \citet{Patton2013} and \citet{Moreno2015}.

\begin{table}
  \begin{center}
    \begin{tabular}{lc|c|c|c|c|} 
        \hline 
             & Prograde (``e") & Polar (``f") & Retrograde (``k") \\
        \hline   \hline 
        \, Primary & &  & \\
       \, \,\,\,\,\,\,\,\,$\phi_1$    & 60$^\circ$  & 60$^\circ$ & -30$^\circ$\\
       \, \,\,\,\,\,\,\,\,$\theta_1$ & 30$^\circ$  & 60$^\circ$ & -109$^\circ$\\
        \hline
        Secondary & &  & \\
      \,  \,\,\,\,\,\,\,\,$\phi_2$    & 45$^\circ$  & 0$^\circ$   & -30$^\circ$\\
      \,  \,\,\,\,\,\,\,\,$\theta_2$ & -30$^\circ$ & 150$^\circ$ & 71$^\circ$\\
       \hline
    \end{tabular}
  \end{center}
\caption{Spin-orbit angles for each of our three orientations (approximately prograde, polar and retrograde) used in this work \citep[and in][]{Patton2013,Moreno2015}, originally drawn from \citet{Robertson2006}. Here, $\theta$ and $\phi$ specify the total angular momentum of each gas disc relative to the orbital plane. See Figure~1 of \citet{Moreno2015} for a schematic.
}
\label{table:angles}
\end{table}
\begin{figure}
\centerline{\vbox{
\vspace{-.12in}
\hbox{
\includegraphics[width=3.5in]{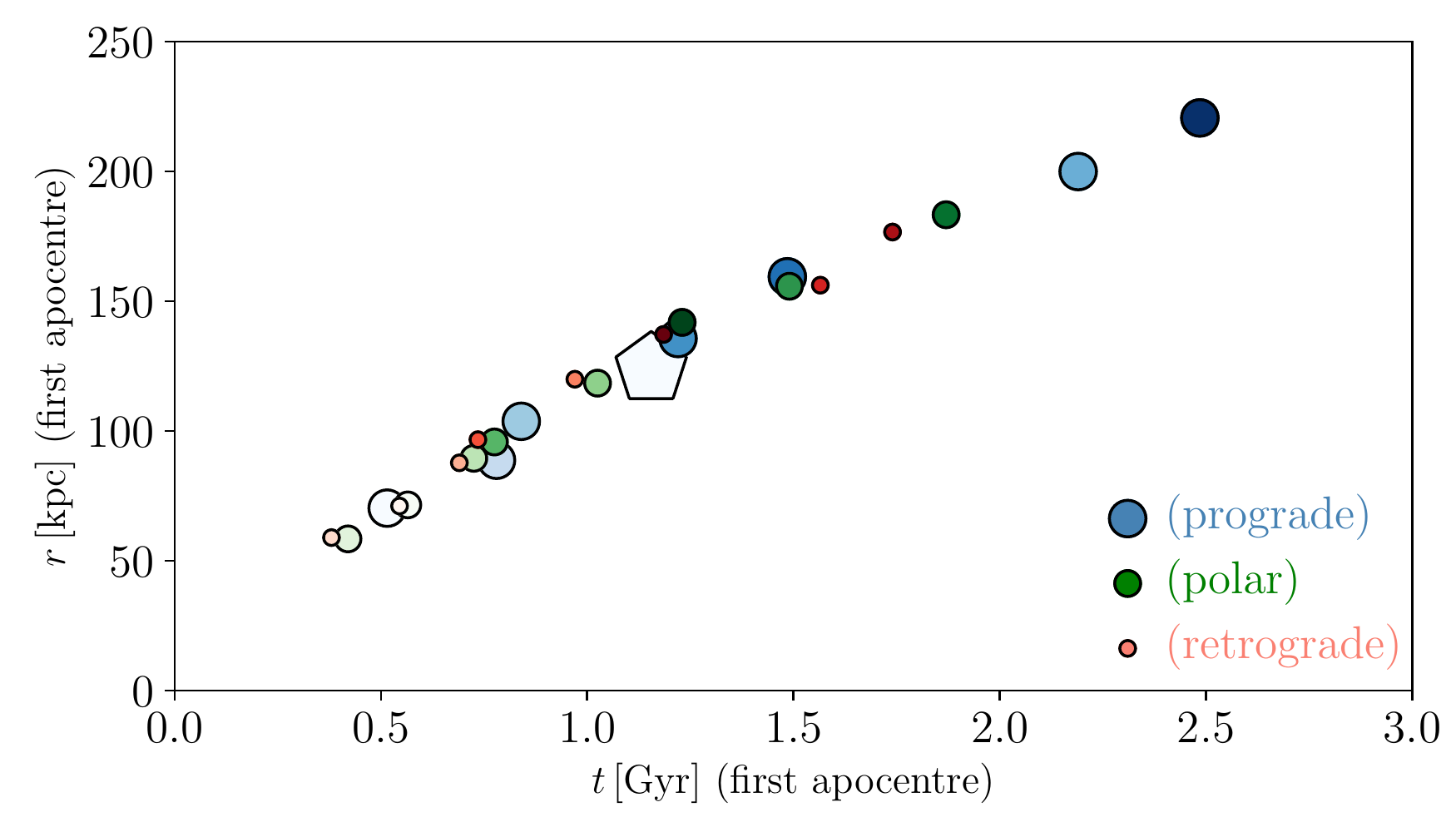}
}
\vspace{-.06in}
\hbox{
\includegraphics[width=3.5in]{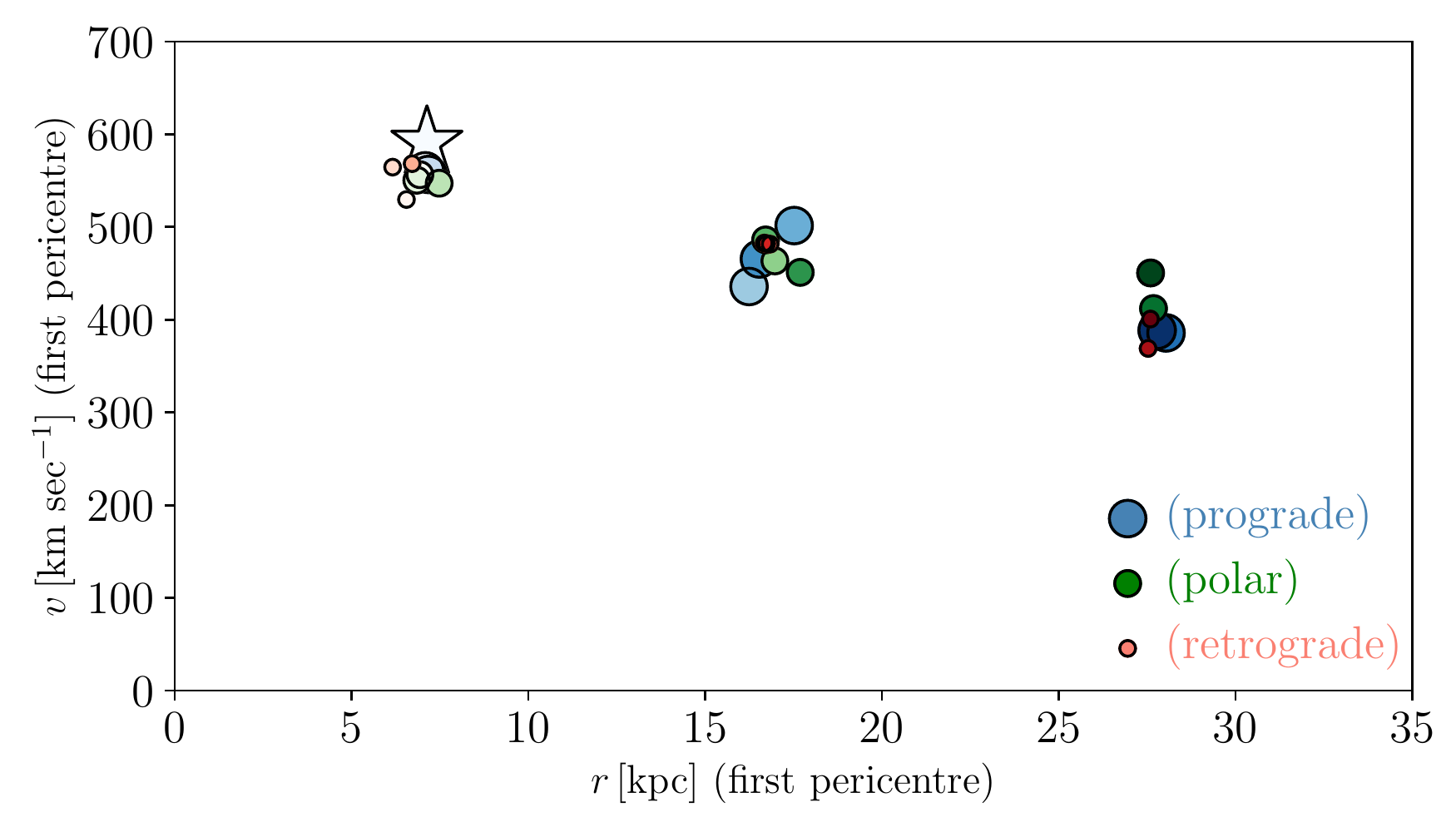}
}
\vspace{-.12in}
}}
\caption{Properties of our galaxy merger suite at first apocentre and pericentre.  Nearly prograde, polar, and retrograde orbits displayed as blue (large), green (intermediate), and orange (small) circles. {\it Top panel:} Separation versus time at first apocentre. {\it Bottom panel:} Relative velocity versus separation at first pericentre. Large blue pentagon (first apocentre) and star (first pericentre) represent our fiducial run (Figure~\ref{fig:fiducialcoverage}).
}
\label{fig:pericoverage}
\end{figure}
\begin{figure}
\centerline{\vbox{
\vspace{-.12in}
\hbox{
\includegraphics[width=3.5in]{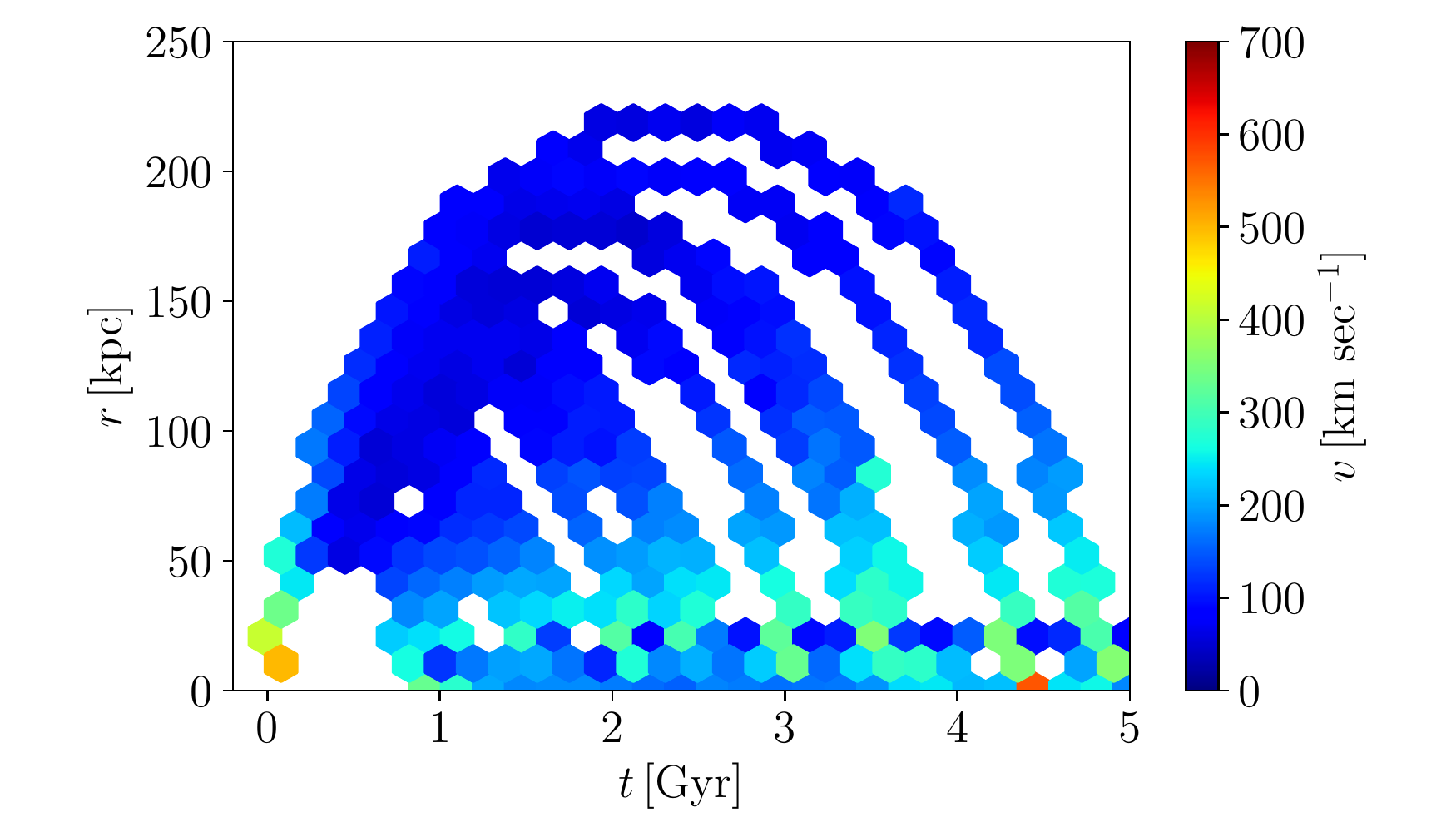}
}
\vspace{-.06in}
\hbox{
\includegraphics[width=3.5in]{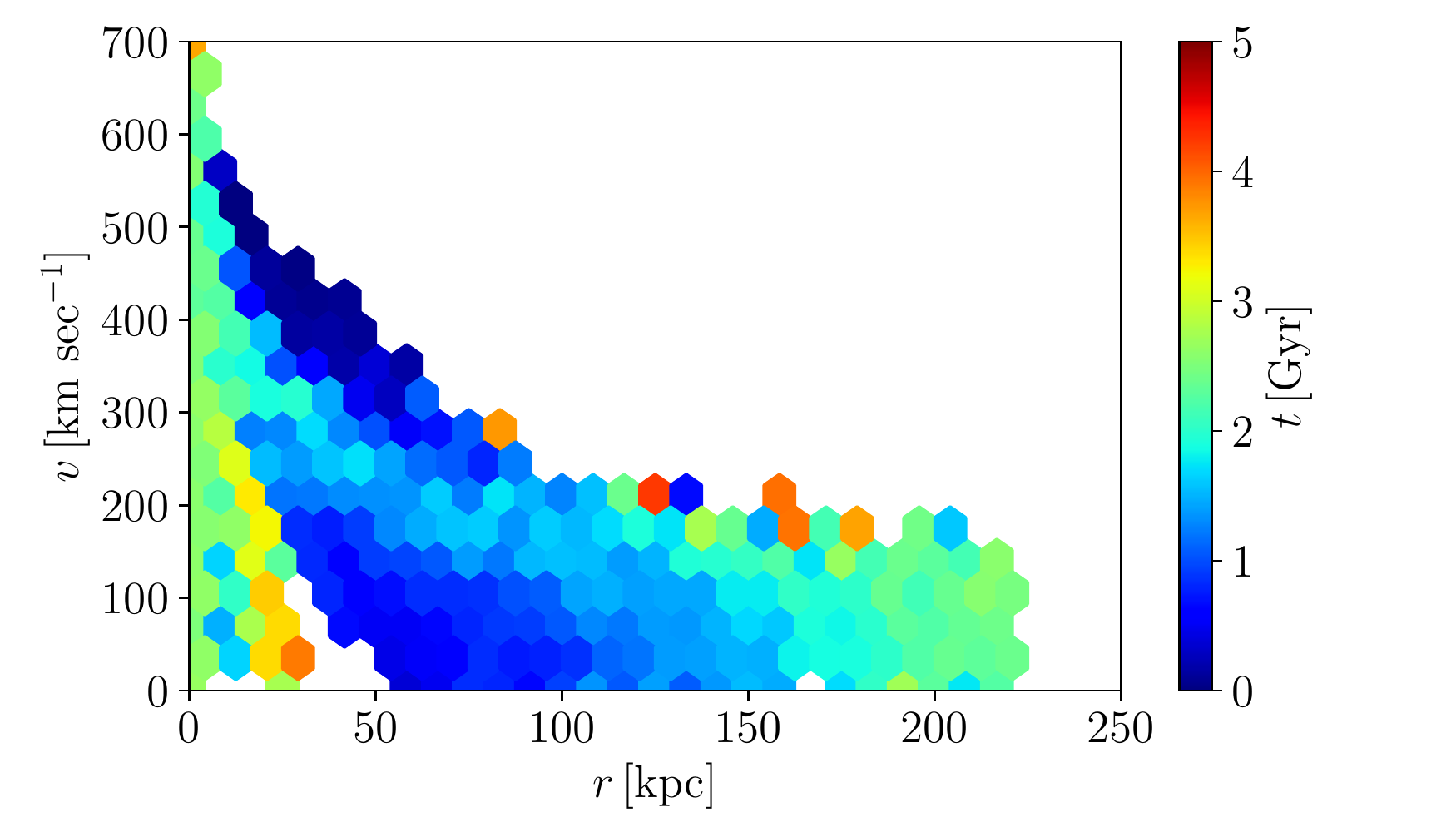}
}
\vspace{-.12in}
}}
\caption{Orbital coverage in our merger suite. {\it Top panel:} Separation versus time since first pericentric passage, colour coded by relative velocity. {\it Bottom panel:} Relative velocity versus separation, colour coded by time since first pericentric passage. Galaxies have high relative velocities at small separations and spend a significant fraction of their time at large separations. 
}
\label{fig:orbitalcoverage}
\end{figure}

The top panel of Figure~\ref{fig:pericoverage} shows separation at apocentre versus time at that location for each merger in our suite. Recall that time is rescaled to zero at first pericentric passage. The separation at apocentre and the time it takes the two galaxies to reach that separation are indicative of the orbit's spatial extent and duration. Large blue circles denote nearly prograde mergers; medium-sized green circles denote nearly polar mergers; and small orange circles denote nearly retrograde mergers. Different shades correspond to different runs within a given orientation. We cover a broad range of separations and timescales. Our fiducial run, represented by the large blue pentagon, is a ``middle-of-the-pack" orbit.
In  \citet[][]{Moreno2015} and \citet{Patton2013}, for each orientation we selected a fixed set of specific eccentricities and impact parameters, consistent with published low-resolution cosmological N-body estimates \citep{KB2006}. With the advent of large cosmological simulations of galaxy formation at higher resolution with baryons, however, it is unclear if those original N-body estimates still hold \citep{Vicente2015}. In this work, we do not fine-tune these orbital parameters. Instead, we tailor our orbits to have certain properties at first pericentric passage, which govern their spatial separation extent and their duration. The bottom panel of Figure~\ref{fig:pericoverage} shows relative velocity versus separation at first pericentric passage for all of our mergers (the large blue star represents our fiducial run). By design, our 24 mergers arrange themselves into three clusters, two containing 9 mergers and one containing 6 mergers -- each with increasing (decreasing) separation (relative velocity) at first pericentric passage:  $\sim$(7 kpc, 560 km sec$^{-1}$), $\sim$(17 kpc, 460 km sec$^{-1}$) and $\sim$(28 kpc, 410 km sec$^{-1}$). Each of these groups can be regarded as a set of strong, moderate and weak encounters  \citep{Donghia2010}. For each orientation, we initially had 9 mergers, but we dropped those orbits with merging times exceeding 5 Gyr (thus reducing our total number of runs from 3$\times$9 = 27 down to 24). This is because it is unlikely for a pair of interacting galaxies to evolve in isolation from the rest of the Universe for such an extended time \citep{Moreno2012,Moreno2013}. Also, our low-redshift initial conditions would no longer be applicable for such long-lasting interactions. This explains why the last cluster at large separations in the bottom panel of Figure~\ref{fig:pericoverage} only contains 6 mergers.

With our range of orbital choices, we are able to cover a broad range of separations and merging timescales. Figure~\ref{fig:orbitalcoverage} shows this. The 2D histograms in this Figure incorporate all times at and after first pericentric passage, including times well past coalescence (recall that most of the paper focuses only on the galaxy-pair period). The top panel shows separation versus time, colour coded by relative velocity. Our orbital choices produce separations as high as 300 kpc, consistent with observations of wide galaxy pairs by \citet{Patton2013,Patton2016}. At large separations, galaxies slow down relative to their companions, thus spending long periods of time near their orbital apocentres. The bottom panel shows relative velocity versus separation, colour coded by time. Soon after first pericentric passage, galaxy pairs have small separations and high velocities (dark blue bins). At later times, as galaxies approach their orbital apocentres, they tend to have larger separations and smaller relative velocities (green bins, bottom right). Short-lived orbits approach their second pericentres with high relative velocities (blue and cyan bins formed in a diagonal with separations $\sim$30-150 kpc). Comparing Figure~\ref{fig:fiducialcoverage} to Figure~\ref{fig:orbitalcoverage} shows that our fiducial run is an `average' orbit. We warn the reader that the use of fixed line-of-sight velocity cuts to select galaxy pairs \citep[e.g., 300 km sec$^{-1}$,][]{SloanClosePairs,Patton2013} may delete true interacting pairs at small separations (upper-left corner).

In summary, we split our suite of merger simulations as follows:  24 mergers $=$ (3 orientations $\times$ 3 first-pericentre separations $\times$ 3 relative velocities at first pericentric passage) $-$ (3 mergers with duration greater than 5 Gyr). 

\subsection{ISM Temperature-Density Regimes}
\label{subsec:gas_phases}

\begin{figure}
\centerline{\vbox{
\vspace{-.12in}
\hbox{
\includegraphics[width=3.5in]{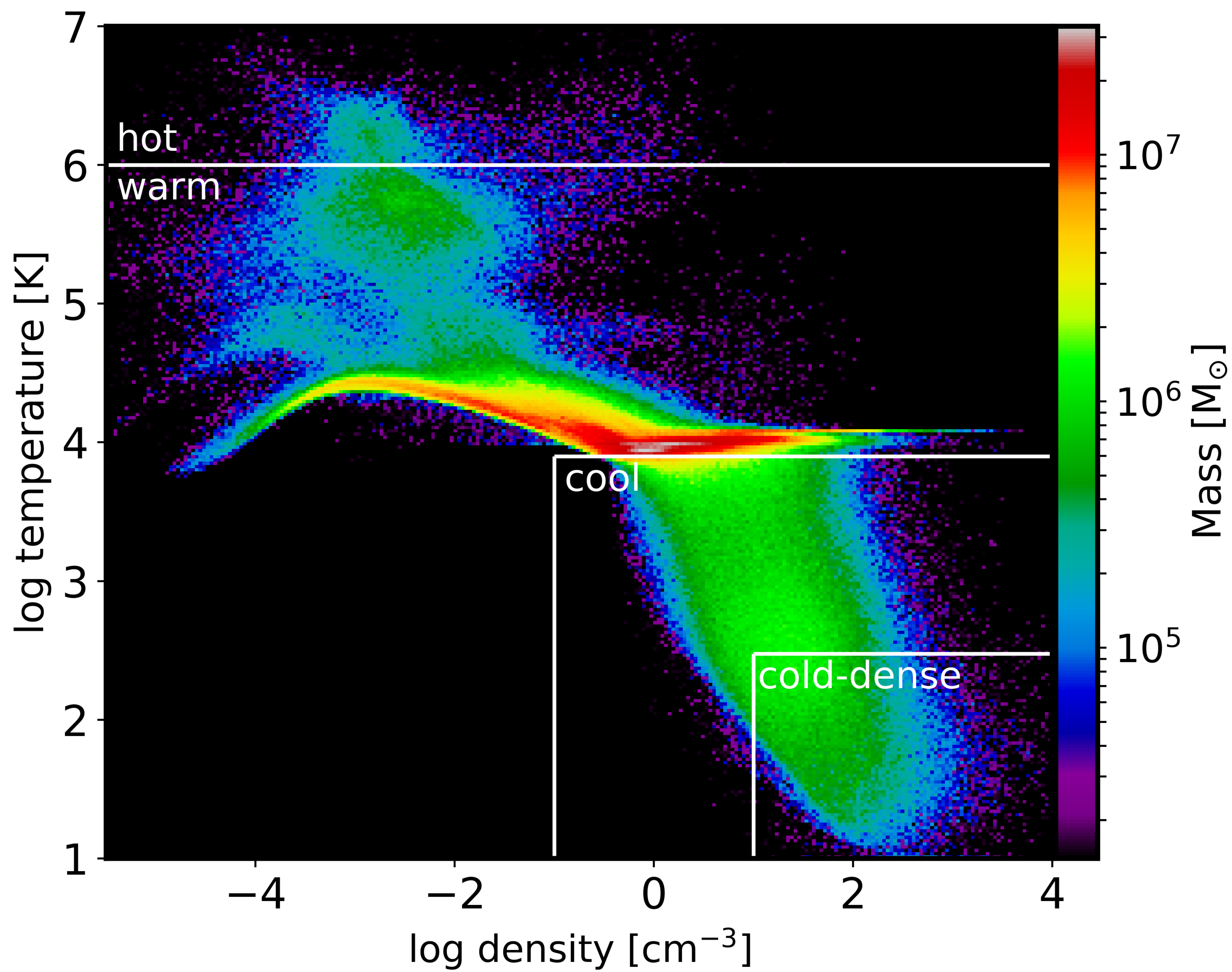}
}
\vspace{-.05in}
}}
\caption{Typical (mass-weighted) temperature-density diagram for gas in our simulations. Horizontal and vertical white lines delineate four temperature-density regions: the {\it hot}, {\it warm}, {\it cool} and {\it cold-dense} regimes.}
\label{fig:phase_diagram}
\end{figure}

To study the structure of the ISM, we split our gas component into four temperature-density regimes: {\it hot}, {\it warm}, {\it cool} and {\it cold-dense}. See Figure~\ref{fig:phase_diagram} and Table~\ref{table:phases} for demarcations. These four temperature-density regimes are motivated by the  phases making up the structure of the ISM in observations: the hot, ionised, atomic and molecular gas phases. In this work, we require {\it hot} gas to have temperature above 1 million Kelvin. Whilst arbitrary, this choice is widely adopted in the literature because it corresponds approximately to the virial temperature of Milky-Way type haloes \citep{Keres2005}. We remind the reader that our initial conditions do not include hot atmospheres, rather our hot component is produced solely as a result of feedback heating up the colder ISM components. Below 1 million Kelvin, we split the ISM into three temperature-density regimes. We expect the {\it warm} regime to be dominated by warm-ionized gas, indicated by the bright band above 8,000 Kelvin. The lower right corner (with $n > 0.1$ and $T < 8000$ K) contains the {\it cool} and {\it cold-dense} regimes. This region has the following structure: a diffuse valley immediately below the 8000-K cut and a cloud peaking at a few hundred Kelvin. For a similar diagram using our model to simulate Milky-Way like galaxies, see Figure~5 of \citet{Guszejnov2017}. We expect this corner to contain a mixture of atomic and molecular gas, with the bottom part of this cloud to host most of the molecular gas  \citep[see e.g.,][]{Mihalas1981,Kulkarni1988,Draine2011}. To split this corner into the cool and cold-dense regimes, we follow \cite{Orr2018} and adopt their conservative cut at 300 K. For alternative cold-dense temperature demarcations, see \cite{Bournaud2015}.

\begin{table}
  \begin{center}
    \begin{tabular}{l|l|l|} 
          \hline

      ISM regimes & Temperature-density demarcations  \\
      \hline \hline
     {warm}       & $(T<10^6\,{\rm K}, n<0.1\, {\rm cm}^{-3})$  \\
       & ${\rm \&} \, \,(8000\,{\rm K} < T < 10^6\,{\rm K},n>0.1\, {\rm cm}^{-3})$  \\
     {cool}       & $(T<8000\,{\rm K}, 0.1 \,{\rm cm}^{-3}<n<10 \,{\rm cm}^{-3})$  \\
          & ${\rm \&} \,\, (300\,{\rm K}<T< 8000\,{\rm K},n<0.1\, {\rm cm}^{-3})$  \\
     {cold-dense}   & $(T<300\,{\rm K}, n>10 \,{\rm cm}^{-3})$  \\
     {hot}            & $(T>10^6\,{\rm K})$   \\
      \hline
     {cold moderately-dense}   & $(T<300\,{\rm K}, 10<n<1000 \,{\rm cm}^{-3})$  \\
     {cold ultra-dense}   & $(T<300\,{\rm K}, n>1000 \,{\rm cm}^{-3})$  \\
      \hline

    \end{tabular}
  \end{center}
\caption{Temperature-density regime demarcations (hot, warm, cool and cold-dense).  The bottom two rows split the cold-dense regime into a cold moderately-dense and a cold ultra-dense component (Section~\ref{subsec:fuelling}, Figure~\ref{fig:molecular_fid}).}
\label{table:phases}
\end{table}

In this paper, we do not attempt to employ sophisticated (and computationally intensive) models of the ionised, atomic and molecular gas directly. For the next generation of simulations by our group, we are currently refining realistic radiative-transfer methods \citep{Krumholz2008,Krumholz2011,Narayanan2017}, coupled with chemical-network solvers, to capture the relative importance of each ISM phase (Lakhlani et al., in prep, Richings et al., in prep). Meanwhile, incipient work aimed to validate our FIRE-2 simulations against the ISM in real galaxies has been conducted already. \citet{Orr2018} demonstrate that their treatment of HI and H$_{\rm 2}$ gas yields reasonable agreement with the observed Kennicutt Law \citep{Kennicutt,Kennicutt2012}. \citet{Guszejnov2017} demonstrates that our model successfully reproduces the GMC mass function in the Milky Way \citep{Rice2016} and the line-width size relation \citep[e.g., the Larson scaling relationship,][]{Larson1981} in our Galaxy \citep{Heyer2009,Heyer2015} and in nearby galaxies \citep{Bolatto2008,Fukui2008,Muraoka2009,Roman2010,Colombo2014,Tosaki2017}. Lakhlani et al. (in prep) is also conducting a through comparison between properties of Milky-Way GMCs in our models and observations in our Galaxy \citep{Dame2001,Miville2017}. Exporting these methods to our galaxy mergers is in the plans. Our main goal here, however, is to explore the behaviour of gas in various temperature-density regimes with a minimum set of assumptions.


\subsection{Caveats and Limitations}
\label{subsec:caveats}

As with any numerical simulation, our work has limitations that impact the extent to which our results should be interpreted and compared with observations. First, it is difficult to compare our various temperature-density ISM regimes directly against observations without employing full radiative-transfer calculations coupled with chemical network solvers. Additionally, our simulations lack feedback from SMBH accretion, which is likely to affect the ISM. For galaxy pairs, this effect is likely to be small \citep{Treister2012}. Likewise, we do not include hot gas atmospheres at the start of our simulations, which might alter the evolution of our hot gas component and other phases. Namely, cooling of hot gas may feed the ionised gas component (and, subsequently, other gas phases) by `hot gas accretion' \citep{Keres2005,Moster2011,Karman2015}. 
Another limitation in our approach is the lack of cosmological context. Interacting galaxies embedded in the cosmic web can increase their cold-gas supply via `cold mode' accretion \citep{Keres2009} and their dynamics can be influenced by third bodies \citep{Moreno2012,Moreno2013}. A promising approach to avoid these problems is to extract merging systems from cosmological simulations \citep[][Patton et al., in prep; Blumenthal et al., in prep]{Tonnesen2012,Sparre2016,Bustamante2018,Hani2018}. Additionally, placing mergers in a cosmological setting helps us identify false pairs (projected interlopers) or contamination by pairs selected before their first approach \citep{Kitzbichler2008,Hayward2013a,Robotham2014}.

Nevertheless, idealised binary merger simulations have significant advantages. First, this method allows us to resolve the ISM to extremely high densities. Secondly, this  `experimental' set up affords us control on the types of progenitors and orbital parameters we employ \citep[e.g.,][]{Cox2006,DiMatteo2007,DiMatteo2008,Moreno2015}, and is ideal for numerical experiments aimed at simulating specific well-known systems \citep{Karl2010,Karl2011,Karl2013,Privon2013,Renaud2015,Lahen2018}. 
In order to maximise the advantages, and mitigate the limitations of the isolated binary-merger method, we take the following steps. First, we avoid long-lived galaxy-galaxy interactions, for which gas accretion from the cosmic web or interactions with external `third' galaxies cannot be ignored. Secondly, in this work we quantify interaction-induced effects by comparing merging systems against isolated `control' galaxies (e.g., by reporting SFR {\it enhancements} and mass {\it excesses}). This approach mitigates the effects caused by other environmental factors. On average, if matched correctly, we expect real non-interacting and interacting galaxies to experience the same level of external effects -- i.e., cosmic web gas fueling, interactions with third bodies, etc. -- as they evolve. Thus, even though real galaxy pairs are embedded in the cosmic web, our idealised binary-merger simulations provide a reasonable approximation as long as we report quantities in terms interacting-to-isolated ratios and avoid long-lived interactions.

\section{Results}
\label{sec:results}

\subsection{Fiducial Run: Star Formation}
\label{subsec:results_fiducial_sfr}

\begin{figure}
\centerline{\vbox{
\vspace{-.12in}
\hbox{
\includegraphics[width=3.5in]{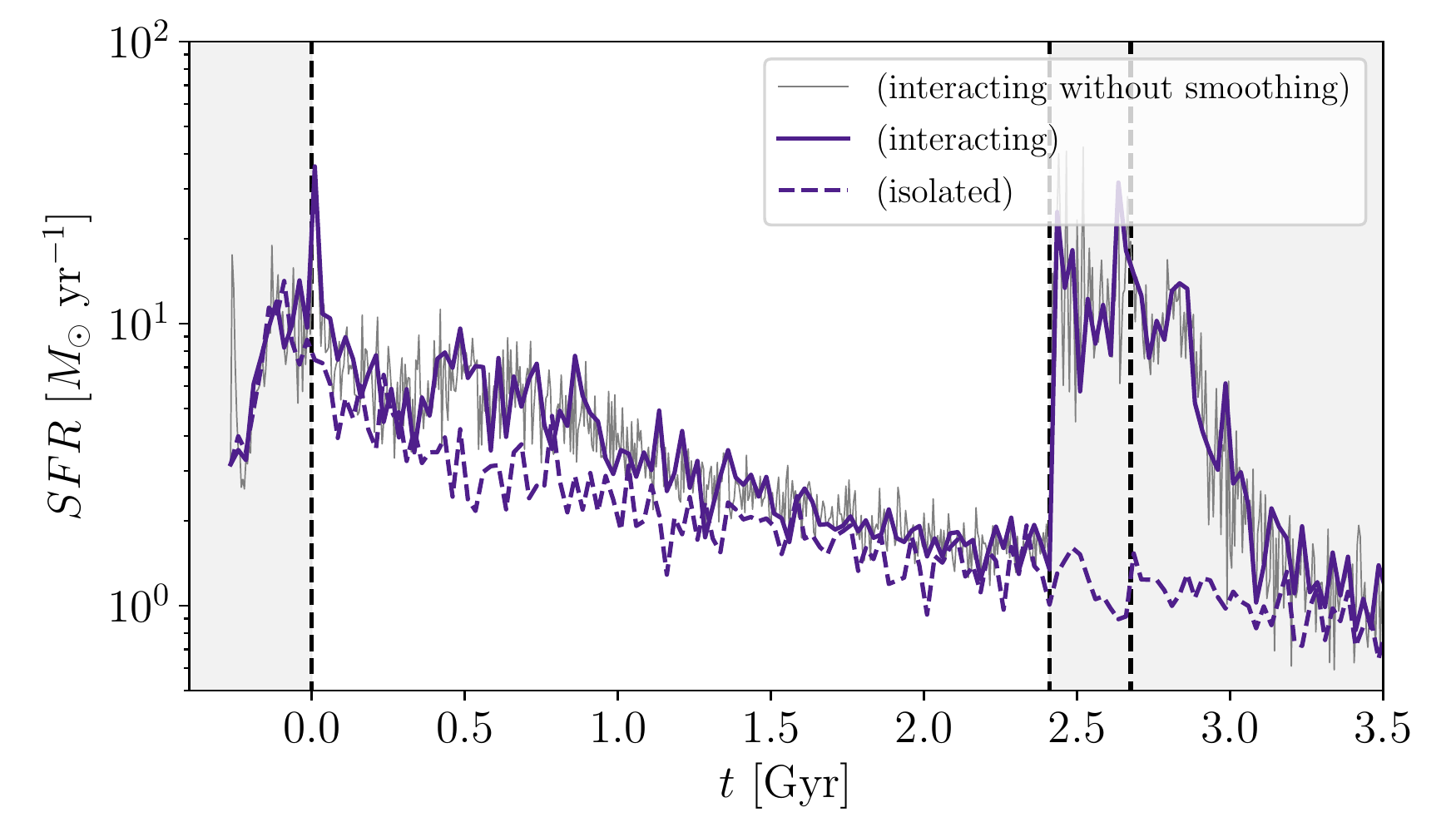}
}
\vspace{-.22in}
\hbox{
\includegraphics[width=3.5in]{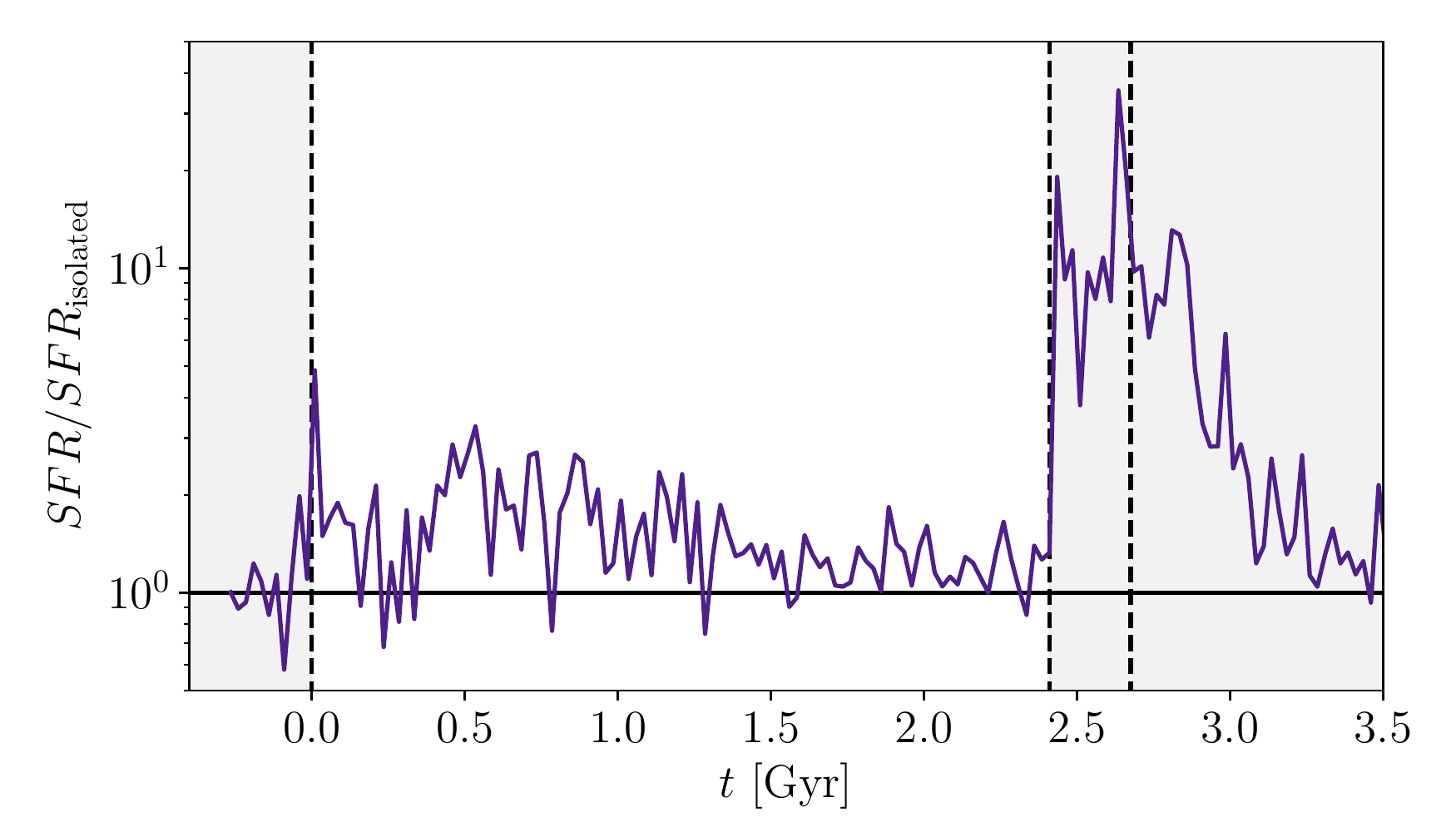}
}
\vspace{-.12in}
}}
\caption{Star formation rate (SFR) in our fiducial run. {\it Top panel:} the sum of SFR in our two interacting galaxies, displayed at every 5 (thin solid grey, without smoothing) and 25 (thick solid purple, smoothed) Myr, and the sum of SFR in the two isolated galaxies (thick dashed purple).  {\it Bottom panel:} Star formation rate enhancement, defined as the star formation in the interacting systems divided by the star formation rate in the isolated galaxies. The dashed vertical lines indicate first and second pericentric passages, plus coalescence (see top panel of Figure~\ref{fig:fiducialcoverage} for definitions). The galaxy-pair period, between first and second pericentric passage (times not masked in light grey), exhibits an extended period of enhanced star formation.
}
\label{fig:sfr_vs_time_fid}
\end{figure}

We summarise the change in SFR experienced by our simulated galaxies during their interaction. To compare interacting galaxies to their `control' isolated counterparts, we define {\it SFR `enhancement'} as the ratio of SFR in the interacting galaxies to the sum of the SFRs in the two isolated galaxies. This mimics what we do in observations \citep{SloanClosePairs,Scudder2012}. The rationale behind this is to tease out the effects caused by interactions from those caused by other environmental effects \citep{Patton2016}. The only difference here is that, instead of looking at individual galaxies, we calculate this ratio for the entire galaxy-pair system. 
The top panel of Figure~\ref{fig:sfr_vs_time_fid} shows the evolution of star formation rate (hereafter SFR) in our fiducial merger. The jagged thin grey solid line shows outputs at every 5 Myr, demonstrating the bursty nature of FIRE galaxies \citep{FIRE,Orr2017bursty,Sparre2017,CAFG2018}-- in contrast to previous works \citep[i.e.,][]{MH96,Torrey2012,Moreno2015}. Hereafter, we interpolate our curves at every 25 Myr for easy display purposes (thick solid purple line). The dashed lines represent the sum of the SFR in the two corresponding isolated galaxies. The vertical dashed lines represent first and second pericentric passages, plus coalescence. Regions outside the galaxy-pair period are masked in light grey.
The bottom panel of Figure~\ref{fig:sfr_vs_time_fid} shows star formation rate enhancement versus time. Overall, interactions elevate SFR in galaxies, in qualitative agreement with previous work \citep{MH96, DiMatteo2007, DiMatteo2008, Teyssier2010, Renaud2013,Moreno2015}. At first pericentric passage, the SFR exhibits a sudden spike, followed by a prolonged period of enhancement (by factors of $\sim$2-3), especially between t$=$0 and $\sim$1.3 Gyr. 
Although not the central focus of the paper, we briefly comment on the sudden rise in SFR and SFR enhancement at second pericentric passage. For our fiducial run, these quantities increase by a factor of $\sim$20. After checking, about half of our runs exhibit similar, albeit generally weaker, upturns. The intensity of SFR and SFR enhancement at second passage and coalescence depends on several factors, including internal properties of the colliding galaxies (their ISM structure) and the geometry of collision at second approach.

\begin{figure}
\centerline{\vbox{
\vspace{-.12in}
\hbox{
\includegraphics[width=3.5in]{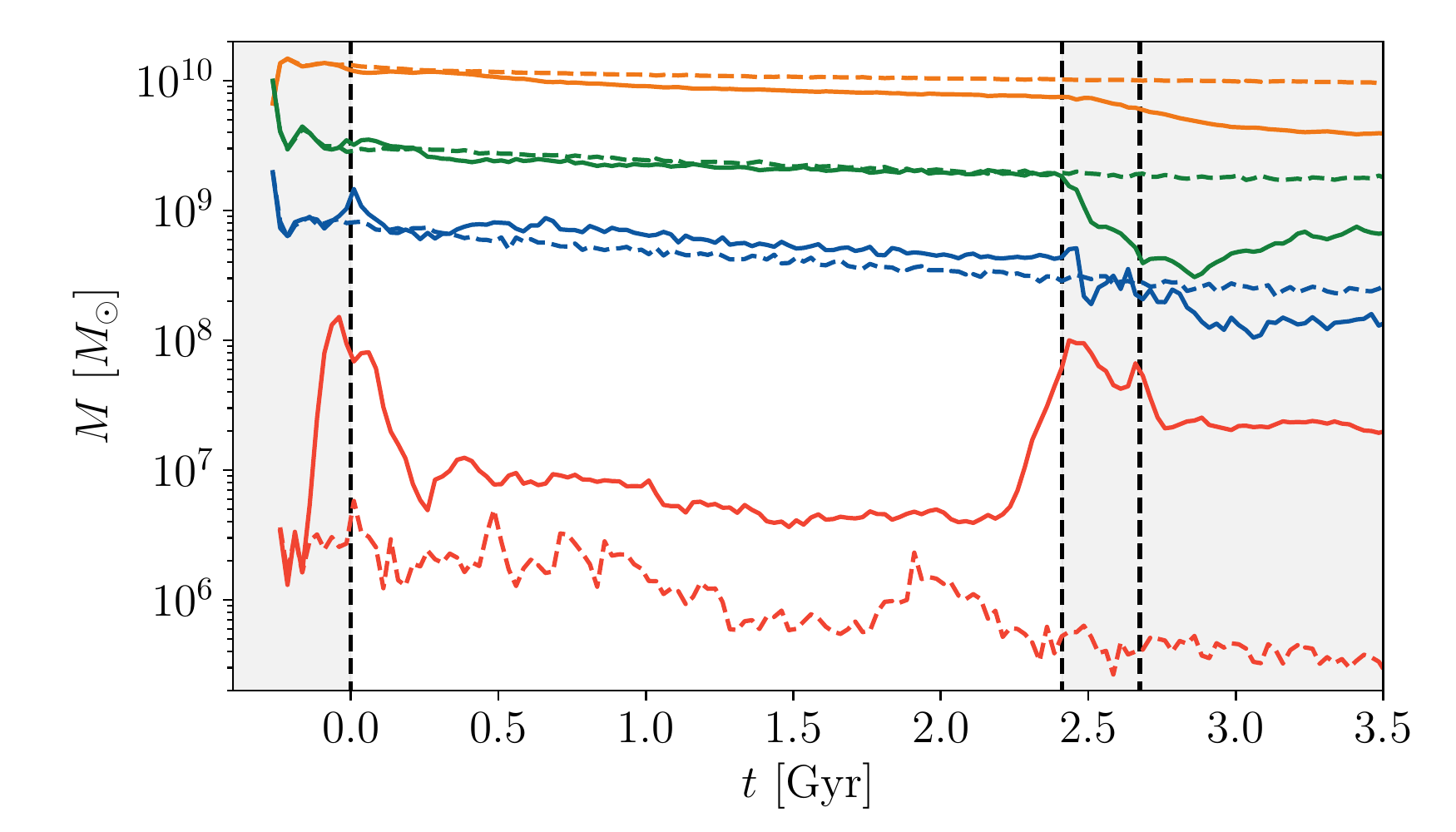}
}
\vspace{-.22in}
\hbox{
\includegraphics[width=3.5in]{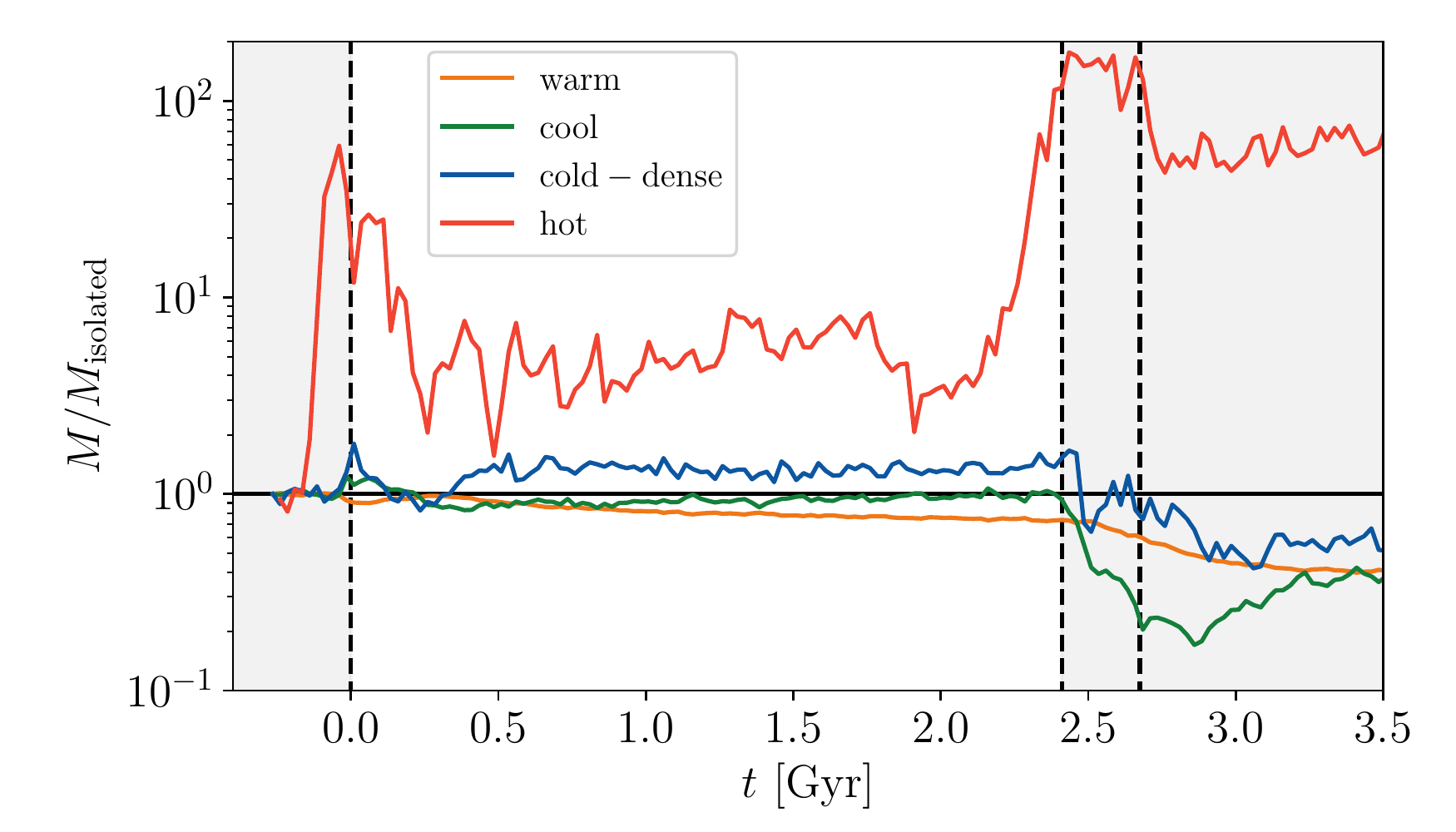}
}
\vspace{-.12in}
}}
\caption{Evolution of gas mass in our fiducial run for each of the following ISM temperature-density regimes: warm (orange), cool (green), cold-dense (blue) and hot (red). {\it Top panel:} Mass versus time. Solid (dashed) lines indicate interacting (isolated) runs. {\it Bottom panel:} Mass excess, defined as the ratio of mass in interacting and isolated galaxies, versus time. Dashed vertical lines indicate first and second pericentric passages, plus coalescence (see top panel of Figure~\ref{fig:fiducialcoverage} for definitions). Times outside the galaxy-pair period are masked in light grey. 
}
\label{fig:phases_vs_time_fid}
\end{figure}

\subsection{Fiducial Run: The Structure of the ISM}
\label{subsec:results_fiducial_phases}


One of the advantages of the {\small FIRE-2} model is its ability to capture the multi-phase structure of the ISM. The central focus of this paper is to study the impact of the encounter on various temperature-density regimes. Table~\ref{table:phases} and Figure~\ref{fig:phase_diagram} contain our adopted demarcations. Figure~\ref{fig:phases_vs_time_fid} shows the evolution of gas mass in our fiducial run, split into four ISM temperature-density regimes: {\it warm} (orange), {\it cool} (green), {\it cold \& dense} (blue) and {\it hot} (red lines) - displayed from top to bottom in decreasing order of fractional contribution to the entire gas mass content. This approximate ordering is maintained for the isolated galaxies (dashed curves), and for most of the simulation time for the interacting system (solid lines). In all cases, the isolated runs are stable and well-behaved on timescales of several Gigayears. The bottom panel shows mass `excesses' for each regime, defined as the ratio of the masses in the interacting system divided by the sum of the masses in the two isolated galaxies. For the interacting runs, the four regimes exhibit the following behaviour:

\begin{itemize}

\item {\bf Warm gas (fiducial run):}\\ Most of the gas mass is in this regime (orange lines). Warm gas is gradually depleted as a function of time in both isolated and interacting galaxies. This depletion is magnified by interactions, especially when the two galaxies are in close proximity (except briefly at $\sim$0.2-0.3 Gyr).\\

\item {\bf Cool gas (fiducial run):}\\  Cool gas (green lines) mass is depleted in both interacting and isolated cases. For the isolated case, this is slow and steady, whereas the interacting case undergoes more dramatic changes: a brief boost, followed by a drop and a long-term steady recovery.\\

\item {\bf Cold-dense gas (fiducial run):}\\ This component (blue lines) experiences depletion over long timescales. The encounter provokes the following behaviour: a brief and sudden spike, followed by a mild and brief suppression. Soon after, a cold-dense gas reservoir is replenished and survives (whilst experiencing slow depletion) until the end of the galaxy-pair period.\\

\item {\bf Hot gas (fiducial run):}\\ This component (red lines) comprises the smallest contribution to the total gas mass budget (recall that we did not include a hot atmosphere in our initial conditions). At both pericentric passages, and at coalescence, the amount of hot gas increases dramatically, possibly due to shock heating. Hot gas excess appears before the actual pericentric passages, indicated by the first two vertical lines, because this process begins as soon as the outer regions of the extended gas component collide with one another. The first image on the third row of Figure~\ref{fig:interaction_sequence} shows this. During the period between first and second pericentric passage, an excess of hot gas is maintained. This excess approximately doubles between t$\sim$1.3-1.9 Gyr. This phase experiences the largest levels of interaction-induced enhancement.\\

\end{itemize}

\subsection{Merger Suite: Star Formation}
\label{subsec:results_suite_sfr}

\begin{figure}
\centerline{\vbox{
\vspace{-.12in}
\hbox{
\includegraphics[width=3.5in]{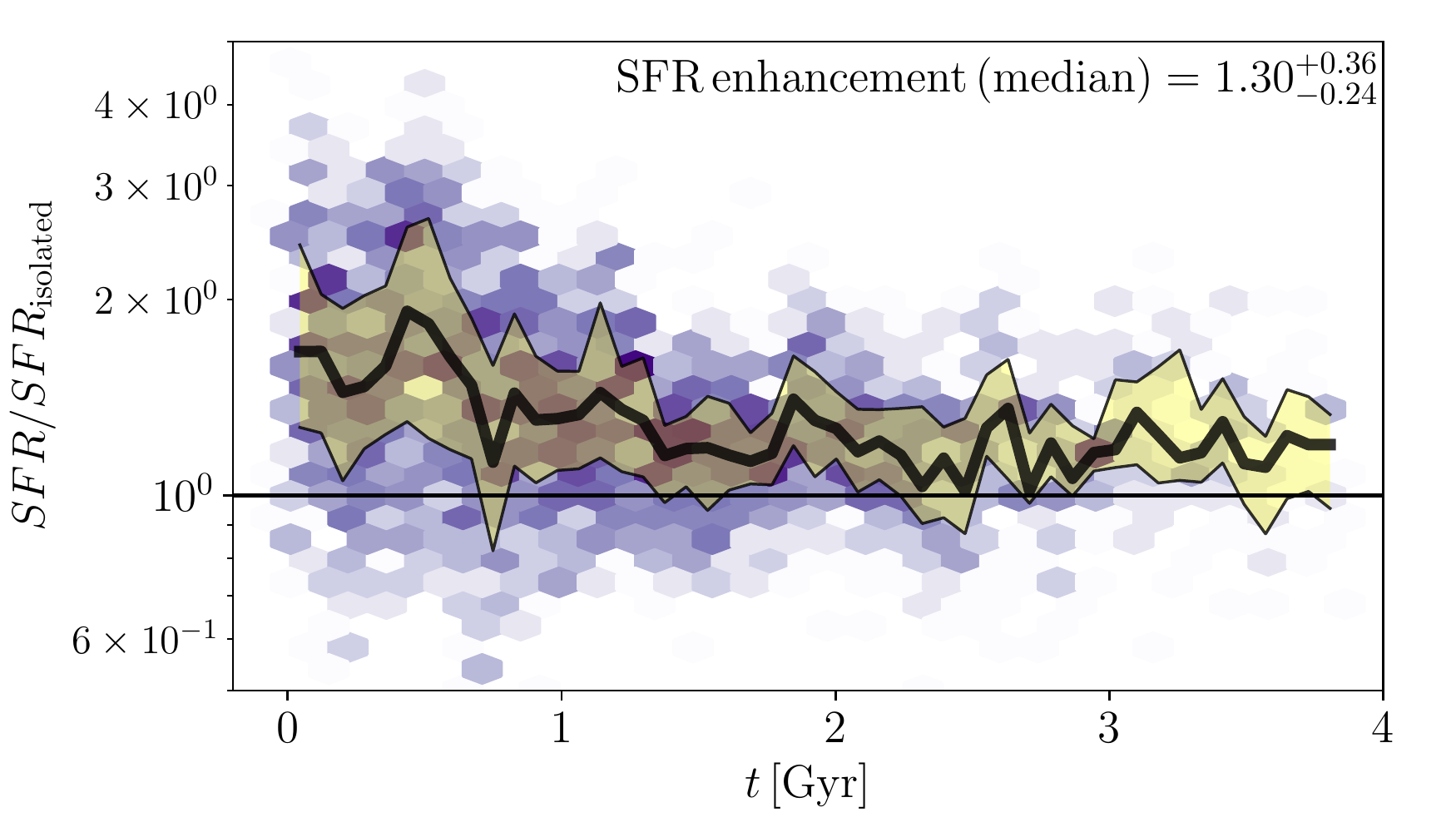}
}
\vspace{-.12in}
}}
\caption{Star formation rate enhancement versus time in our merger suite. Only galaxy-pair periods are included: between first and second pericentric passage, with time set to $t=0$ Gyr at first passage. Darker hexagons indicate higher incidence of interacting systems per bin, displayed on a logarithmic colour scale. Thick (thin) solid lines indicate the median (top and bottom quartiles). On average, galaxy interactions enhance star formation. The index in the upper-right corner represents the sample median and the lower-to-upper quartile range. See equation~(\ref{eqn:index}) for format.}
\label{fig:sfr_vs_time_suites}
\end{figure}

\begin{figure}
\centerline{\vbox{
\vspace{-.12in}
\hbox{
\includegraphics[width=3.5in]{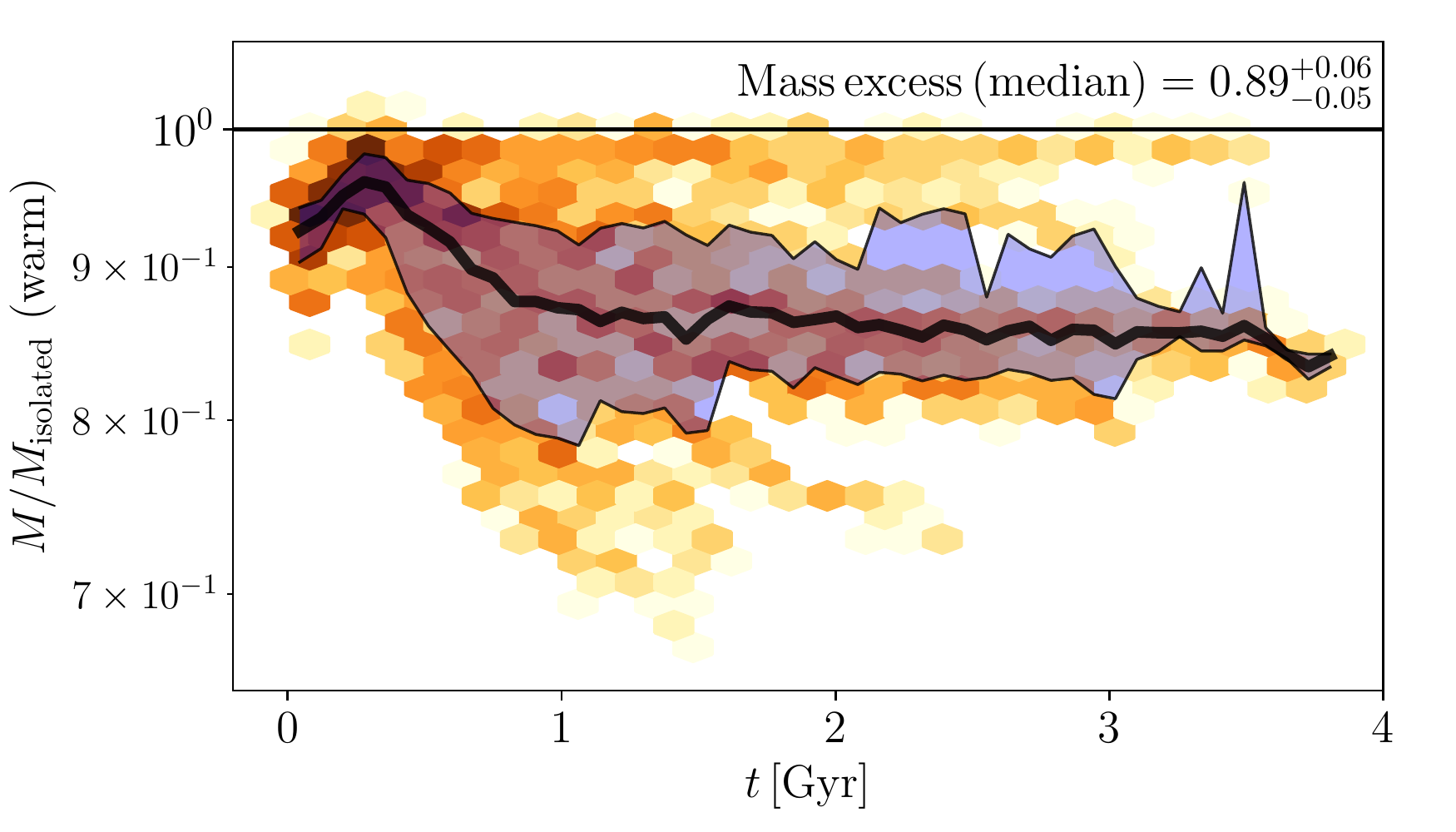}
}
\vspace{-.18in}
\hbox{
\includegraphics[width=3.5in]{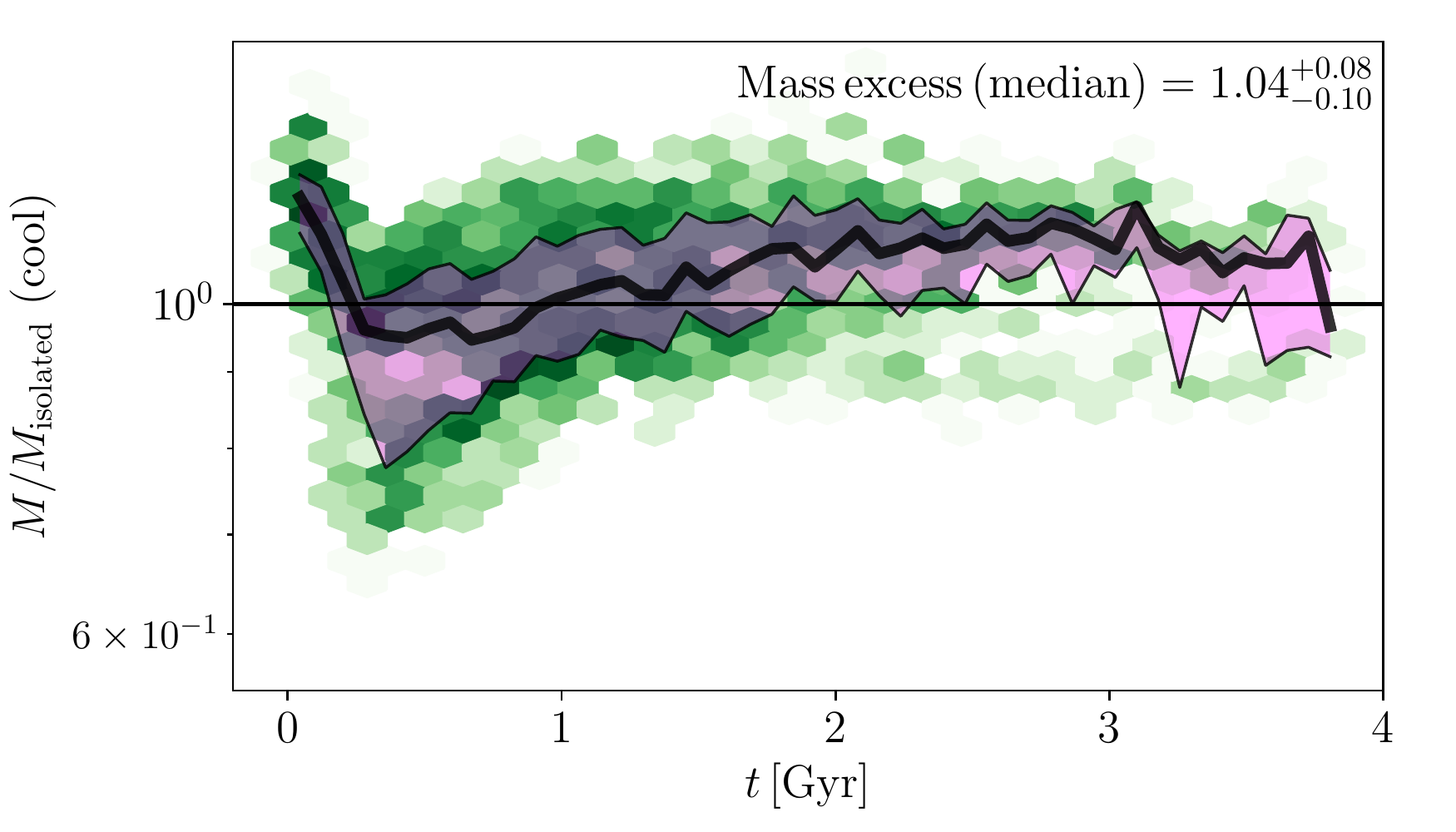}
}
\vspace{-.18in}
\hbox{
\includegraphics[width=3.5in]{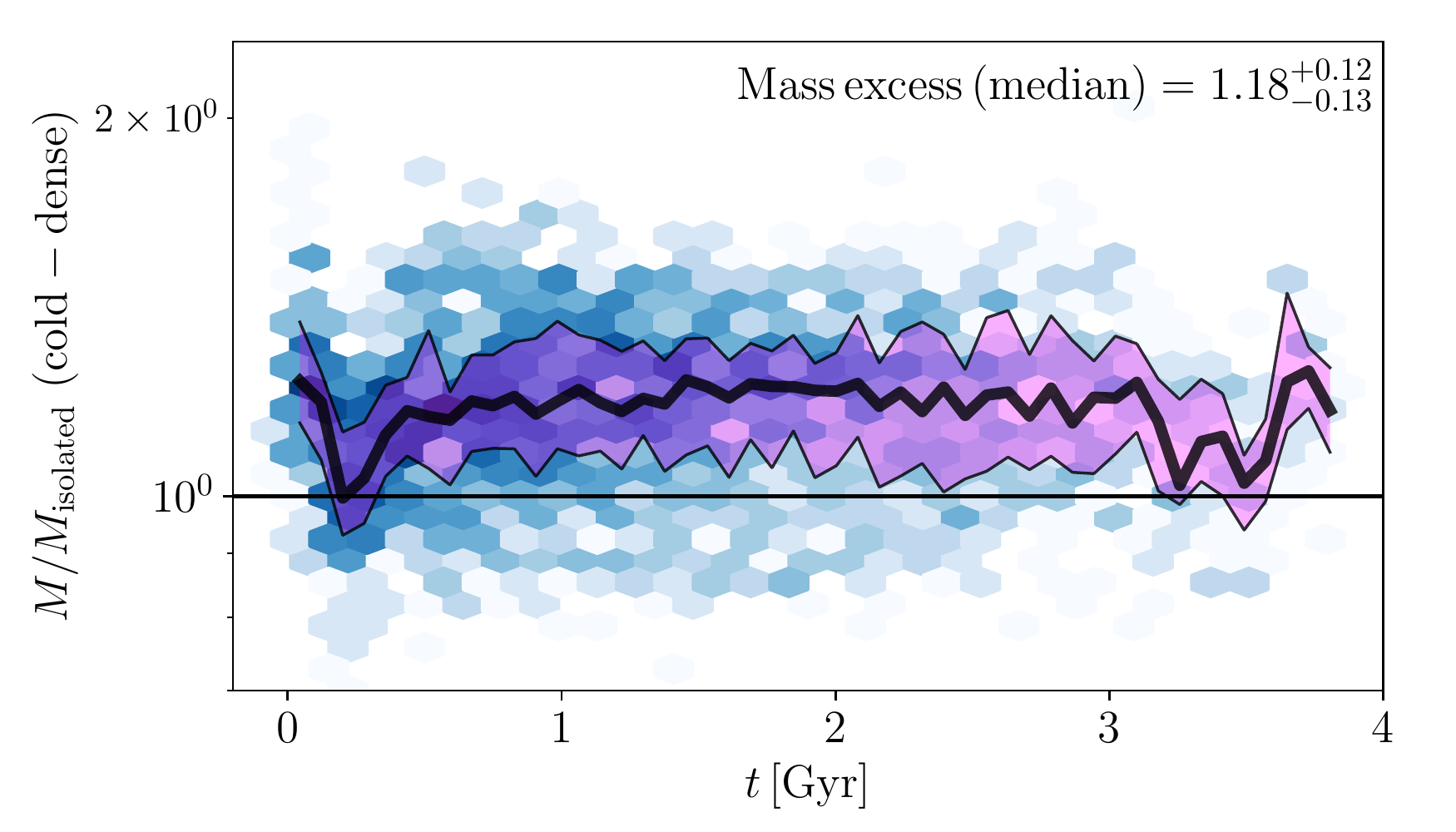}
}
\vspace{-.18in}
\hbox{
\includegraphics[width=3.5in]{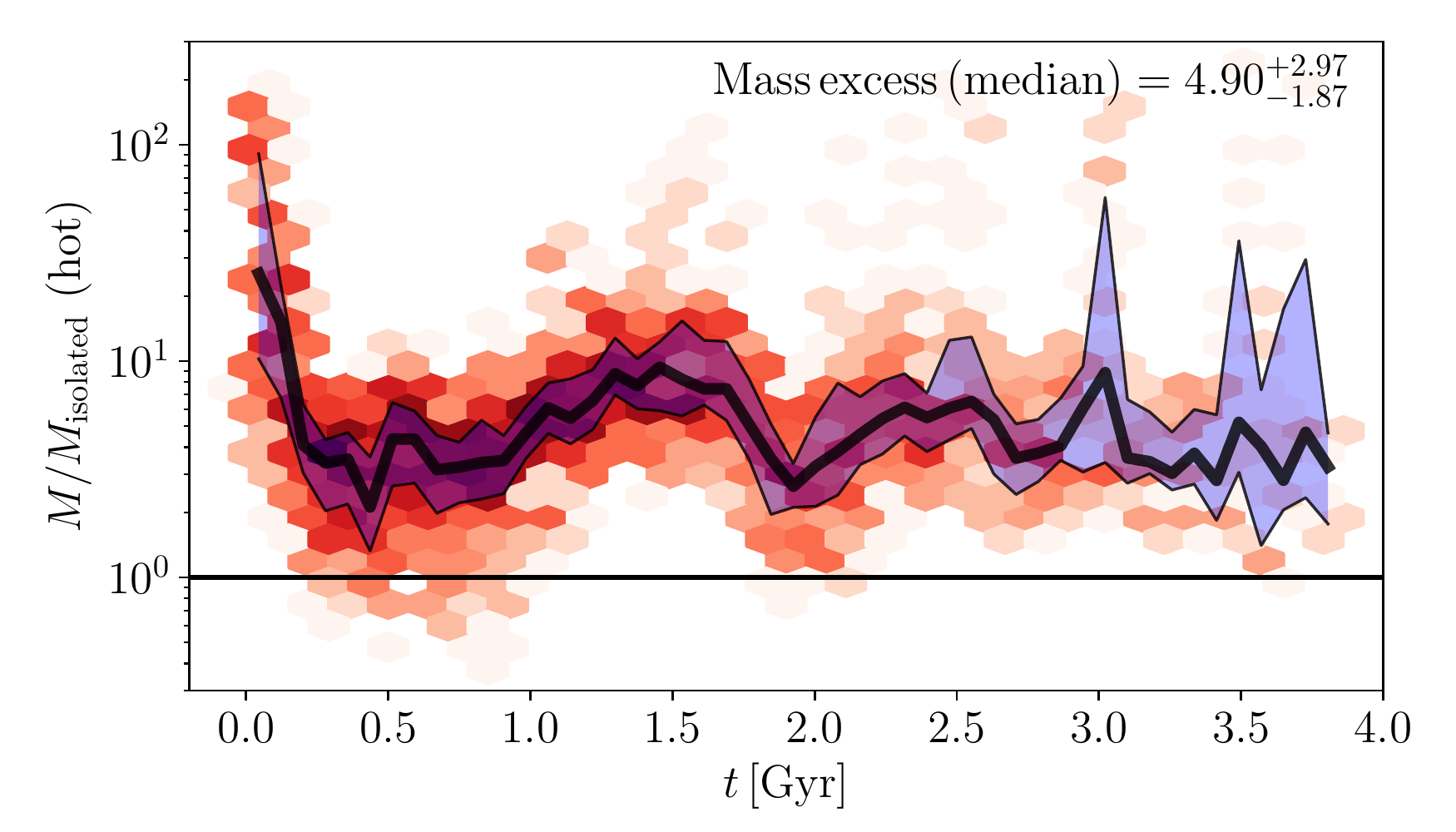}
}
\vspace{-.12in}
}}
\caption{Time evolution of mass excesses in various ISM temperature-density regimes. {\it Top to bottom:} warm (orange), cool (green), cold-dense (blue) and hot (red) gas. Thick (thin) solid lines indicate the median (top and bottom quartiles). Darker hexagons indicate higher incidence of interacting systems per bin, displayed on a logarithmic colour scale. The index in the upper-right corner of each panel represents the sample median and the lower-to-upper quartile range (equation~\ref{eqn:index}).}
\label{fig:phases_vs_time_suites}
\end{figure}

Figure~\ref{fig:sfr_vs_time_suites} shows star formation rate enhancement versus time for the entire merger suite. We only show the galaxy-pair period, between first and second pericentric passage, whose duration varies from merger to merger. This is a generalised version of Figure~\ref{fig:sfr_vs_time_fid} (excluding times masked in light grey). Darker hexagons indicate higher incidence of interacting systems per bin, displayed on a logarithmic colour scale. The thick solid black line indicates the median, and the thin solid lines encompass the top and bottom quartiles. On average, star formation is enhanced across our merger suite, in qualitative agreement with observations \citep{Woods2006,SloanClosePairs,Lambas2012, Scott2014,Patton2013,Patton2016} and earlier simulations \citep{BH96,DiMatteo2007,DiMatteo2008,Teyssier2010,Renaud2013,Patton2013,Moreno2015}. The level of enhancement, and the scatter around the median, diminishes with time.
We report the median SFR enhancement, and the corresponding upper and lower quartiles, across the entire sample (including all times within the galaxy-pair period) in the upper-right corner of the figure. Hereafter, we display this information using an index with this format:

\begin{equation}
\label{eqn:index}
{\rm index}={x_{50}}^{ x_{75}-x_{50}}_{ x_{25}-x_{50}},\,\,{\rm where}\,\,x_{q}={\rm the \,\,}q^{\rm th}\,\,{\rm percentile}.
\end{equation}
To illustrate, our median SFR enhancement is represented as

\begin{equation}
\label{eqn:index_sfr}
{\rm index} \Big{|}_{\rm \small SFR \, enhancement} = {1.30}^{+0.36}_{-0.24},
\end{equation}
where $x_{50}$ $=$ 1.30, with upper quartile at $x_{75}$ $=$ 1.66 $=$ $x_{50}$ $+$ 0.36 and lower quartile at $x_{25}$ $=$ 1.06 $=$ $x_{50}$ $-$ 0.24. We report similar indices to indicate gas mass excesses in the ISM (Figures~\ref{fig:phases_vs_time_suites} and \ref{fig:dense_molecular_suite}).
Our presentation combines several mergers with different time durations between first and second pericentric passage. At late times, these results are dominated by those mergers with longer durations. A few alternative ways to present our results include: (1) using `merging stages' \citep[i.e., as in][]{Veilleux2002,Haan2011,BarreraBallesteros2015,Hung2015,Larson2016,Smith2018}; (2) rescaling time by dividing by the time between first and second passage \citep{Privon2014}; and (3) using three-dimensional or randomly-selected projected separations \citep{Patton2013}. We elect to present our results as a function of time to display how these physical processes unfold more clearly.

\subsection{Merger Suite: The Structure of the ISM}
\label{subsec:results_suite_sfr}

Figure~\ref{fig:phases_vs_time_suites} shows the time evolution of gas mass excess in various temperature-density regimes in our merger suite (galaxy-pair period only). It generalises the bottom panel of Figure~\ref{fig:phases_vs_time_fid} (excluding regions masked in light grey). From top to bottom, panels are organised by regime: warm (orange), cool (green), cold-dense (blue) and hot (red). Thick (thin) solid lines refer to the median (top and bottom quartiles). Vertical axes on each panel have different ranges for display purposes. The index in the upper-right corner shows the median and quartiles (equation~\ref{eqn:index}). Each phase exhibits the following behaviour:

\begin{itemize}

\item {\bf Warm gas (merger suite):}\\
In general, interactions suppress the warm gas mass content (sample median $=$ 0.89). The intensity and duration of this depletion varies from merger to merger. \\

\item {\bf Cool gas (merger suite):}\\
Cool gas suppression is followed by a slow and steady recovery. By t$\sim$1 Gyr and after, most mergers in our suite exhibit mild cool gas excess ($\lesssim$10\% above unity). This depletion plus recovery results in a negligible overall mass excess (sample median $=$ 1.04). \\

\item {\bf Cold-dense gas (merger suite):}\\
Interacting galaxies in our suite exhibit an excess in cold-dense gas at all times (sample median $=$ 1.18). After a brief dip, the establishment of a reservoir of cold-dense gas occurs, which survives for several Gyrs. \\

\item {\bf Hot gas (merger suite):}\\
This component experiences the highest levels of mass excess (sample median $=$ 4.90), especially at pericentric passages. After the large boost at first pericentre, the hot gas excess stabilises to values generally above unity ($\sim$2-4). Dramatic spikes are driven by hot gas boosts at second pericentre for those mergers experiencing second approach within that window of time. (We remind the reader that our initial conditions do not include a hot gas atmosphere.)

\begin{figure}
\centerline{\vbox{
\vspace{-.08in}
\hbox{
\includegraphics[width=3.5in]{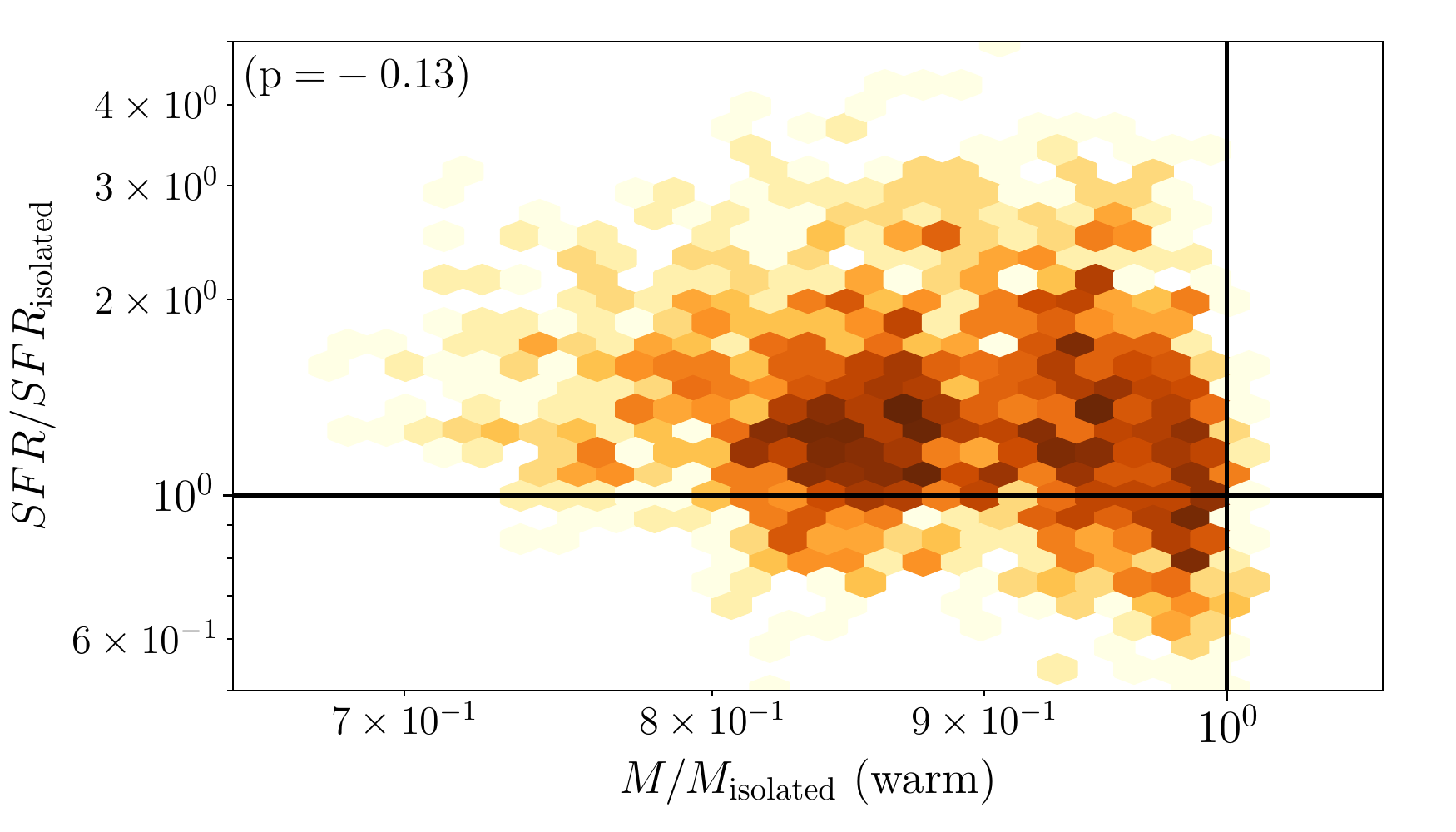}
}
\vspace{-.08in}
\hbox{
\includegraphics[width=3.5in]{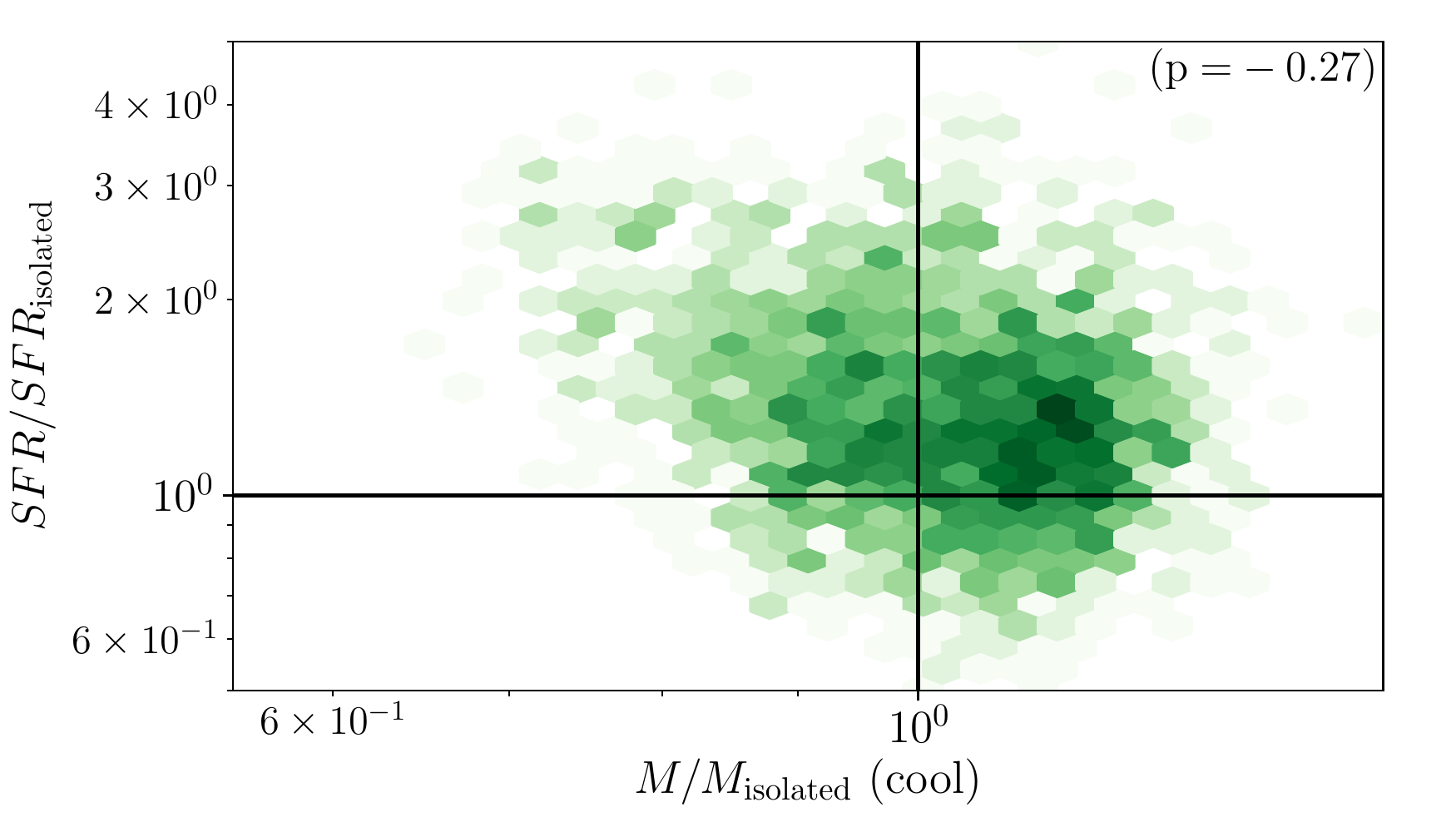}
}
\vspace{-.08in}
\hbox{
\includegraphics[width=3.5in]{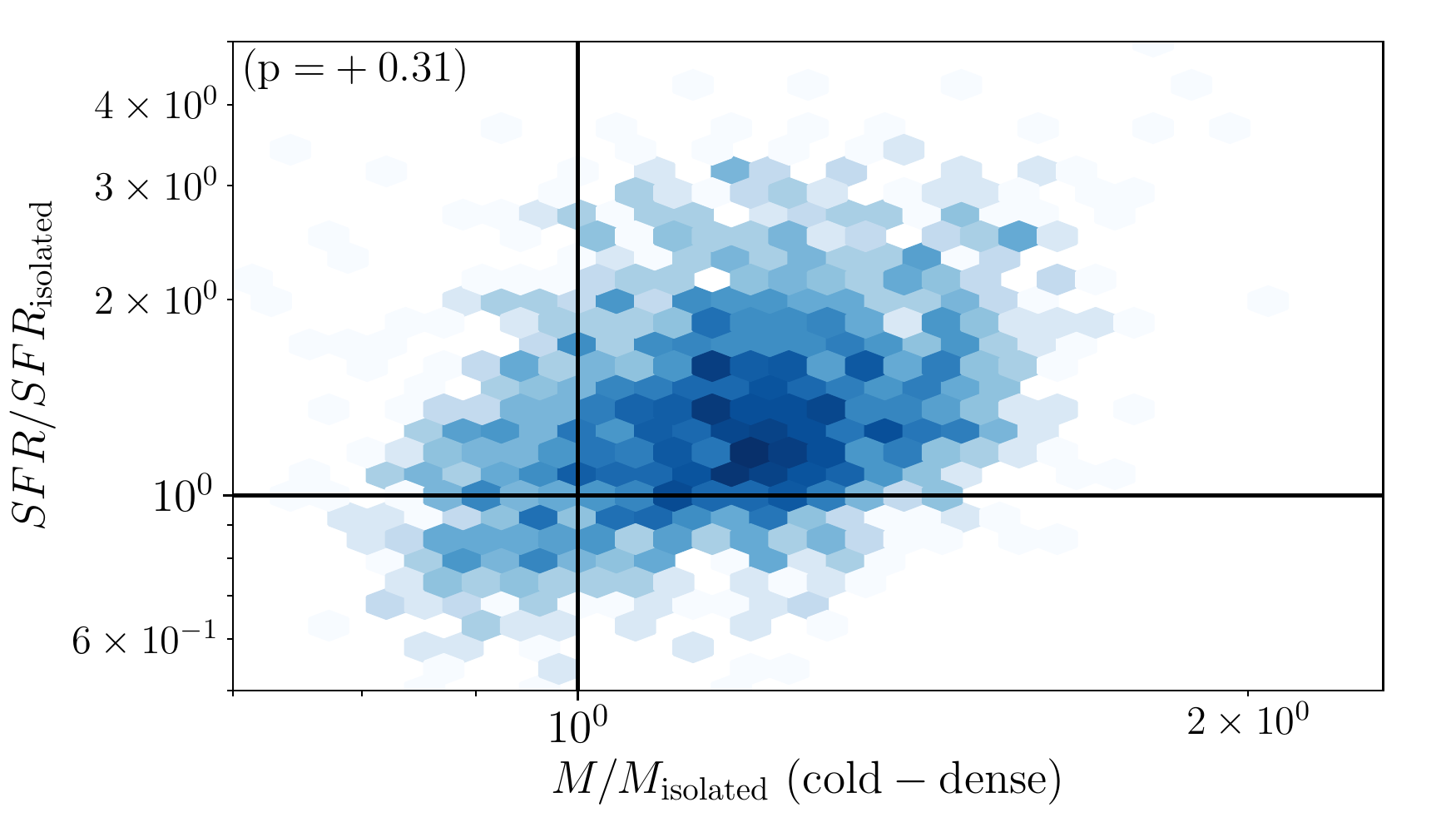}
}
\vspace{-.08in}
\hbox{
\includegraphics[width=3.5in]{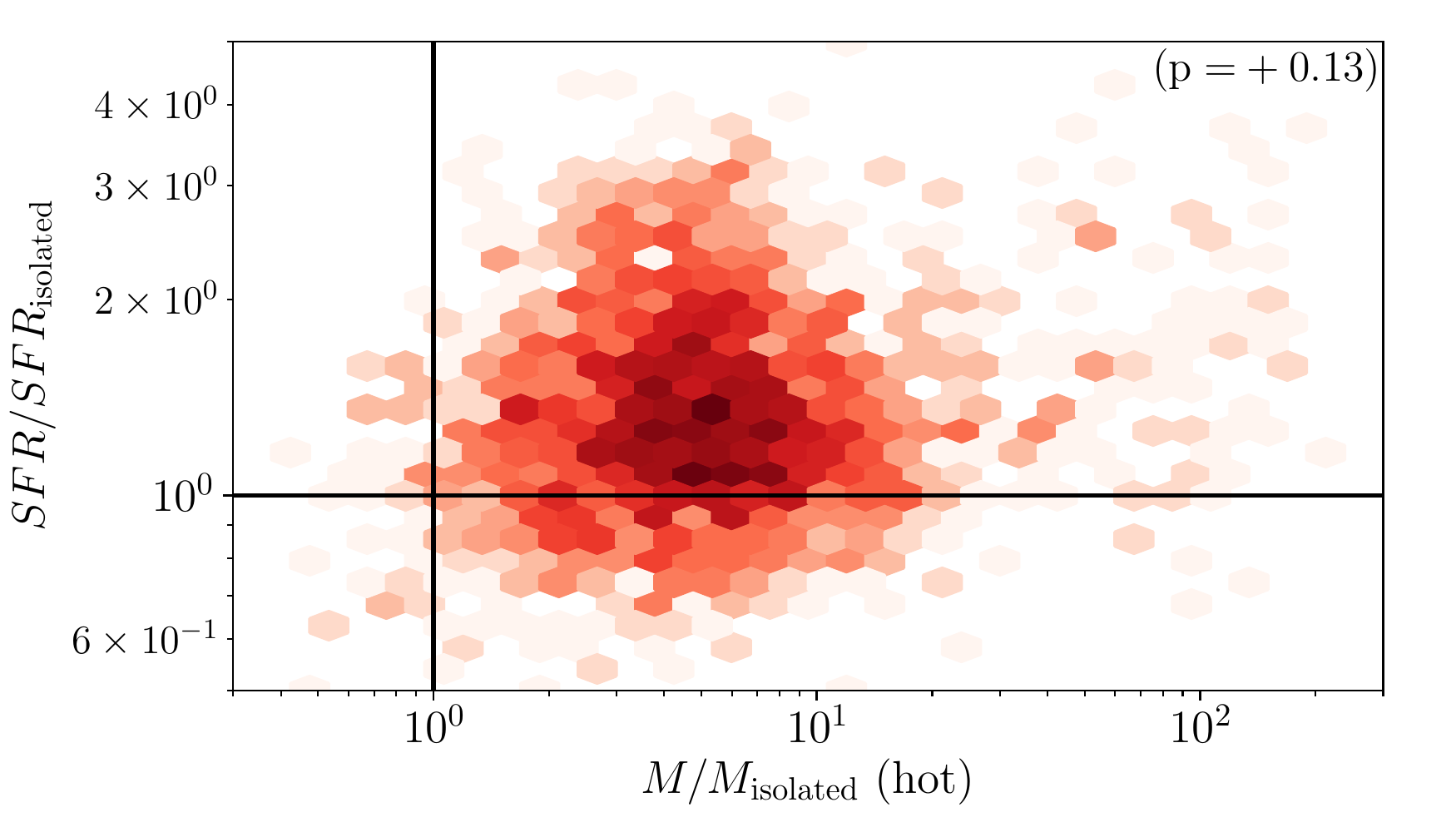}
}
\vspace{-.12in}
}}
\caption{SFR enhancement versus gas mass excess for each ISM temperature-density regime in our merger suite (galaxy-pair period only). {\it Top to bottom:} warm (orange), cool (green), cold-dense (blue) and hot (red). Darker hexagons indicate higher incidence of interacting systems per bin, displayed on a logarithmic colour scale. The Pearson correlation coefficient (equation~\ref{eqn:pearson}) is displayed on the top corner of each panel.}
\label{fig:sfr_vs_mass_pearson}
\end{figure}

\subsection{Merger Suite: Star Formation and its Connection to the ISM}
\label{subsec:interplay}

This section explores possible correlations between star formation and the evolution of the various temperature-density regimes in the ISM. Our goal is quantify potential correlations between SFR enhancement and gas mass enhancements or suppressions. Figure~\ref{fig:sfr_vs_mass_pearson} shows SFR enhancement versus mass excess for each regime (top-to-bottom): warm (orange), cool (green), cold-dense (blue) and hot (red). We display the corresponding Pearson correlation coefficient on the top corner of each panel. This is defined as
\begin{equation}
\label{eqn:pearson}
p_{X,Y} = \frac{{\rm cov}(X,Y)}{\sigma_X\sigma_Y},
\end{equation}
where cov$(X,Y)$ is the covariance between $X$ and $Y$, and $\sigma_X$ and $\sigma_Y$ are the standard deviations of $X$ and $Y$. Here, $\{p=1,-1,0\}$ refer respectively to perfect correlation, perfect anti-correlation, and zero correlation. We find the following connections to SFR enhancement:
\begin{itemize}
\item No correlation with warm gas mass suppression (upper-left panel, Pearson coefficient $=$ -0.13). 
\item Weak anti-correlation with cool gas mass excess (upper-right panel, Pearson coefficient $=$ -0.27). 
\item Weak correlation with cool-dense gas mass excess  (lower-left panel, Pearson coefficient $=$ +0.31).
\item No correlation with hot gas mass excess (lower-right panel, Pearson coefficient $=$ +0.13). 
\end{itemize}

\end{itemize}

On a final note, we comment on the bimodal structure of the 2D histogram in Figure~\ref{fig:sfr_vs_mass_pearson} corresponding to the warm gas component (top panel). We find (not shown) that the peak to the right of $M/M_{\rm isolated}$=0.9 is dominated by retrograde mergers, whilst the peak to the left of $M/M_{\rm isolated}$=0.9 is dominated by prograde and polar mergers. In other words, our simulations suggest that spin-orbit orientation governs the effectiveness of warm gas depletion. Older binary merger simulations show that the spin-orbit orientation plays a role in establishing the following: (1) the level of star formation enhancement \citep{DiMatteo2007,Moreno2015}; (2) the spatial extent of star formation \citep{Moreno2015}; and (3) the likelihood of the remnant to be a fast or a slow rotator \citep{Bois2011}. Recently, cosmological simulations have probed the physical role of orbital parameters in merging systems. \citet[][\small{EAGLE}]{Lagos2018} show that co-rotating mergers are more conducive to the enhancement of the stellar specific angular momentum, whilst \citet[][{Illustris}]{Li2018} find that most prolate galaxies are produced by nearly radial orbits. We defer a systematic study on the dependance of orbital parameters to a future paper.

\subsection{Inter-Regime Transition Rates}
\label{subsec:ecology}

To obtain a better physical picture, we investigate in detail how the various gas regimes feed and drain one another as star formation unfolds. To achieve this, in this section we explore the behaviour of mass transition rates, and how these build up each gas component. Here, we focus only on the fiducial case because we have shown that this run successfully captures the most relevant trends in our merger suite (Figures~\ref{fig:sfr_vs_time_fid} and \ref{fig:phases_vs_time_fid} versus Figures \ref{fig:sfr_vs_time_suites} and \ref{fig:phases_vs_time_suites}). Focusing on a single run allows us to provide a detailed narrative of some of the changes each regime experiences; although the details of these will be different for different mergers, the salient characteristics can be found throughout our merger suite.

Figures~\ref{fig:ionized_rate_vs_time_fid}, \ref{fig:atomic_rate_vs_time_fid}, \ref{fig:molecular_rate_vs_time_fid} and \ref{fig:hot_rate_vs_time_fid} show mass transition rates relevant to each of our four temperature-density regimes in the ISM: the warm, cool, cold-dense and hot gas component respectively. Vertical dashed lines indicate first and second pericentric passages, plus coalescence. Periods of time outside the galaxy-pair period are masked in light grey. We use $M_{\alpha}(t)$ to represent mass as a function of time (top panel of Figure~\ref{fig:phases_vs_time_fid}), where Greek indices refer to our four temperature-density gas regimes, plus new stars. The time derivatives, ${\rm d}M_\alpha(t)/{\rm d}t$, are shown in the top (bottom) panels as thick solid (dashed) lines fluctuating about zero for the interacting (isolated) case. These fluctuations produce upturns and downturns in $M_{\alpha}(t)$. Each derivative can expressed as the sum of net inter-regime transition rates:
\begin{equation}
\label{eqn:dmdt}
\frac{{\rm d}M_\alpha(t)}{{\rm d}t}=\sum_{\beta\neq\alpha}{\cal R}_{\alpha 
\leftrightarrow \beta}(t),
\end{equation}
where ${\cal R}_{\alpha \leftrightarrow \beta}$ represents the net rate at which two temperature-density regimes ``$\alpha$" and ``$\beta$" (labelled ``$\alpha$" $\leftrightarrow$ ``$\beta$" in the Figures) exchange mass. By definition,
\begin{equation}
\label{eqn:ralphabeta}
{\cal R}_{\alpha 
\leftrightarrow \beta}(t) = R_{\alpha \rightarrow \beta}(t) - R_{\beta \rightarrow \alpha}(t),
\end{equation}
where
\begin{equation}
\label{eqn:ralphatobeta}
R_{\alpha \rightarrow \beta}(t) = \frac{{\rm d}M(t)}{{\rm d}t}\big{|}_{\alpha \rightarrow \beta}
 \end{equation}
is the mass transition rate from regime ``$\alpha$" to regime ``$\beta$". 
In practice, to calculate these rates, we employ particle IDs across consecutive snapshots (separated by 5 Myr) and match gas elements that covert from one gas regime into another, or into new stellar particles. Here, net transition rates are shown as thin solid (dashed) lines in the top (bottom) panel for the interacting (isolated) case. For display purposes, we smooth over 25-Myr timescales. Different time binning changes the `jaggedness' of the curves, but does not alter our main results. Positive (negative) net transitions correspond to curves above (below) zero. Note that net transition rates are anti-symmetric: ${\cal R}_{\alpha \leftrightarrow \beta} = - {\cal R}_{\beta \leftrightarrow \alpha}$; i.e., vertical mirror images of the inverse transition. For example, the green line in Figure~\ref{fig:ionized_rate_vs_time_fid} (top panel, labelled ``warm $\leftrightarrow$ cool") is the negative (vertical mirror image) of the orange line in Figure~\ref{fig:atomic_rate_vs_time_fid} (top panel, labelled ``cool $\leftrightarrow$ warm").

Before describing how each gas regime exchanges mass with others in detail, a few comments are in order. First, we emphasise that our figures only display {\it net} transition rates, not regular rates -- i.e., ${\cal R}_{\alpha \leftrightarrow \beta}$ in equation~(\ref{eqn:ralphabeta}), as opposed to ${R}_{\alpha \rightarrow \beta}$ in equation~(\ref{eqn:ralphatobeta}). Secondly, whilst it is true that flows between two gas regimes occurs in both directions, our simulations show that inter-regime transitions tend to favour a preferred direction (thin lines rarely cross the horizontal zero line). We discuss specific cases where halting and reversal occurs, resulting in interesting behaviour. Thirdly, we note that the main effect caused by encounters is the amplification of net transition rates: solid lines in top panels (corresponding to interacting galaxies) versus dashed lines bottom panels (corresponding to isolated galaxies). We devote most of the discussion to merging galaxies during this period of time. Lastly, we note that the large fluctuations before first pericentric passage (left of the first dashed vertical line, top panel of Figure~\ref{fig:phases_vs_time_fid}) are caused primarily by the ISM of the two galaxies coming to contact before the two centres (marked by the positions of the two supermassive black holes) reach their minimum separation for the first time. The first two panels on the third row of Figure~\ref{fig:interaction_sequence} illustrate this. A secondary effect is the fact that, for the very first few snapshots, the galaxies in our simulation are becoming stable (i.e., spikes on the left-end of the bottom panels). We verified that stabilization-driven fluctuations disappears before the two galaxies come into first contact.

\begin{itemize}

\begin{figure}
\centerline{\vbox{
\hbox{
\includegraphics[width=3.4in]{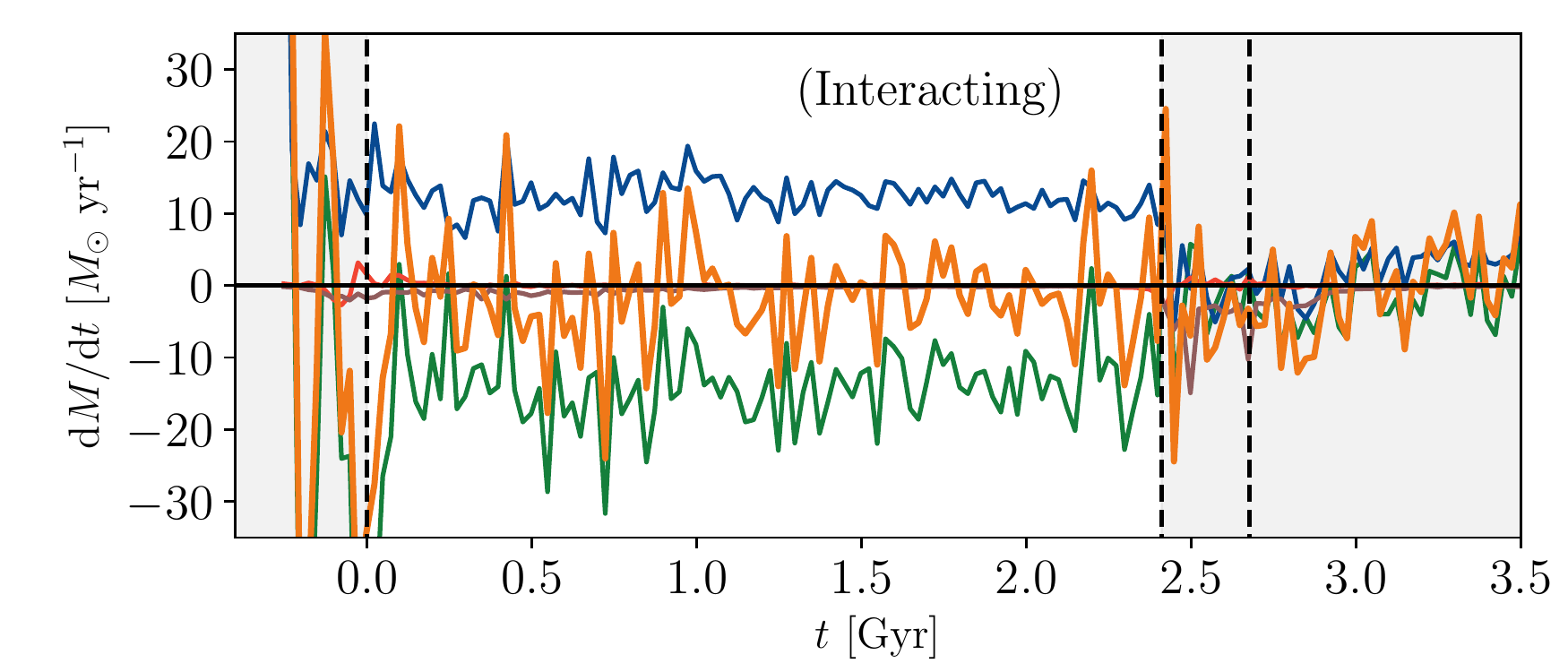}
}
\vspace{-.15in}
\hbox{
\includegraphics[width=3.4in]{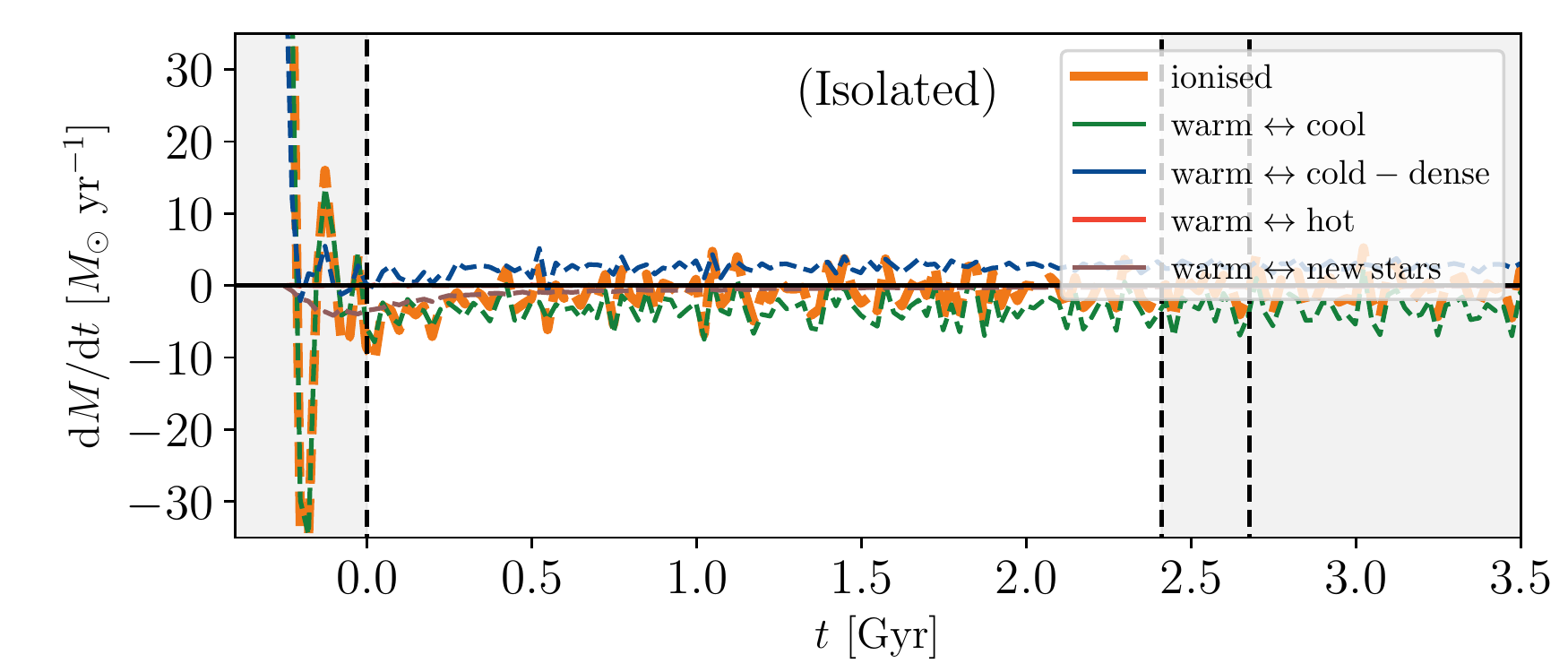}
}
\vspace{-.1in}
}}
\caption{
The interplay between warm gas and other gas temperature-density regimes. {\it Top panel:} Rate of change of warm mass versus time (thick orange) and contributions from net transitions with other regimes: warm $\leftrightarrow$ cool (thin green line); warm $\leftrightarrow$  cold-dense (thin blue line); warm $\leftrightarrow$ hot (thin red line); and warm  $\leftrightarrow$ new stars (thin brown line). {\it Bottom panel:} Same as middle panel, but for the isolated case. Dashed vertical lines indicate first and second pericentric passages, plus coalescence (see top panel of Figure~\ref{fig:fiducialcoverage} for definitions). Periods outside the galaxy-pair period are masked in light grey. Positive (negative) values indicate net gains (loses) from (into) other gas regimes. The evolution of the warm gas component is governed primarily by net influxes from cold-dense gas competing with net outfluxes into cool gas. See equations~(\ref{eqn:dmdt})-(\ref{eqn:ralphatobeta}) for definitions. In general, interactions amplify the magnitude of inter-regime transition rates.
} 
\label{fig:ionized_rate_vs_time_fid}
\end{figure}


\item {\bf Warm gas (transition rates, fiducial run):}\\
Figure~\ref{fig:ionized_rate_vs_time_fid} shows the evolution of the warm gas component and its interplay with other ISM regimes. In the top panel, the thick solid orange line represents the the time-derivative of mass in this regime as a function of time. This quantity is governed primarily by the competition of two inter-regime transitions: a net gain caused by cold-dense gas converting into warm gas (thin blue line) and a net loss caused by warm gas converting into cool gas (thin green line). In general, {\it the rate at which warm gas turns into cool gas exceeds the rate at which cold-dense gas turns into warm gas, resulting in an overall slow depletion of warm gas} (top panel).
Deviations from this general behaviour are caused by special circumstances. For instance, the brief recovery at t$\sim$0.3 Gyr (dashed and solid orange lines touching, top panel of Figure~\ref{fig:phases_vs_time_fid}) is driven by periods in which the net transformation of warm gas into cool gas is halted or reversed (green lines in top panel crossing into the positive side). Namely, this recovery is caused by brief periods where both the cool and cold-dense gas become warm gas -- possibly due to intense stellar feedback. \\\\

\begin{figure}
\centerline{\vbox{
\hbox{
\includegraphics[width=3.4in]{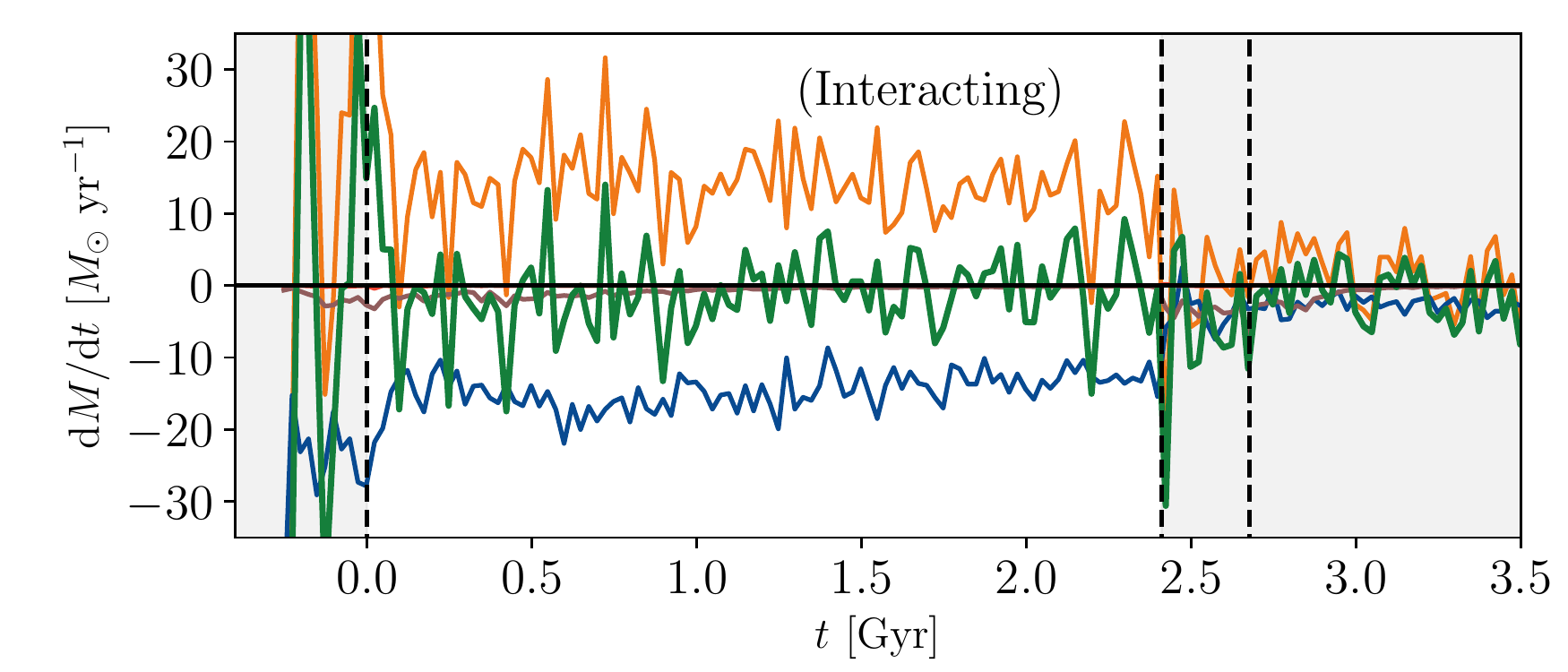}
}
\vspace{-.15in}
\hbox{
\includegraphics[width=3.4in]{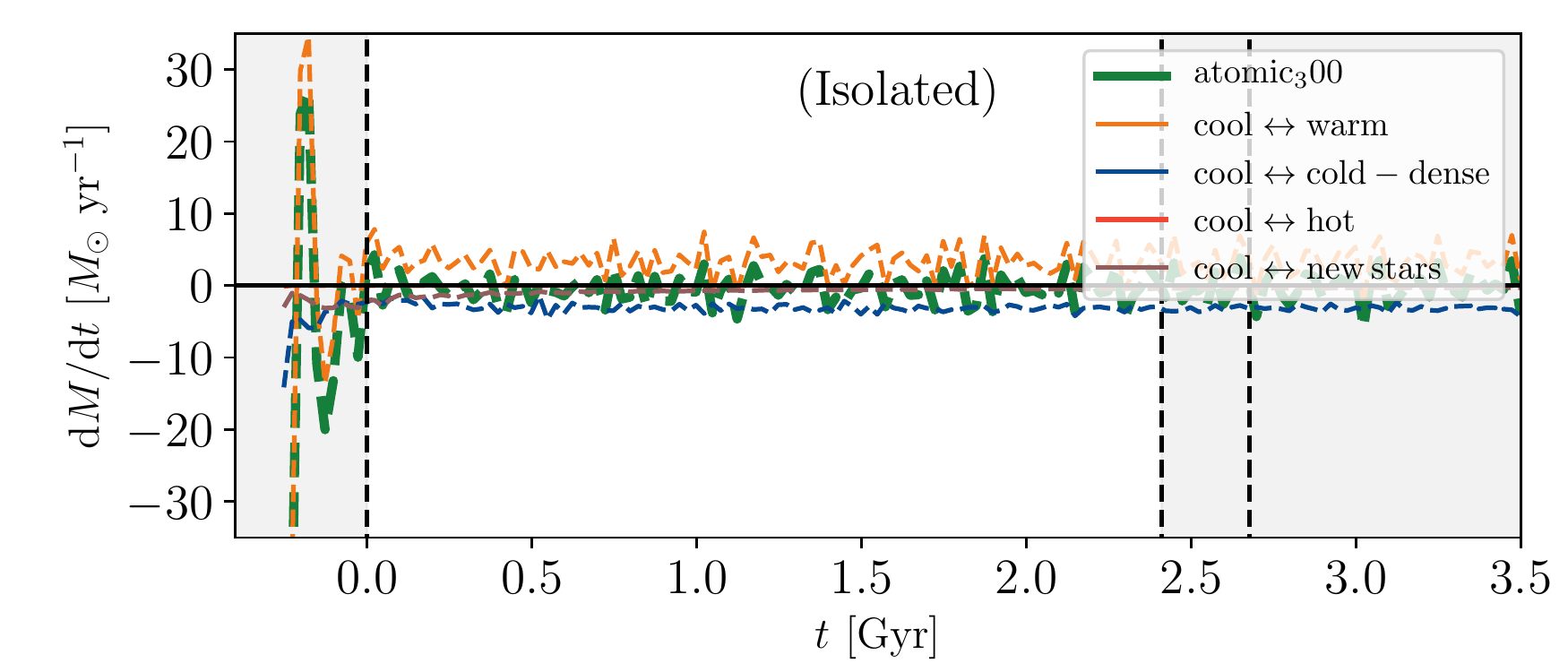}
}
\vspace{-.1in}
}}
\caption{
The interplay between cool gas and other gas temperature-density regimes. {\it Top panel:} Rate of change of cool mass versus time (thick green line) and contributions from net transitions with other regimes: cool $\leftrightarrow$ warm (thin orange line); cool $\leftrightarrow$ cold-dense (thin blue line); cool $\leftrightarrow$ hot (thin red line); and cool $\leftrightarrow$ new stars (thin brown line). {\it Bottom:} Same as middle panel, but for the isolated case. Dashed vertical lines indicate first and second pericentric passages, plus coalescence (see top panel of Figure~\ref{fig:fiducialcoverage} for definitions). Periods outside the galaxy-pair period are masked in light grey.  Positive (negative) values indicate net gains (loses) from (into) other gas regimes. The evolution of the cool gas component is governed primarily by net influx from warm gas competing with net outflux into cold-dense gas. See equations~(\ref{eqn:dmdt})-(\ref{eqn:ralphatobeta}) for definitions. In general, interactions amplify the magnitude of inter-regime transition rates.
}
\label{fig:atomic_rate_vs_time_fid}
\end{figure}

\item {\bf Cool gas (transition rates, fiducial run):}\\
Figure~\ref{fig:atomic_rate_vs_time_fid} shows the evolution of the cool gas component and its interplay with other ISM regimes. In the top panel, the thick solid green line represents the the time-derivative of mass in this regime as a function of time. This quantity is governed primarily by the competition of two inter-regime transitions: a net gain caused by warm gas converting into cool gas (thin orange line) and a net loss caused by cool gas converting into cold-dense gas (thin blue line).  In general, {\it the rate at which cool gas turns into cold-dense gas exceeds the rate at which warm gas turns into cool gas, resulting in a slow depletion of cool gas} (top panel).
Deviations from this general behaviour are caused by special circumstances. The boost in cool gas right after first passage (solid green line, top panel of Figure~\ref{fig:phases_vs_time_fid}) is caused by a high net influx of warm gas that is not compensated by a high-enough net loss into cold-dense gas (top panel, green spikes at t$\sim$0 Gyr). Also, strong dips in cool gas are caused by periods in which the net transformation of warm gas into cool gas is halted or reversed (top panel, orange lines crossing into the negative side).\\

\begin{figure}
\centerline{\vbox{
\hbox{
\includegraphics[width=3.4in]{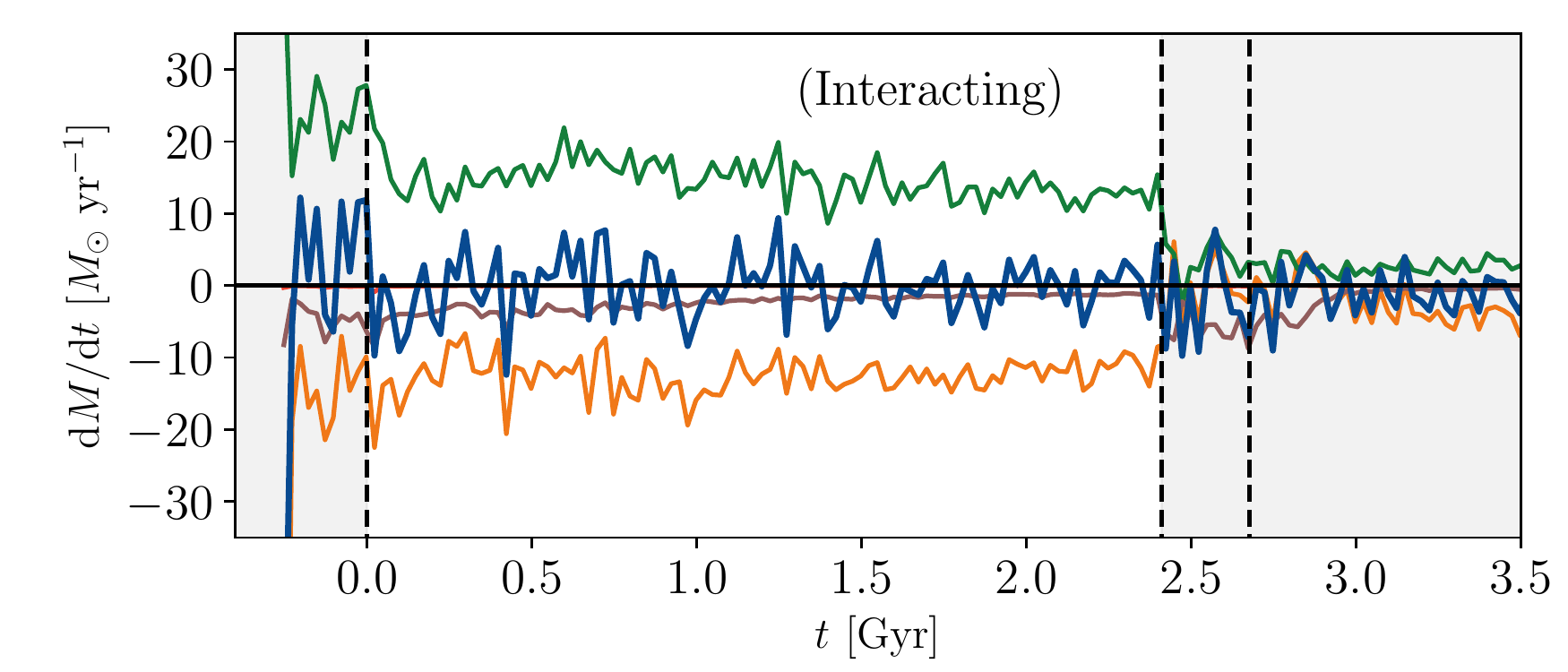}
}
\vspace{-.15in}
\hbox{
\includegraphics[width=3.4in]{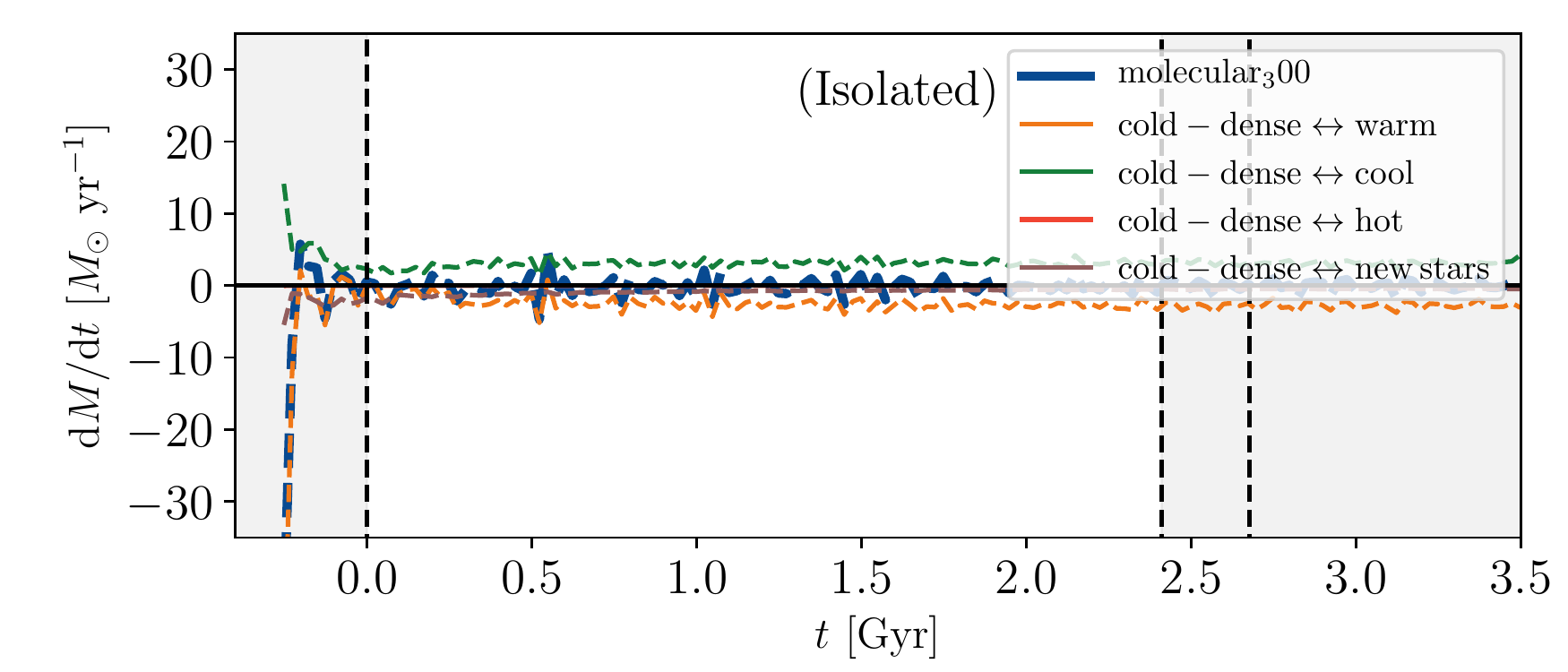}
}
\vspace{-.1in}
}}
\caption{
The interplay between cold-dense gas and other gas temperature-density regimes. {\it Top panel:} Rate of change of cold-dense mass versus time (thick blue line) and contributions from net transitions with other regimes: cold-dense $\leftrightarrow$ warm (thin orange line); cold-dense $\leftrightarrow$ cool (thin green line);  cold-dense $\leftrightarrow$ hot (thin red line); and cold-dense $\leftrightarrow$ new stars (thin brown line). {\it Bottom panel:} Same as middle panel, but for the isolated case.  Dashed vertical lines indicate first and second pericentric passages, plus coalescence (see top panel of Figure~\ref{fig:fiducialcoverage} for definitions). Periods outside the galaxy-pair period are masked in light grey.  Positive (negative) values indicate net gains (loses) from (into) other gas regimes. The evolution of the cold-dense gas component is governed primarily by net influx from cool gas competing with net outflux into warm gas and new stars. See equations~(\ref{eqn:dmdt})-(\ref{eqn:ralphatobeta}) for definitions. In general, interactions amplify the magnitude of inter-regime transition rates.
}
\label{fig:molecular_rate_vs_time_fid}
\end{figure}
 
\item {\bf Cold-dense gas (transition rates, fiducial run):}\\
Figure~\ref{fig:molecular_rate_vs_time_fid} shows the evolution of the cold-dense gas component and its interplay with other ISM regimes. In the top panel, the thick blue line represents the the time-derivative of mass in this regime as a function of time. This quantity is governed primarily by the competition of three transitions: a net gain caused by cool gas converting into cold-dense gas (thin green line), a net loss caused by cold-dense gas converting into warm gas (thin orange line), and the consumption of cold-dense gas into new stars (thin brown line). In general, {\it the net rate at which cold-dense gas turns into warm gas and new stars exceeds the net rate at which cool gas turns into cold-dense gas, resulting an overall slow depletion of cold-dense gas} (top panel). Note that depletion of cold-dense into warm gas tends to dominate over consumption of cold-dense gas into new stars.
Deviations from this general behaviour are caused by special circumstances. The boost in cold-dense gas right after first pericentric passage (solid blue line, top panel of Figure~\ref{fig:phases_vs_time_fid}) is caused by a high net influx of both cool and warm gas, possibly because gas is being compressed by the encounter (green and yellow peaks near the first vertical dashed line, middle panel). 
Another example occurs at $\sim$0.4 Gyr after first pericentric passage. For a brief period of time, {\it the conversion of cold-dense into warm gas is diminished, whilst the conversion of cool into cold-dense gas increases, causing the build up of a cold-dense gas reservoir} (orange lines approaching zero from below, middle panel).\\

\begin{figure}
\centerline{\vbox{
\hbox{
\includegraphics[width=3.4in]{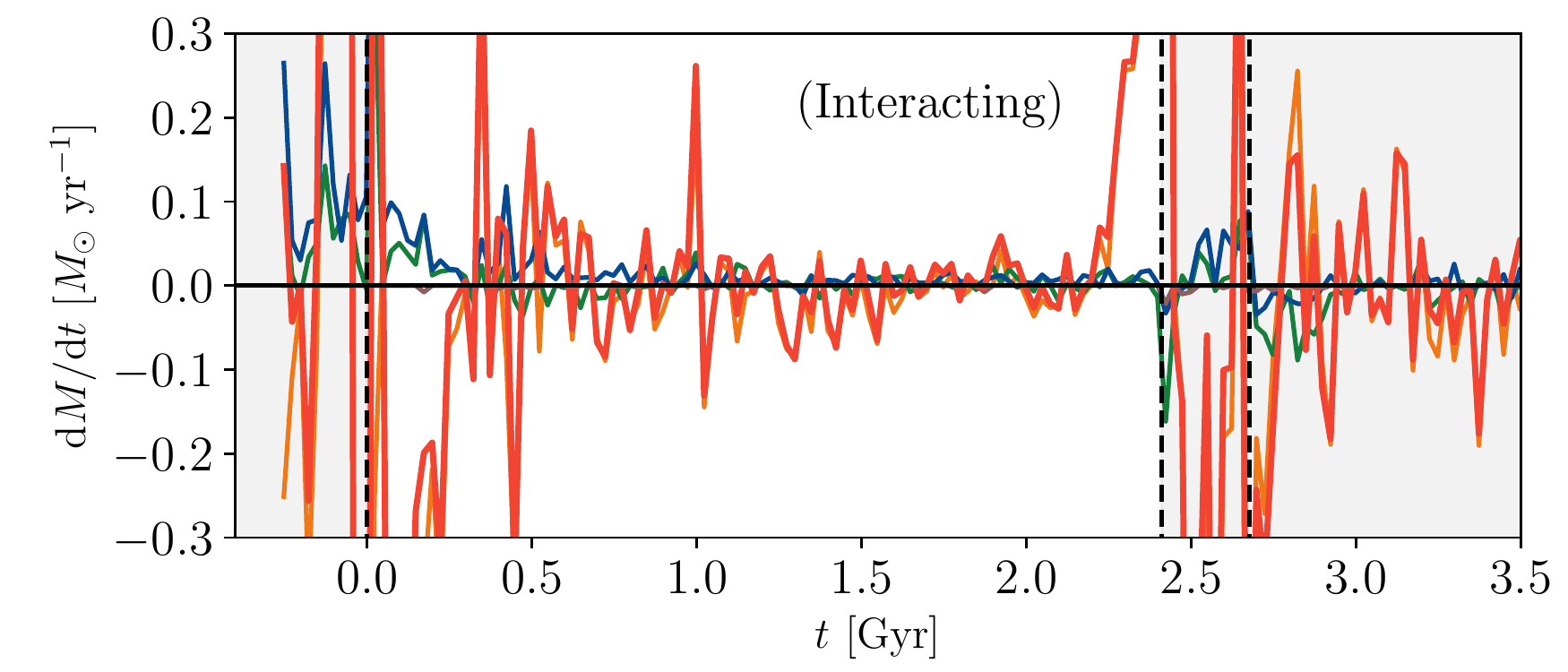}
}
\vspace{-.15in}
\hbox{
\includegraphics[width=3.4in]{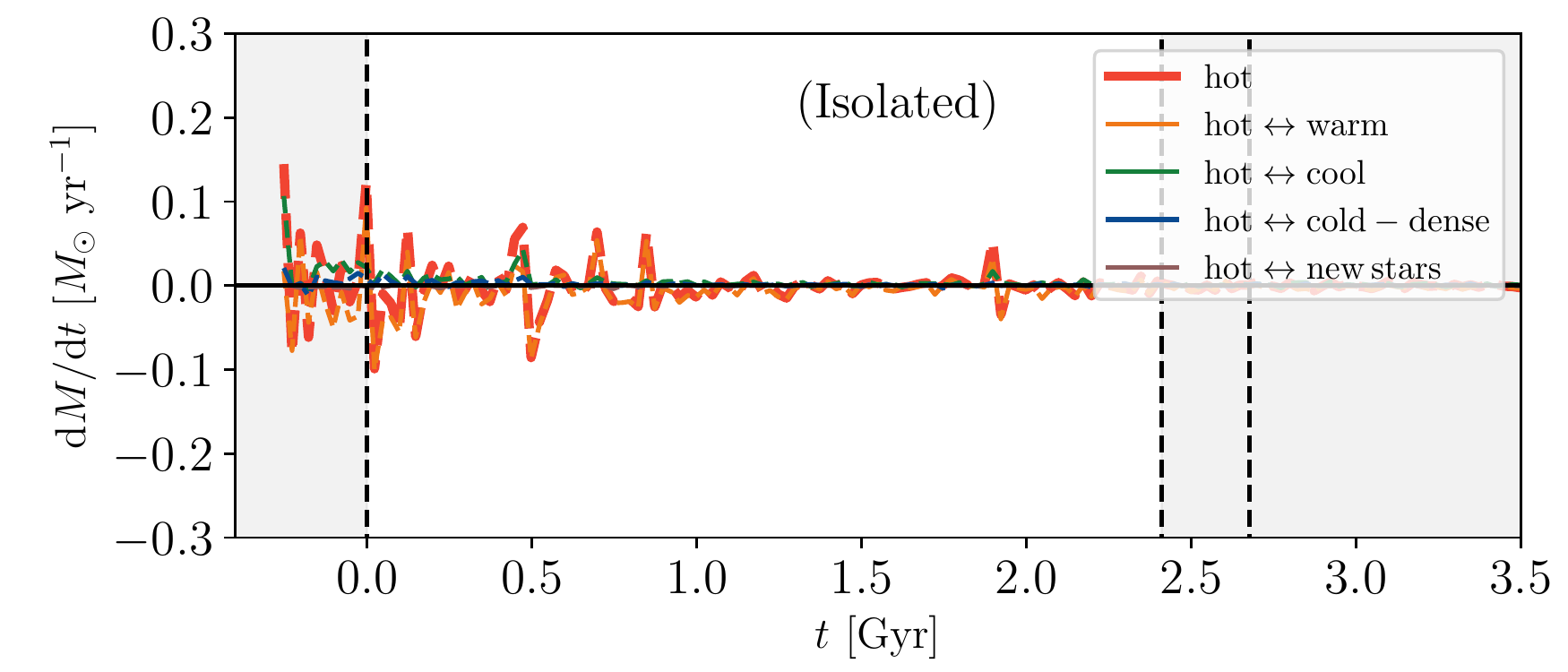}
}
\vspace{-.1in}
}}
\caption{
The interplay between hot gas and other gas temperature-density regimes. {\it Top panel:} Rate of change of hot mass versus time (thick blue) and contributions from net transitions with other regimes: hot $\leftrightarrow$ warm (thin orange line);  hot $\leftrightarrow$ cool (thin green line); hot $\leftrightarrow$ cold-dense (thin blue line); and hot $\leftrightarrow$ new stars (thin brown line). {\it Bottom panel:} Same as middle panel, but for the isolated case. Dashed vertical lines indicate first and second pericentric passages, plus coalescence (see top panel of Figure~\ref{fig:fiducialcoverage} for definitions). Periods outside the galaxy-pair period are masked in light grey.  Positive (negative) values indicate net gains (loses) from (into) other gas regimes. The evolution of the hot component primarily driven by influx from and outflux into warm gas, with minor influxes from cold-dense and cool gas. See equations~(\ref{eqn:dmdt})-(\ref{eqn:ralphatobeta}) for definitions. In general, interactions amplify the magnitude of inter-regime transition rates.
}
\label{fig:hot_rate_vs_time_fid}
\end{figure}

\item {\bf Hot gas (transition rates, fiducial run):}\\
Figure~\ref{fig:hot_rate_vs_time_fid} shows the evolution of the hot component and its interplay with other ISM regimes. In the middle panel, the thick solid red line represents the the time-derivative of mass in this regime as a function of time. This quantity is primarily governed by the net mass interchange between the hot and warm regimes. In particular, {\it the boosts and drops in hot gas right before and after pericentric passages are caused by large net exchanges with the warm gas component}. In other words, the hot gas ``follows" the warm gas. Because the mass in hot gas is orders of magnitude smaller than the mass in warm gas, the warm gas component is largely unaffected by the fluctuations experienced by the hot gas component. Exceptions to this behaviour in which warm gas ``controls" the hot gas take place in the early stages of interaction. During the first $\sim$0.4 Gyr after first passage, the sudden drop in hot gas (caused by a net transformation into warm gas) is mildly mitigated by small gains caused by the net conversion of some cold-dense and cool gas into hot gas (solid red line, top panel of Figure~\ref{fig:phases_vs_time_fid}).  We speculate that this is driven by feedback associated with enhanced star formation. These net losses do not affect the cold-dense and cool components because the hot gas mass budget is significantly smaller than the budget contributions from those two components.
The amplitude of phase transitions involving hot gas are generally small - except near pericentric passages, where hot gas is produced as gas located in the outskirts of galaxies is shock heated by the encounter.\\

\end{itemize}

\section{Discussion}
\label{sec:discussion}

\begin{figure}
\centerline{\vbox{
\vspace{-.12in}
\hbox{
\includegraphics[width=3.5in]{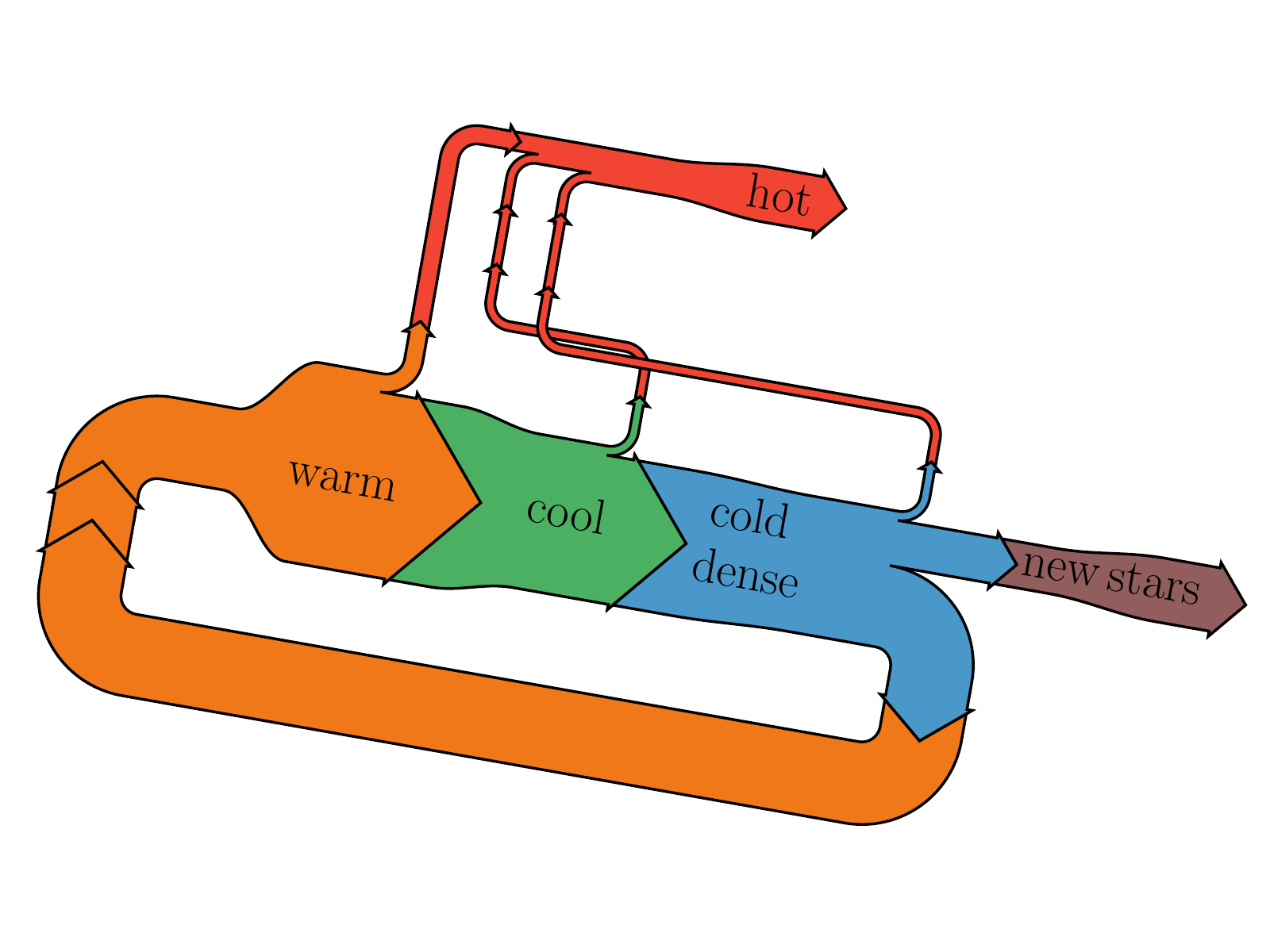}
}
\vspace{-.12in}
}}
\caption{Schematic of the inter-regime mass exchange pipeline in our simulated galaxies. From top-left to bottom-right: warm (orange) $\rightarrow$ cool (green) $\rightarrow$ cold-dense (blue) $\rightarrow$ \rm new stars (brown). This pipeline loops back, showing that transitions from cold-dense back to warm gas dominate over star formation. All three temperature-density gas regimes feed the hot component (red), though this process is dominated by mass exchanges with the warm component. Channel thickness scales approximately with the relative importance of each transition (not to scale). Arrows indicate typical net transition rate directions (equation~\ref{eqn:ralphabeta}), which can be reversed occasionally. See Figures~\ref{fig:ionized_rate_vs_time_fid}, \ref{fig:atomic_rate_vs_time_fid} \ref{fig:molecular_rate_vs_time_fid} and \ref{fig:hot_rate_vs_time_fid} for specific net transition rates in our fiducial run. }
\label{fig:flowchart}
\end{figure}

\subsection{An Emerging Picture}
\label{subsec:ecology}

The central goal of this paper is to investigate how various temperature-density regimes in the ISM interact with one another, and their connection to star formation. The {\small FIRE-2} model, which has the ability to resolve the multi-phase structure of the ISM, is ideal to achieve this goal. By investigating net transition rates between ISM temperature-density regimes, we learn that the content in each regime is governed by the competition of net transitions between each regime and two other ``neighbouring" regimes:
\begin{itemize}
\item cold-dense $\rightarrow$ warm  \,\,\,\,\,\,\,\,\,\,\,\,\,$\rightarrow$ cool,
\item warm \,\,\,\,\,\,\,\,\,\,\,\,$\rightarrow$ cool \,\,\,\,\,\,\,\,\,\,\,\,\,\,\,\,\,$\rightarrow$ cold-dense,
\item cool \,\,\,\,\,\,\,\,\,\,\,\,\,\,\,\,$\rightarrow$ cold-dense $\rightarrow$ warm ($+$ new stars),
\end{itemize}
with the hot component feeding from the other regimes:
\begin{itemize}
\item warm ($+$ cool and cold-dense) $\rightarrow$ hot.
\end{itemize}
We illustrate this with a Sankey flow diagram in Figure~\ref{fig:flowchart}. The thickness of each pipe approximately represents typical transition rate amplitude (not drawn exactly to scale). At the end of the pipeline, a fraction of the cold-dense gas turns into new stars, but most of it loops back into warm gas. 
We find that galaxy-galaxy interactions are responsible for the following interesting effects:
\begin{itemize}
\item Warm gas depletion is amplified.
\item Cool gas depletion is amplified, especially early.
\item A reservoir of extra cold-dense gas is enhanced.
\end{itemize} 
Inter-regime net transition rates offer a physical explanation. Namely, whilst the direction of flow in this pipeline is approximately steady and one-directional, interactions have the potential to accelerate, halt, or even reverse the direction of these inter-regime transitions.

\begin{figure}
\centerline{\vbox{
\hbox{
\includegraphics[width=3.4in]{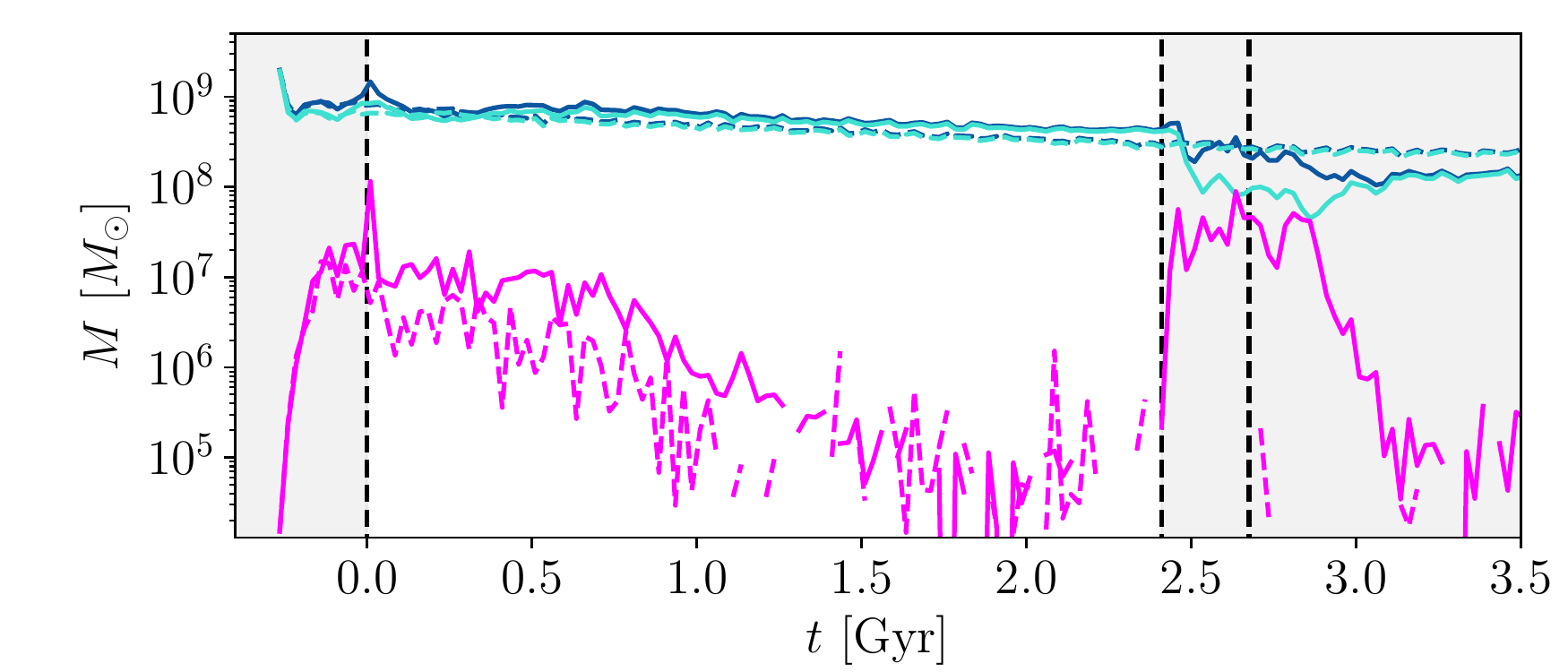}
}
\vspace{-.15in}
\hbox{
\includegraphics[width=3.4in]{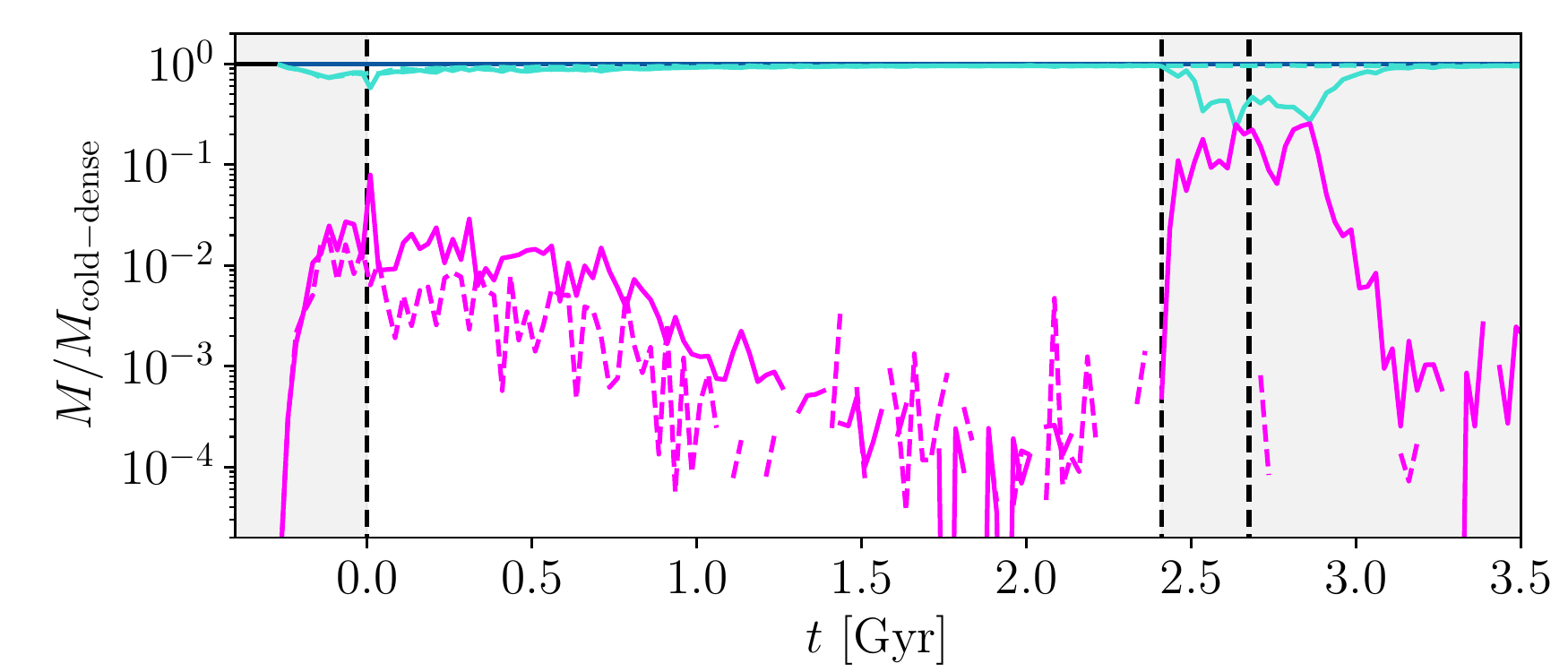}
}
\vspace{-.15in}
\hbox{
\includegraphics[width=3.4in]{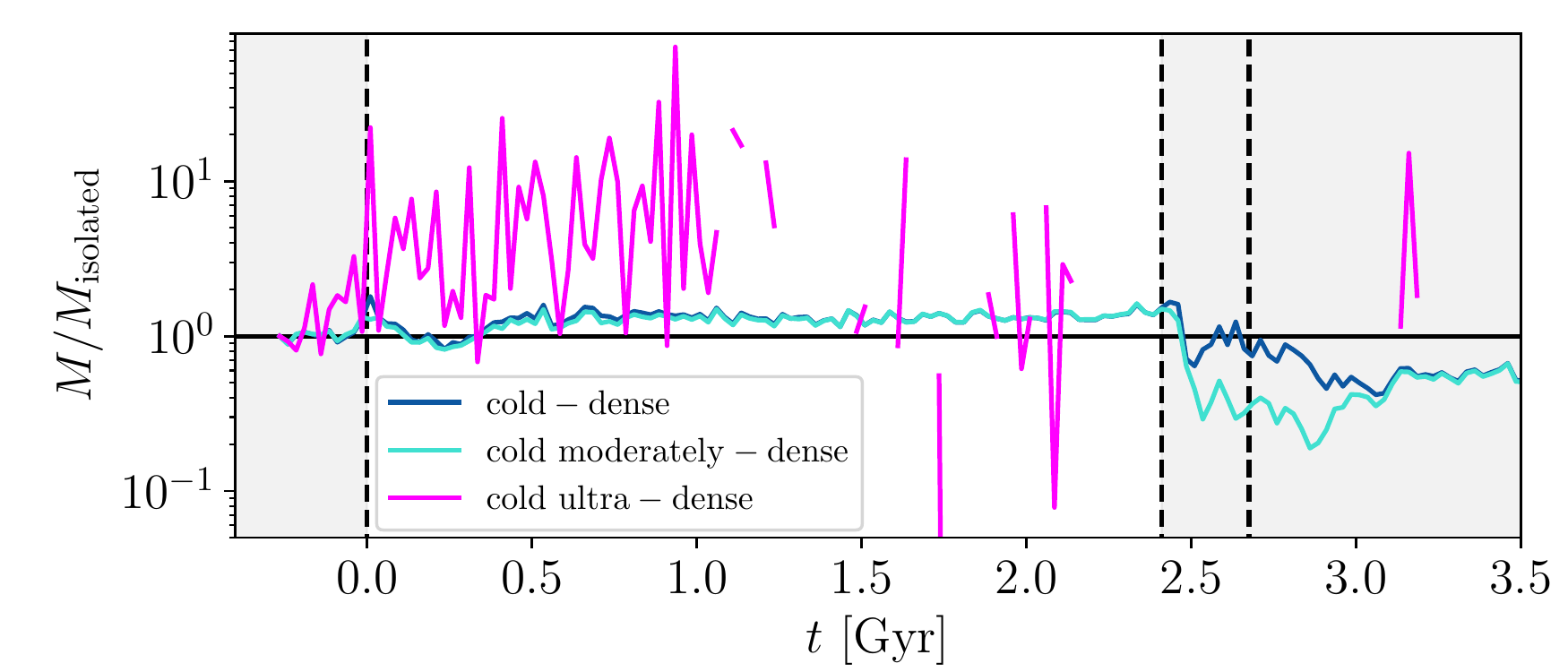}
}
\vspace{-.1in}
}}
\caption{
The evolution of cold-dense gas mass in our fiducial run, further split by density. The entire cold-dense gas content (blue) is split into cold moderately-dense ($n=10-1000 \,\,{\rm cm}^{-3}$, cyan) and cold ultra-dense ($n>1000 \,\ {\rm cm}^{-3}$, magenta) gas. {\it Top panel:} Mass versus time. Solid (dashed) lines indicate interacting (isolated) runs. {\it Middle panel:} Fraction of cold-dense gas in each density subset versus time. {\it Bottom panel:} Mass excess versus time. Dashed vertical lines indicate first and second pericentric passages, plus coalescence. Times outside the galaxy-pair period are masked in light grey. Most of mass in the cold-dense gas reservoir consists of cold moderately-dense gas; cold ultra-dense gas accounts for only, at most, a few percent of the entire budget (during the interacting periods). Excess in cold ultra-dense gas is a better tracer of SFR enhancement than excess in the entire cold-dense or the cold moderately-dense gas component.
}
\label{fig:molecular_fid}
\end{figure}
\begin{figure}
\centerline{\vbox{
\vspace{-.05in}
\hbox{
\includegraphics[width=3.5in]{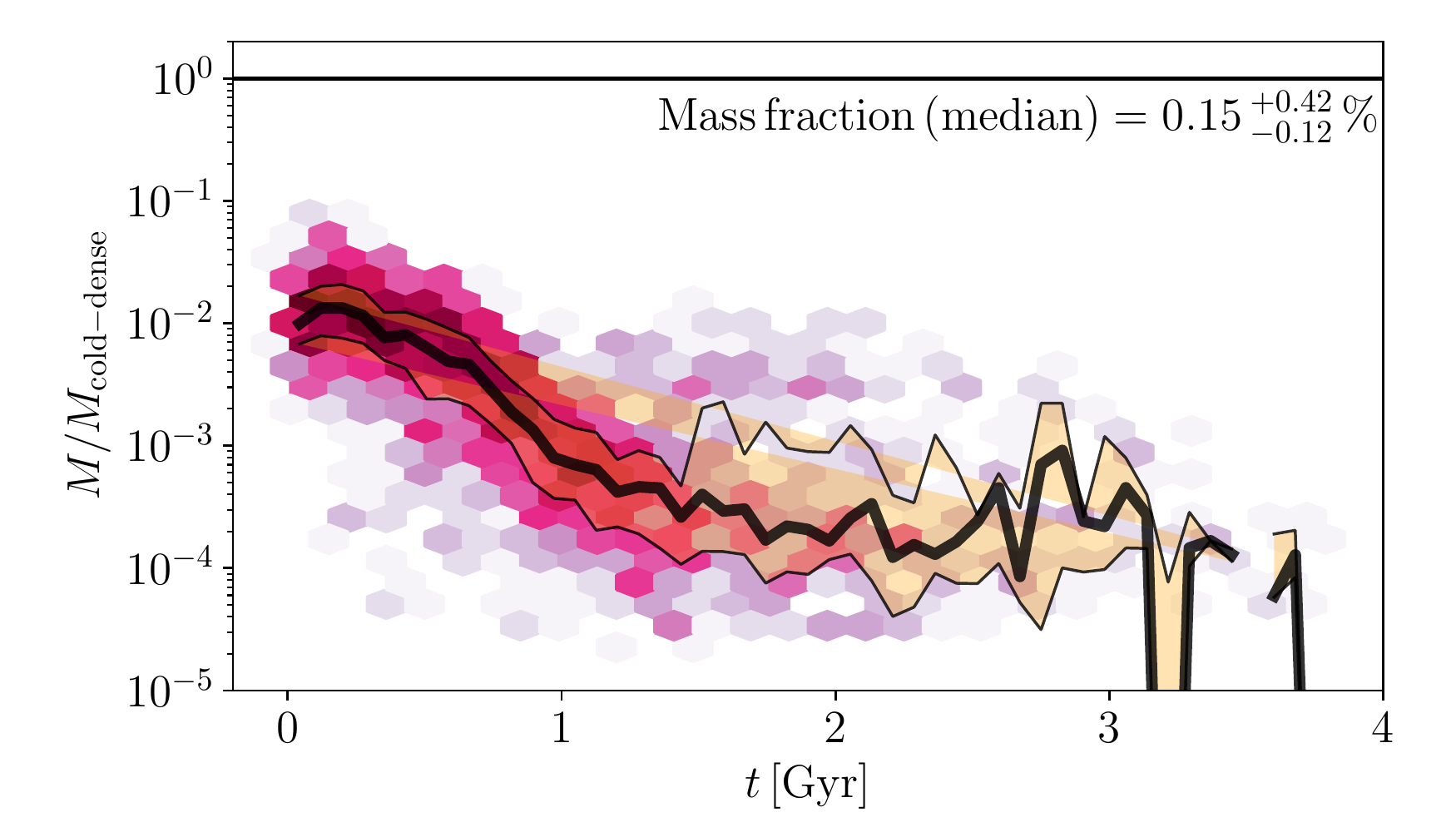}
}
\vspace{-.18in}
\hbox{
\includegraphics[width=3.5in]{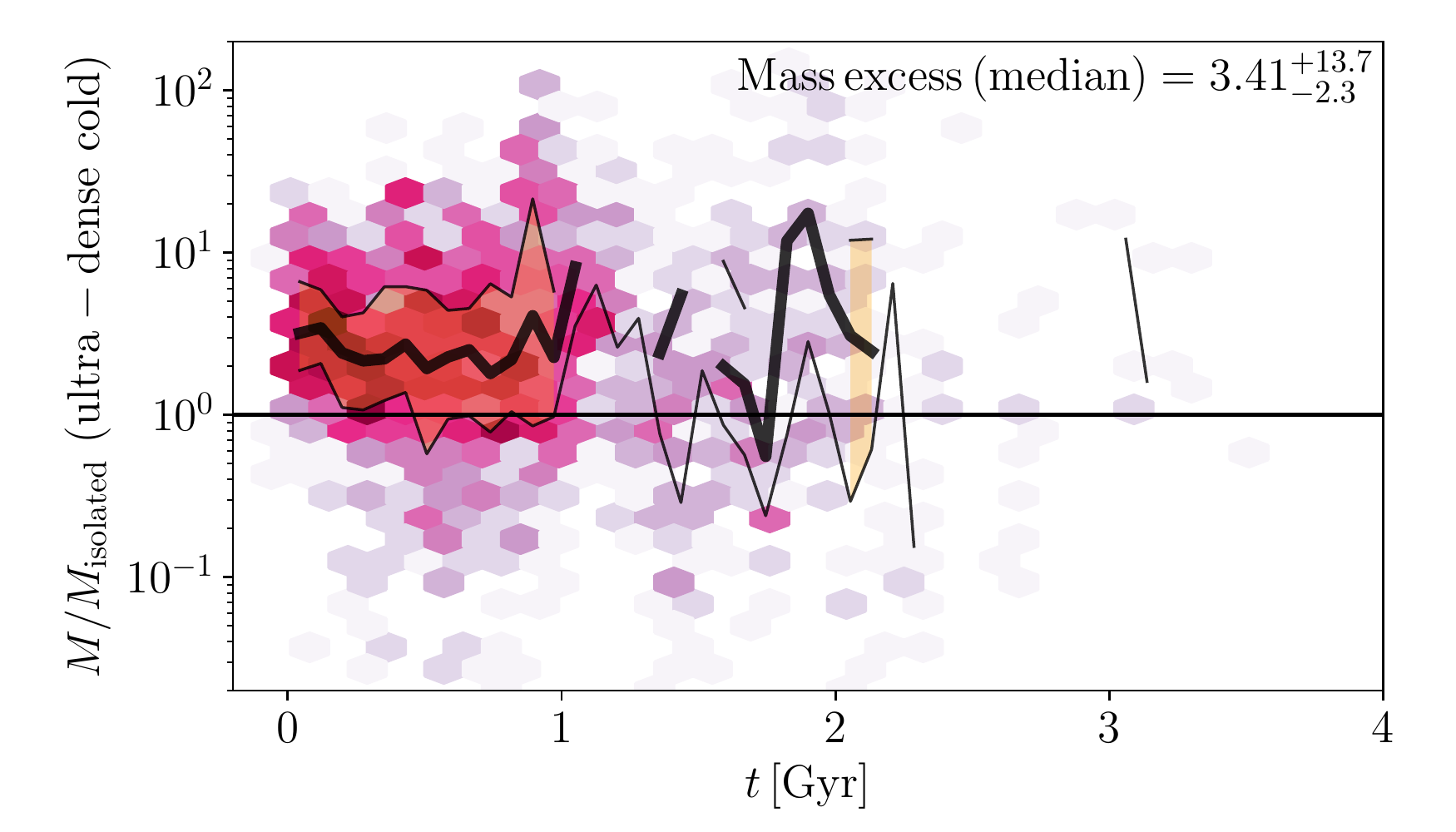}
}
\vspace{-.05in}
\hbox{
\includegraphics[width=3.5in]{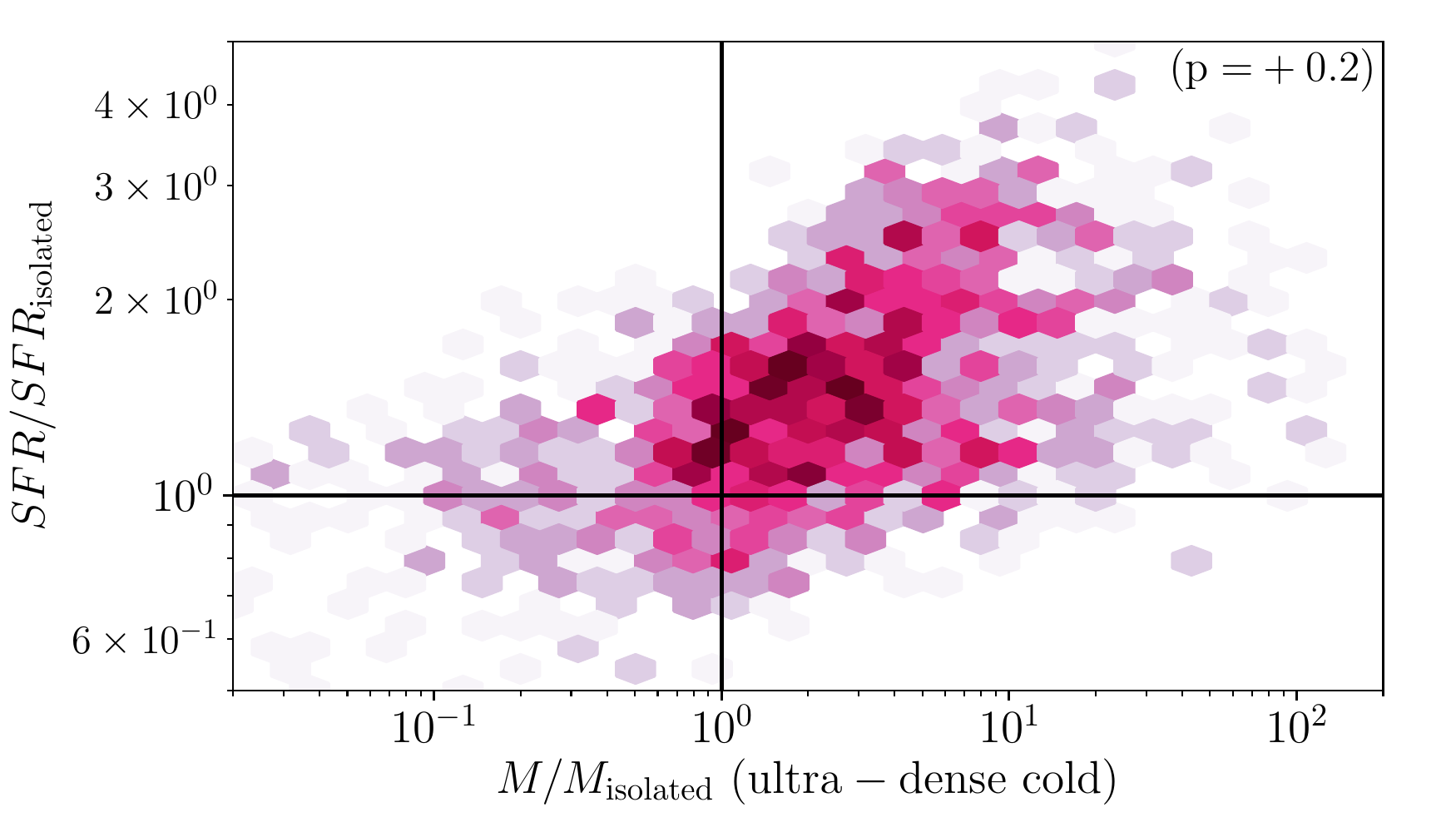}
}
\vspace{-.08in}
}}
\caption{Cold ultra-dense gas ($n>1000 \,\,{\rm cm}^{-3}$) versus time in our merger suite (galaxy-pair period only). {\it Top panel:} Fraction of cold-dense gas with densities greater than 1000 ${\rm cm}^{-3}$ versus time. {\it Middle panel:} Cold ultra-dense gas mass excess versus time. Darker hexagons indicate higher incidence of interacting systems per bin, displayed on a logarithmic colour scale. Thick (thin) solid lines indicate the median (top and bottom quartiles). The index in the upper-right corner represents the sample median and the lower-to-upper quartile range (top and middle panels). See equation~(\ref{eqn:index}) for format. {\it Bottom panel:} SFR enhancement versus excess in cold ultra-dense gas. Only galaxy-pair periods are included: between first and second pericentric passage, with time set to t = 0 Gyr at first passage. The Pearson correlation coefficient (equation~\ref{eqn:pearson}) is displayed on the top corner. On average, galaxy interactions strongly enhance the amount of cold ultra-dense gas mass. We find no significant correlation between enhanced SFR and excess in cold ultra-dense gas mass. }
\label{fig:dense_molecular_suite}
\end{figure}

\subsection{Cold Ultra-Dense Gas}
\label{subsec:fuelling}

Our simulations show that the gradual depletion of cold-dense gas is driven primarily by a net conversion into warm gas. Occasionally, depletion is halted and reversed, resulting in the replenishment of a cold-dense gas reservoir, which is depleted on much longer timescales. This build up of cold-dense gas is driven by a brief period in which the transition from cool to cold-dense gas dominates. Figure~\ref{fig:sfr_vs_mass_pearson} suggests that cold-dense gas excess and SFR enhancement in interacting galaxies are mildly correlated. By construction, SFR in the {\small FIRE-2} model occurs in only the very densest portions of the cold-dense gas budget, as expected by observations \citep{Lada2012}. The focus of this section is to discuss the behaviour of cold ultra-dense gas (with $n>1000 \, {\rm cm}^{-3}$) in our simulations. 

Figure~\ref{fig:molecular_fid} (top panel) shows the evolution of cold-dense gas (blue) in our fiducial run, split into {\it cold moderately-dense} gas  ($n=10-1000 \,{\rm cm}^{-3}$, cyan) and {\it cold ultra-dense} gas ($n>1000 \,\, {\rm cm}^{-3}$, magenta). The density cut at $n=1000 \, {\rm cm}^{-3}$ captures all the star-forming cold-dense gas (by construction). Note that the ultra-dense gas component between apocentre and second passage is close to our mass resolution limit ($1.4\times10^{4}$ $M_{\odot}$, Table~\ref{table:specs}), resulting in broken lines. The middle panel shows the fraction of cold-dense gas in the cold moderately-dense (cyan) and cold ultra-dense (magenta) components. Overall, during the galaxy-pair period, cold ultra-dense gas only accounts for, at most, a few percent of the entire budget. After apocentre, it drops below a thousandth. The bottom panel shows interaction-induced mass excesses. Whilst the cold-dense and cold moderately-dense budgets are elevated by factors of only $\sim$1-2, the cold ultra-dense gas mass increases by a factor higher than $\sim$10. 

Figure~\ref{fig:dense_molecular_suite} explores the behaviour of the cold ultra-dense  gas content across our entire merger suite. The top panel shows the fraction of cold-dense gas with densities higher than $n=1000 \,\, {\rm cm}^{-3}$. We find that, on average, only $\sim$0.15\% of the cold-dense gas mass in our simulations achieves such high densities. The middle panel shows mass excess. We find that interactions strongly elevate the amount of ultra-dense gas in galaxies, by a factor of $\sim$3.41 on average. The bottom panel explores the possibility of a correlation between SFR enhancement and excess of ultra-dense cold gas in interacting galaxies. By construction, and motivated by observations \citep{Lada2012}, cold ultra-dense gas, as selected here, corresponds to star-forming gas. However, we only find a mild correlation (Pearson coefficient $=$ 0.2) when `enhanced'/`excess' quantities are considered. We speculate that this scatter is driven by the fact that we are exploring the high-density tail of the gas density function, in a regime where our simulated galaxies contain only a handful (if any) of gas elements at those densities. Higher resolution is required to accurately investigate this correlation.

\subsection{Connection to Observations}
\label{subsec:observations}

Our suite of simulations shows that {\it close interactions enhance star formation in galaxies} by $\sim$30\% on average (Figure~\ref{fig:sfr_vs_time_suites}), confirming predictions by earlier simulations \citep{MH96, DiMatteo2007, DiMatteo2008,Teyssier2010,Renaud2013,Hayward2014,Moreno2015}, which are qualitatively consistent with observations \citep{Woods2006,SloanClosePairs,Lambas2012,SloanClosePairs, Ellison2013, Patton2013, Scott2014}. 
Beyond this, the central goal of this paper is to connect star formation enhancement with the behaviour of the various temperature-density ISM regimes in interacting galaxies. Because of its ability to capture the multi-phase structure of the ISM, {\small FIRE-2} is well positioned for this purpose \citep{FIRE,FIRE2}. Below we discuss connections between our simulation work and observations.\\

\begin{itemize}

\item {\bf Cool gas (connection to observations):}\\
We predict that {\it close interactions do not elevate or suppress cool-gas mass levels in galaxies significantly during the galaxy-pair period} (median $\sim$4\% increase -- Figure~\ref{fig:phases_vs_time_suites}, upper-right panel). This median value is the combination of a brief period of mass suppression during the first Gyr after first pericentric passage, followed by a recovery period in the cool gas mass content. Beyond $\sim$1.5 Gyr after first passage, the majority of our mergers exhibit elevated cool gas mass content. To connect the evolution of this component with observations, HI is the standard tracer of cool-gas in galaxies. 

There are currently conflicting indications in the observational literature concerning the HI fraction in merging galaxies. Some studies \citep[e.g.,][]{Braine1993,Ellison2015,Zuo2018} have found no difference in the HI fractions of merging and control galaxies, whereas others \citep[][Ellison et al. 2018]{Casasola2004,Huchtmeier2008,Jaskot2015,Janowiecki2017} find enhanced gas fractions. One of the main observational challenges in determining the HI content in observations is that the vast majority of data are obtained with single dish telescopes. Such facilities have very large beams, often several arcminutes in diameter. Resolving the two components in an interacting pair is therefore often not possible. Although corrections can be attempted for this blending \citep[e.g.,][]{Zuo2018}, it is nonetheless an uncertainty in the analysis. Along this vein, cosmological simulations show that this effect can overestimate the neutral-hydrogen content in galaxies, especially in satellites \citep[][IllustrisTNG]{Stevens2018}. Alternatively, the blending problem can be mitigated with high spatial resolution interferometric observations. Some individual galaxies have been mapped in detail this way \citep[e.g.,][]{Hibbard1994,Koribalski2004}. Whilst such studies can reveal the detailed structure of HI in mergers, they cannot be used for statistical determination of changes in gas fractions. \citet{Scudder2015} presented a larger sample of close pairs observed with the Very Large Array (VLA), in order to investigate how star formation rate enhancement correlates with gas fraction. However, the lack of a suitable control sample meant that \citet{Scudder2015} were not able to quantify how enhanced/depleted these galaxies were compared to non-interacting galaxies.\\

\item {\bf Cold-dense gas (connection to observations):}\\
Our simulations predict that {\it close interactions elevate cold-dense gas mass content in galaxies during the galaxy-pair period} (median $\sim$18 \% increase -- Figure~\ref{fig:phases_vs_time_suites}, lower-left panel). If we follow other published works in the literature \citep[e.g.,][]{Bournaud2015,Orr2018} and treat our cold-dense regime is a proxy for molecular gas, our results are qualitatively in line with observations. Using a survey of CO(1-0) and CO(2-1) observations with the IRAM 30 meter telescope, \citet{Braine1993} finds that CO luminosity is enhanced in tidally disturbed galaxies. \citet{Combes1994} find similar results using a sample of IRAS-detected galaxies in binary systems. These authors also find a strong correlation between normalised CO and FIR luminosities, suggesting that enhancement in molecular gas content is responsible for the triggering of star formation in interacting galaxies. More recently, \citet{Violino2018} used IRAM 30-m CO(1-0) observations of 11 galaxies with close companions to show that interactions boost the {H$_{2}$} gas content in these systems. They find an increase of 0.4 dex, slightly higher than our excess of $\sim$18\% (Figure~\ref{fig:phases_vs_time_suites}, lower-left panel). \citet{Kaneko2013}, and \citet{Ueda2014} find similar results.

Although there is a broadly consistent qualitative agreement between our simulations, and observations, that the cold-dense gas component (traced by the molecular phase) is enhanced by the interaction, several caveats are necessary on both sides. For example, having defined the cold-dense component (Figure~\ref{fig:phase_diagram}), it is trivial for us to recover the mass in this regime. However, observations must convert an observed CO luminosity to a molecular gas mass, via the adoption of a conversion factor \citep[$\alpha_{CO}$,][]{Narayanan2013,Bolatto2013}. The value of $\alpha_{CO}$ is notoriously different in `normal' disk galaxies and ULIRGs, and the choice of an appropriate value for mergers is germane \citep{Narayanan2011,Narayanan2012}. Secondly, our experimental design, which involves evolving the interacting galaxies also in isolation, makes a quantitative assessment of the merger straightforward. Observations need to carefully construct their control samples in order to avoid biases.  Several recent papers \citep[e.g.,][Pan et al., submitted]{Violino2018}  have made important progress in recognizing these vital issues. On the simulation side, a more direct comparison would require full radiative-transfer calculations of the $\alpha_{CO}$ factor. Work in this direction is currently in progress by our group (Bueno et al., in prep). Our simulations also show that excess in the cold-dense gas mass content is mildly correlated with enhanced SFR (Pearson coefficient $=$ +0.31, Figure~\ref{fig:sfr_vs_mass_pearson}). Our fiducial run also show connection between these two quantities: interactions simultaneously boost star formation and form a reservoir of cold-dense gas (Figures~\ref{fig:sfr_vs_time_fid} and \ref{fig:phases_vs_time_fid}). However, even when star formation enhancement peters out, this reservoir remains. In general, this is accompanied by a slow recovery of a cool gas reservoir (Figure~\ref{fig:phases_vs_time_suites}). In other words, as in the case of cool gas \citep[and HI-gas in observations,][]{Ellison2018}, our simulations suggest that {\it the cold-dense (or cool) gas content is not exhausted (or expelled) after episodes of elevated star formation in interacting galaxies}. 

Lastly, we also break down our cold-dense reservoir by density and find that it is dominated by gas that is {\it too diffuse to form stars} -- i.e., only $\sim$0.15\% of the cold-dense gas content achieves densities exceeding 1000 cm$^{-3}$ (Figures~\ref{fig:molecular_fid} and Figure~\ref{fig:dense_molecular_suite}). Emission line ratios, such as $L_{\rm CO(3-2)}/L_{\rm CO(1-0)}$ and $L_{\rm HCN}/L_{\rm CO(1-0)}$ \citep{Gao2004,Lada2012,Kauffmann2017,Onus2018}, are good tracers of the fraction of cold-dense gas in the ultra-dense regime. \cite{Hopkins2013dense} shows that, when applied to isolated galaxies, our model -- 100\% star formation efficiency in self-gravitating (at resolution scale), self-shielded gas at densities exceeding 1000 cm$^{-3}$ -- is in reasonable agreement with observations. Additional measurements of these emission-line tracers have the potential to constrain the relative fraction of cold ultra-dense to cold-dense gas in interacting galaxies.  It is plausible that the turbulent nature of the ISM, especially during and soon after intense periods of star formation, may make the ISM stable against collapse, even in the presence of an abundant gas supply \citep{Alatalo2015,Smercina2018}. This can potentially explain why molecular gas reservoirs in galaxy pairs are enhanced.\\

\item {\bf Hot gas (connection to observations):}\\
Very few observations focused on the hot gas content in interacting galaxies have been performed to date. \citet{Henriksen1999} analyse a sample of 52 galaxy pairs from the Catalog of Paired Galaxies \citep{Karachentsev1972}, 25 of which are spiral-spiral systems. These authors find that, at a fixed luminosity in the B-band, X-ray luminosity in spiral-spiral pairs is enhanced relative to normal spirals. Similarly, using published publicly-available catalogues \citep{Vorontsov1959,Arp1966,Arp1987}, \citet{Casasola2004} find that X-ray luminosities from diffuse gas is higher in interacting galaxies than in normal ones. Our simulations predict that {\it close galaxy interactions elevate the hot gas mass component in galaxies significantly} (median $\sim$400\% -- Figure~\ref{fig:phases_vs_time_suites}, lower-right panel), in line with the aforementioned observations. \citet{Smith2018} use archived {\it Chandra} data to measure the diffuse X-ray luminosity, $L_{\rm X}$(gas), from a sample of 49 equal-mass galaxy pairs at various stages of interaction. They find that, for galaxies SFR $>$ 1 M$_{\odot}$yr$^{-1}$,  $L_{\rm X}$(gas)$/$SFR is not correlated with SFR or interacting stage. These authors do not report ratios of $L_{\rm X}$(gas)$/$SFR versus isolated controls, so a quantitative comparison is not possible. These authors claim that their results suggest that $\sim$2\% of the total energy output from supernovae and stellar winds converts into X-ray flux. We remind the reader that our initial conditions do not include a hot gas atmosphere, which might alter our predictions.\\

\end{itemize}

\subsection{Connection to Other Simulations}
\label{subsec:theories}

Many previous works investigate SFR enhancements and (total) gas depletion due to galaxy mergers using simulations, whereas fewer studies consider the effects of mergers on specific ISM regimes, usually because of limitations of the sub-grid models employed. We mention a few of the most relevant works here. Using two different numerical methods, \citet{DiMatteo2008} find that simulated low-redshift major mergers exhibit only modest SFR enhancements (rarely more than a factor of 5 during the coalescence-phase starbursts, and even less during the pre-coalescence phases). \citet{Cox2008} presents the results from a suite of simulations of low-redshift galaxy mergers to investigate the dependence of starburst strength on mass ratio. They find that the amplitude of the SFR enhancement at coalescence decreases sharply with increasing mass ratio, and even for the equal-mass mergers, the elevation in SFR during the pre-coalescence phase is less than a factor of a few. Various authors have also find that the strength of the SFR enhancement depends on the progenitor galaxies' gas fractions. For example, \citet{Fensch2017} argue that mergers are less efficient at high redshift, owing to galaxies typically having higher gas fractions than at low redshift.

The conclusions about how mergers enhance the SFR and alter the phase balance of the ISM are unfortunately sensitive to both resolution and the implementation of stellar feedback, thus complicating the interpretation of simulation results. \citet[][see also \citealt{Bournaud2008,Bournaud2011,Perret2014}]{Teyssier2010}
demonstrate that how well the clumpy structure of the ISM is resolved can have important consequences for how galaxy mergers drive starbursts. They argue that at sufficiently high resolution, interactions lead to enhanced fragmentation into cold clouds and that this enhanced fragmentation \citep[rather than nuclear inflows, as in previous works that employed lower resolution and effective equation of state ISM models,][]{Torrey2012,Patton2013,Moreno2015} is responsible for the SFR enhancement in mergers. \citet{Teyssier2010} find that galaxy-galaxy interactions cause the gas density probability density function (PDF) to shift to higher densities, more so in their higher-resolution simulation. This also occurs in our simulations, as evidenced by the enhancement in cold-dense gas -- and especially the even stronger enhancement in the ultra-dense cold gas content (bottom panel of Figure~\ref{fig:molecular_fid}). Since specific conclusions may be sensitive to the details of the stellar feedback model used in the simulations \citep[e.g.][]{Hopkins2013,Perret2014}; it would be consequently worthwhile to revisit the analysis of \citet{Teyssier2010} using our {\small FIRE-2} simulations in future work.

The above caveats aside, most simulations -- both low- and high-resolution and with more or less sophisticated models for stellar feedback -- find that mergers have relatively modest effects on the SFR and gas phase structure during most of the pre-coalescence phase, consistent with the results of our work.

\section{Conclusions}
\label{sec:conclusions}

We present results from a large suite of idealised (non-cosmological) high-resolution (parsec-scale) galaxy merger simulations constructed with the ``Feedback In Realistic Environments" (Version 2) model \citep[{\small FIRE-2},][]{FIRE2}, a feedback-regulated galaxy formation framework capable of capturing the multi-phase structure of the interstellar medium (ISM). Our suite consists of 24 simulations, including sets of (nearly) prograde, polar and retrograde mergers, covering a broad range of spatial extents and durations (Figures~\ref{fig:pericoverage} and \ref{fig:orbitalcoverage}). Our goal is to investigate how various temperature-density regimes of the ISM fuel and drain one another as galaxy-galaxy interactions induce episodes of enhanced star formation. We consider the following ISM regimes: {\it warm}, {\it cool}, {\it cold-dense} and {\it hot}, motivated by the observed ionised, atomic, molecular and hot gas phases in galaxies. See Table~\ref{table:phases} and Figure~\ref{fig:phase_diagram} for details.

This papers focuses on the `galaxy-pair period' of merging, defined as the period of time between first and second pericentric passage. Our main results are as follows:
\begin{itemize}
\item {\bf Interactions enhance star formation in galaxies}\\ (average enhancement $=$ 30\%).
\item {\bf Interactions do not alter the cool gas content}\\ (average mass excess  $=$ 4\%). 
\item {\bf Interactions build up a cold-dense gas reservoir}\\ (average mass excess  $=$ 18\%). 
\item {\bf Interactions significantly elevate the ultra-dense gas content} (average mass excess  $=$ 240\%).
\item {\bf Most cold-dense gas is too diffuse to create stars}\\ (only $\sim$0.15\% has densities $>$ 1000 cm$^{-3}$).
\end{itemize} 
The first point is qualitatively in line with observations \citep{Woods2006,SloanClosePairs,Lambas2012,Ellison2013, Patton2013,Scott2014} and predictions by earlier simulations \citep{MH96, DiMatteo2007, DiMatteo2008, Teyssier2010, Renaud2013,Hayward2014, Moreno2015}. Regarding the second and third points: one might na\"{i}vely expect intense star-forming episodes to either consume the cold-dense gas content (and deplete the cool gas as this transforms into cold-dense gas), or expel this gas via feedback produced by new stars. Instead, our feedback-regulated model allows for the recovery of cool gas after brief depletion -- and keeps most of the cold-dense gas in a diffuse state (fifth bullet point), thus maintaining the presence of a cold-dense gas reservoir as star formation unfolds, in line with observations of interacting galaxies with molecular gas mass excess \citep{Violino2018}.

Regarding the other two regimes in the ISM, we find:
\begin{itemize}
\item {\bf Interactions enhance the depletion of warm gas}\\ (average suppression $=$ 11\%).
\item {\bf Interactions strongly elevate the hot gas content}\\ (average mass excess $=$ 390\%). 
\end{itemize} 
In particular, X-ray observations suggest that galaxy encounters augment the hot gas mass budget \citep{Henriksen1999,Casasola2004}. However,  because our simulations do not include initial hot gas atmospheres, our results are only indicative of partial contributions to the hot gas budget by heating of the galactic ISM.

Future work includes: (1) investigations of how our results depend on the details of orbital configurations and mass ratios; (2) assessment of the spatial of distribution star formation and the structure of the ISM; (3) perform radiative-transfer calculations, coupled with chemical network solvers, to construct mock surveys for better comparisons with existing observations; (4) deeper investigations of the physical mechanisms driving the fate of cold-gas gas, which might either stay moderately-dense, become ultra-dense and be consumed by star formation, or be cycled back into the warm or hot gas budget by stellar feedback; (5) and conduct similar studies at and after coalescence.

\section*{Acknowledgements} 

The computations in this paper were run on the Odyssey cluster supported by the FAS Division of Science, Research Computing Group at Harvard University. Support for JM is provided by the NSF (AST Award Number 1516374), and by the Harvard Institute for Theory and Computation, through their Visiting Scholars Program. The Flatiron Institute is supported by the Simons Foundation. DK is supported by the NSF (AST Award Number 1715101), and by a Cottrell Scholar Award from the Research Corporation for Science Advancement. We thank Jillian Scudder and George Privon for illuminating discussions, Jason Brown for helping us create our galaxy mergers website, and the two anonymous reviewers, whose insightful suggestions improved the quality of this paper. JM thanks Dra. Nicole Cabrera Salazar (Movement Consulting) for co-mentoring a large group of outstanding undergraduate students of colour, including MB. We honour the invaluable labour of the maintenance and clerical staff at our institutions, whose contributions make our scientific discoveries a reality. This research was conducted on Tongva-Gabrielino Indigenous land.

\bibliographystyle{mnras}
\bibliography{paper}

\end{document}